\documentclass{IEEEtran}
\usepackage{cite}
\usepackage{amsmath,amssymb,amsfonts}
\usepackage{graphicx}
\usepackage{subcaption} %
\usepackage{textcomp,nicefrac}

\usepackage{longtable} 
\usepackage{ltablex} 

\DeclareRobustCommand{\mychar}{%
  \begingroup\normalfont
    \includegraphics[height=0.35cm,width=0.35cm]{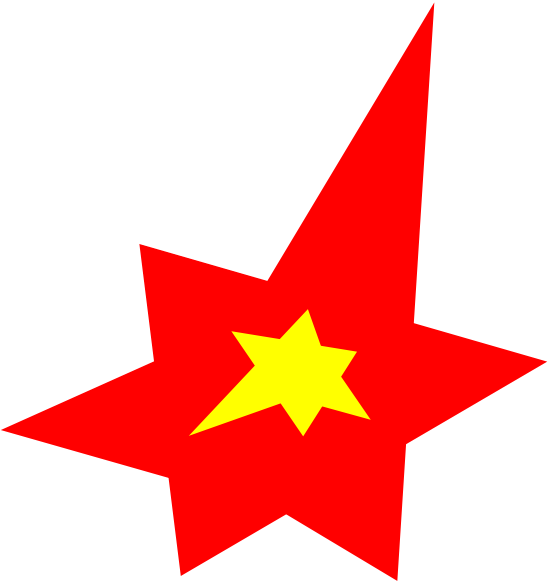}%
  \endgroup
}

\def\BibTeX{{\rm B\kern-.05em{\sc i\kern-.025em b}\kern-.08em
T\kern-.1667em\lower.7ex\hbox{E}\kern-.125emX}}
\begin{document}
\title{Needs, trends, and advances in scintillators for radiographic imaging and tomography}
\author{Zhehui Wang, \IEEEmembership{Senior Member, IEEE}, Christophe Dujardin, Matthew S. Freeman, Amanda E. Gehring, James F. Hunter, Paul Lecoq, \IEEEmembership{Fellow, IEEE}, Wei Liu, 
Charles L. Melcher, \IEEEmembership{Life Fellow, IEEE}, C.  L. Morris, Martin Nikl, Ghanshyam Pilania, Reeju Pokharel, Daniel G. Robertson, Daniel J. Rutstrom, \IEEEmembership{Student Member, IEEE}, Sky K. Sjue, Anton S. Tremsin,  
S. A. Watson, Brenden W. Wiggins, Nicola M. Winch, and Mariya Zhuravleva, \IEEEmembership{Member, IEEE} 
\thanks{Submitted to IEEE TNS for review on Dec. xx, 2022 as a part of the special issue for the 16th International Conference on Scintillating Materials \& their Applications (SCINT22), September 19-23, 2022, Santa Fe, NM, USA. 
``This work, Los Alamos report number LA-UR-22-32994, was supported in part by the U.S. Department of
Energy under the Contract No. 89233218CNA000001.'' 
Z. Wang, M. S. Freeman, A. E. Gehring, J. F. Hunter, C. L. Morris, G. Pilania, R. Pokharel, S. K. Sjue, 
S. A. Watson, B. W. Wiggins  and N. M.  Winch are with the Los Alamos National Laboratory, Los Alamos, NM 87545, USA (e-mails: zwang@lanl.gov, msf@lanl.gov, gehring@lanl.gov, jhunter@lanl.gov, cmorris@lanl.gov, gpilania@lanl.gov, reeju@lanl.gov, sjue@lanl.gov, 
scottw@lanl.gov, bww@lanl.gov, nwinch@lanl.gov).}
\thanks{C. Dujardin is with Institut Lumière Matière
UMR5306 CNRS
Université Claude Bernard Lyon1, France (email:christophe.dujardin@univ-lyon1.fr)}
\thanks{P. Lecoq is with the European Organization
for Nuclear Research (CERN), CH-1211 Geneva, Switzerland (e-mail: Paul.Lecoq@cern.ch).}
\thanks{W. Liu and D. Robertson are with Mayo Clinic, Phoenix, AZ 85054  (e-mails: Liu.Wei@mayo.edu, Robertson.Daniel@mayo.edu).}
\thanks{M. Nikl is with Institute of Physics, AS CR
Cukrovarnicka 10
16200 Prague, Czech Republic (email: nikl@fzu.cz).}
\thanks{D. Rutstrom, C. L. Melcher, and M. Zhuravleva are with
the University of Tennessee, Knoxville, TN 37996, USA (e-mails: drutstro@vols.utk.edu, cmelcher@utk.edu, mzhuravl@utk.edu).}
\thanks{A. S. Tremsin is with  the Space Sciences Laboratory, University of California at Berkeley, Berkely, CA 94720, USA (e-mail: astr@berkeley.edu).}
}

\maketitle

\begin{abstract}
Scintillators are important materials for radiographic imaging and tomography (RadIT), when ionizing radiations such as X-rays, energetic charged particles or neutrons are used to reveal optically opaque internal structures of matter. Since its discovery and invention by R\"ontgen, RadIT now come in many forms or modalities such as phase contrast X-ray imaging, coherent X-ray diffractive imaging, high-energy X- and $\gamma-$ray radiography at above 1 MeV, X-ray computed tomography (CT), proton imaging and tomography (IT), neutron IT, positron emission tomography (PET), high-energy electron radiography, muon tomography, {\it etc}. High spatial, temporal resolution, high sensitivity, and radiation hardness, among others, are common metrics for RadIT performance, which are enabled by, in addition to scintillators, advances in particle sources especially high-luminosity accelerators and high-power lasers, photodetectors especially complementary metal-oxide-semiconductor (CMOS) pixelated sensor arrays, and lately data science. Medical imaging, non-destructive testing, nuclear safety and safeguards are traditional applications for RadIT. Examples of rapidly growing or emerging applications include space, additive manufacturing (AM), machine vision, and virtual reality or `metaverse'. Scintillator metrics such as light yield, decay time and radiation hardness are discussed in light of RadIT metrics. More than 160 kinds of scintillators and applications are presented during the SCINT22 conference. Some new trends include inorganic and organic scintillator composites or heterostructures, liquid phase synthesis of perovskites and single-crystal micrometer-thick films, use of multi-physics models and data science lately to guide scintillator development and discovery, structural innovations such as photonic crystals, nano-scintillators enhanced by the Purcell effect, heterostructural scintillating fibers, and multilayer configurations. Plenty of new opportunities exist in optimization of RadIT performance with reduced radiation dose, data-driven measurements, photon/particle counting and tracking methods supplementing time-integrated measurements, multimodal RadIT, and novel applications of RadIT for scintillator development and discovery.
\end{abstract}

\begin{IEEEkeywords}
Data-driven discovery, dose, fast timing, high energy physics (HEP), inorganic scintillator, ionizing radiation, multimodal imaging, nanomaterial, photodetectors, photonic crystal, Purcell effect, radiographic imaging, radiographic tomography, radiography, scintillation, structured scintillators.
\end{IEEEkeywords}

\section{Introduction}
\label{sec:introduction}
\IEEEPARstart{S}{cintillators} are important materials for radiographic imaging and tomography (RadIT), when ionizing radiations such as X-rays, energetic charged particles (e.g., electrons, positrons, protons, and $\alpha$-  [$^4$He] particles), neutrons, and others are used to penetrate through optically opaque objections and to reveal their internal material structures. RadIT started with R\"ontgen's discovery of X-rays and invention of X-ray radiography in 1895~\cite{Ron:1896}, which predated the discoveries of electrons and atomic nuclei as the elementary building blocks of matter. By the 1930s, quantum mechanical interpretation of atomic structures and fundamental forces paved the way towards understanding of material properties such as crystal or periodic lattice structures, polycrystalline structures, high-entropy materials, defects, and phase transition~\cite{BLP:1971}, and in principle, also provided the theoretical framework to interpret X-ray radiographs resulting from X-ray interactions with the electronic structures of materials. Nuclear interactions with X-rays, except for X-ray energies above 1 MeV, are usually ignored due to the predominant mass of the nuclei over the mass of an electron. 

 Since their initial use by R\"ontgen, Crookes and other pioneers, there is now an enormous number of scintillators to choose from for X-rays, RadIT and other applications. It is hardly exaggerating to say that a scintillator can be found, if not already, in each phase and form of matter. More than 160 kinds of scintillators and their applications were reported in the 16th International  Conference on Scintillating Materials \& their Applications, Santa Fe, NM, Sept. 19-23, 2022 (`SCINT22')  conference, see Table.~\ref{scint:tab} in the appendix for a summary. It is clear from the table that the majority of the scintillators belong to inorganic chemicals. The SCINT conference series dated back to 1992 has accordingly addressed the inorganic scintillator science and technology predominantly~\cite{DAB:2018}. Some additional reviews on scintillators relevant to ionizing radiation detection, high energy physics (HEP), medical imaging can be found, for example, in~\cite{Moses:2001, Weber:2002, NY:2015, Yana:2018, KLM:2021}. 
 
 Starting with the four common phases of matter, there are solid scintillators, liquid scintillators, gas scintillators and plasma scintillators.  Stand-alone, chemically stable, solid-state scintillators are by far the most convenient to use. Scintillators have also been classified according to their elemental composition, namely organic scintillators, inorganic scintillators, oxides, garnets, halids, rare-earth scintillators, {\it etc.}  In addition to doped halides such as NaI:Tl, CsI:Tl, rare earth inorganic scintillators such as cerium-doped lutetium-yttrium oxyorthosilicate (LYSO), garnets such as cerium-doped lutetium aluminium garnet  (LuAG) are among the popular scintillator choices in X-ray and RadIT applications today. New formulations of rare-earth doping of inorganic scintillators using europium (Eu), praseodymium (Pr), ytterbium (Yb), and others remain an exciting discovery frontier for faster scintillation decay time, minimal afterglow, higher light yield in the desired wavelengths, more flexible emission wavelength tuning, and other performance improvements. 
 
 Scintillators may still be classified according to their material structures, such as single-crystal scintillators, polycrystal scintillators, perovskite scintillators, glass scintillators, ceramic scintillators, plastic scintillators, hetero-structured or composite scintillators, nanoscintillators, {\it etc}. In addition to new scintillator discoveries, a new trend is to combine existing organic and inorganic scintillators in the same system, driven by `higher information yield'  from a radiation field such as particle identification (e.g., neutron/$\gamma$-ray discrimination), higher energy resolution, finer spatial resolution in imaging and particle tracking, pico-second (ps) time and/or timing resolution in time-of-flight (TOF), up to 4$\pi$ detection solid angle, larger detection volume, and lower cost. With the emergence of liquid- or solution-based synthesis of scintillators, and additive manufacturing (AM) technology, scintillators may also be classified according to their synthesis and manufacturing methods. Liquid-phase synthesis of perovskites have enjoyed some phenomenal success in the recent years~\cite{ZCB:2021}. For example, lead-free low dimensional perovskite-like metal halides such as ternary copper(I) halides were found to have very high photoluminescence quantum yields, $\sim 90$ thousand photons/MeV (kph/MeV), and large Stokes shift,  in addition to their photophysical properties and stability~\cite{YYN:2019}. CsPbBr\textsubscript{3} reported a light yield
of 50 kph/MeV and 1 ns decay time  at 7 K~\cite{MKK:2020}.  AM technology for scintillator fabrication remains at its infancy. Advances are still needed to 3D print some most common polymer bases such as polystyrene and polyvinyltoluene (PVT). 
 
 Scintillator light yield and X-ray stopping power (or X-ray attenuation mean free path equivalently) are the first two material properties to be considered when selecting a scintillator for X-ray detection, including X-ray imaging and tomography (IT). Light yield is a measure of the number of optical photons per unit X-ray energy (1 MeV  or 1 keV, for example) deposited in the scintillator. Stopping power is a measure of scintillator thickness for effective attenuation of X-rays. Both the light yield and stopping power affect the X-ray detection efficiency. Meanwhile, the growing adoption of and continuous advances in X-ray radiography technology and its variants such as X-ray microscopy, X-ray phase contrast imaging,  X-ray diffractive imaging, X-ray tomography, X-ray  ptychography motivated new scintillator discoveries and development~\cite{Snig:1995, Wilkins:1996,  MTI:1996, TCZ1998, HPS:2000, HHM:2012, Miao:2015}, since light yield and stopping power alone are no longer sufficient for the diversified applications. In Sec.~\ref{sec:pr} below, we shall expand upon the discussion of different aspects of X-ray IT, such as spatial and temporal resolutions, that require consideration of other scintillator properties besides the light yield and the attenuation length. 
 
An overview of scintillator applications in RadIT is given in Sec.~\ref{sec:css}, with additional references included for interested readers and developers. Besides X-rays, different particle IT such as proton IT, neutron IT,  electron IT, positron emission tomography (PET), have also been invented, which have not only greatly enriched the field of RadIT, but also motivated scintillator development and new scintillator properties due to the different particle interaction physics.  Some highlights of the recent scintillator development are given in Sec.~\ref{scin:new}. Table.~\ref{scint:tab} in the appendix gives more information including the contacts. Scintillator-based RadIT are also closely correlated with advances in light and particle sources such as particle accelerators~\cite{Ekd:2001,PS:xx}, detector technologies, esp. 2D photodetectors and more recently data science~\cite{Wang:2022}. Besides topics such as scintillators for space applications, scintillator development and discovery are poised to enter a new phase through big data mining, multiphysics models, new experimental information derived from automated, high throughput,  and in-situ RadIT, including multi-modal RadIT, as discussed in Sec.~\ref{sec:TO}.  It is clear that the interdisciplinary marriage of SCINT and RadIT science and technology offers many exciting opportunities for innovation in the coming decade, as summarized at the end in Sec.~\ref{sec:conc}. 


\section{RadIT and Scintillator requirements \label{sec:pr}}

\begin{figure}[htb!]
\centering\includegraphics[width=0.9\linewidth]{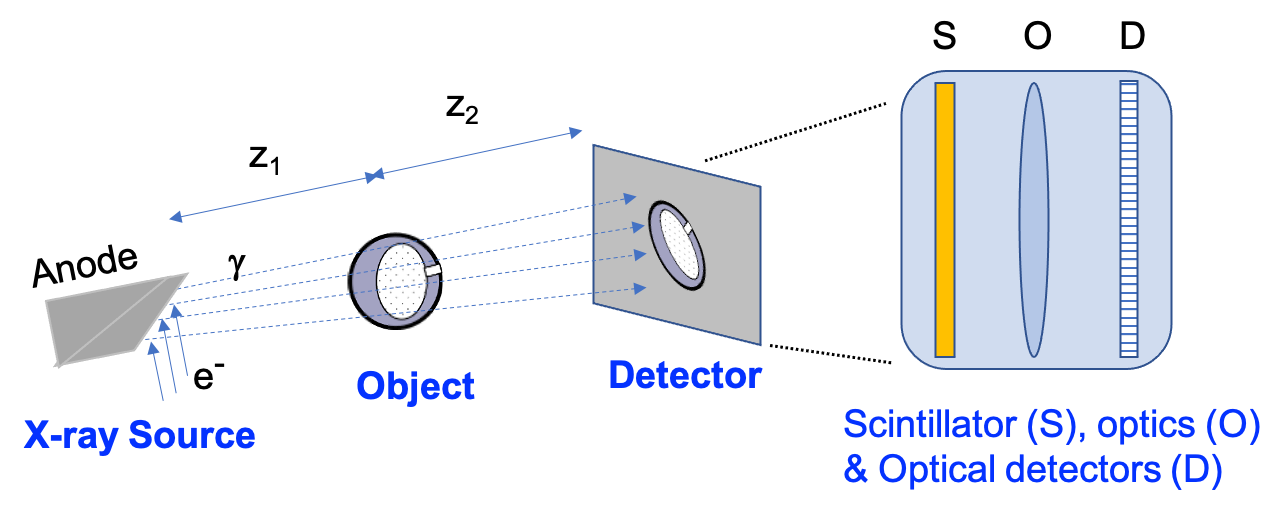}
\caption{The basic setup of X-ray radiography consists of an X-ray source, the object, and a detector. Scintillators, together with the optics and optical array detectors such as a CCD camera, are often used for radiography using high-energy X-rays. X-ray image size may be estimated with geometric or ray optics model. The magnification, for example, is given by ($z_1+z_2)/z_2$ for a point source. $z_1$ is the distance between the point source and the object, $z_2$ the distance between the object and the detector. Image blurs usually occur due to the finite source size, X-ray scattering, finite scintillator thickness, optical blur and finite detector pixel size.}
\label{fig:rad1}
\end{figure}

The most common setup of X-ray radiography has essentially remained the same as in R\"ontgen's original work~\cite{Ron:1896}, as illustrated in Fig.~\ref{fig:rad1}, which consists of an X-ray source, the object to be radiographed, and a detector that captures the two-dimensional (2D) projection of the object. R\"ontgen's experiments used two independent methods to detect X-rays: 1.) Barium platinocyanide [BaPt(CN)$_4$ or C$_4$BaN$_4$Pt] fluorescent screens and others such as calcium sulfide (CaS) to convert X-rays into visible light emissions to be seen by eyes, and 2.) Photographic plates to record the X-ray radiographs though X-ray-induced chemistry. In addition to a scintillator, a modern detector assembly may include optics (optional but preferred) and photodetectors such as a CCD or a CMOS camera for digital X-ray radiography.

In X-ray radiography, X-ray attenuation is commonly used to generate images and contrast in objects. The transmitted X-ray intensity through an object ($I$) that reaches a scintillator is attenuated from the source intensity ($I_0$) by the line integrated density or areal density, see for example~\cite{xdb:2009}, 
\begin{equation}
I = I_0 \exp (- \sum_i \int_0^L dl \rho w_i \sigma_i / A_i M_0).
\label{eq:att1}
\end{equation} 
Here the object thickness that X-ray traverses is $L$. The object is a material compound of multiple elements represented by $i$. $\rho$ is the mass density of the compound. $w_i = g_iA_i/\sum g_iA_i$ is the fraction by weight of the $i$th atom in the compound molecule.  $g_i$ is the number of $i$th atoms in the compound molecule. $A_i$ is the corresponding atomic mass number. $M_0$ is the atomic mass unit. $\sigma_i$ the total X-ray attenuation cross section corresponds to the $i$th element. $\sigma_i/ A_i M_0 \equiv \mu_i $ is also called the mass attenuation coefficient, which varies with the type of element in the periodic table but does not dependent on the density. The integral formula Eq.~(\ref{eq:att1}) is also applicable to materials with a mixture of compounds or position-dependent mass density $\rho = \rho(l)$, which we shall not elaborate further for algebraic simplicity. For materials with a uniform density $\rho$, Eq.~(\ref{eq:att1}) reduces to $\displaystyle{I = I_0 \exp( - \sum_i L \rho w_i \sigma_i / A_i M_0)}$. 

Neglecting smaller probability events such as photonuclear processes, the total X-ray attenuation cross section in most imaging and tomography (IT) settings can be approximated by a sum of four cross sections, 
\begin{equation}
\sigma_i  = \sigma_i^{pe} + \sigma_i^{coh} + \sigma_i^{inc}  + \sigma_i^{pair}.
\end{equation} 
Here $\sigma_i^{pe}$, $\sigma_i^{coh}$, $\sigma_i^{inc}$, and $\sigma_i^{pair}$ are photoelectric absorption, coherent or Rayleigh scattering, incoherent or Compton scattering, and electron-positron pair production cross section, respectively. For each element $Z_i$ in the periodic table, the photoelectric cross section dominates at low X-ray energies up to a threshold ($\lesssim$ 0.1 MeV, lower thresholds for low-Z atoms such as hydrogen and carbon), $\displaystyle{\sigma_i^{pe} \propto \frac{1}{A_i} \frac{Z_i^\alpha}{E^\beta}} \propto \frac{Z_i^{\alpha -1}}{E^\beta}$, with $\alpha \sim 4-6$ and $\beta \sim 3-3.5$~\cite{Lang:2017,BLP:1971}.  The X-ray photoelectric attenuation cross section is a strong function of the atomic number $Z_i$ and decreases rapidly with the X-ray energy ($E$). At above $\sim$ 0.1 MeV and depending on $Z_i$, the incoherent scattering cross section becomes dominant until the electron-position pair production becomes important. The electron-positron pair production
threshold is at twice the electron mass energy $2m_ec^2 = 1.022$ MeV. The pair-production becomes significant only above $\sim 3$ MeV and depends on $Z_i$ linearly~\cite{Att:2008}. 

\begin{figure}[htb!]
\centering\includegraphics[width=0.9\linewidth]{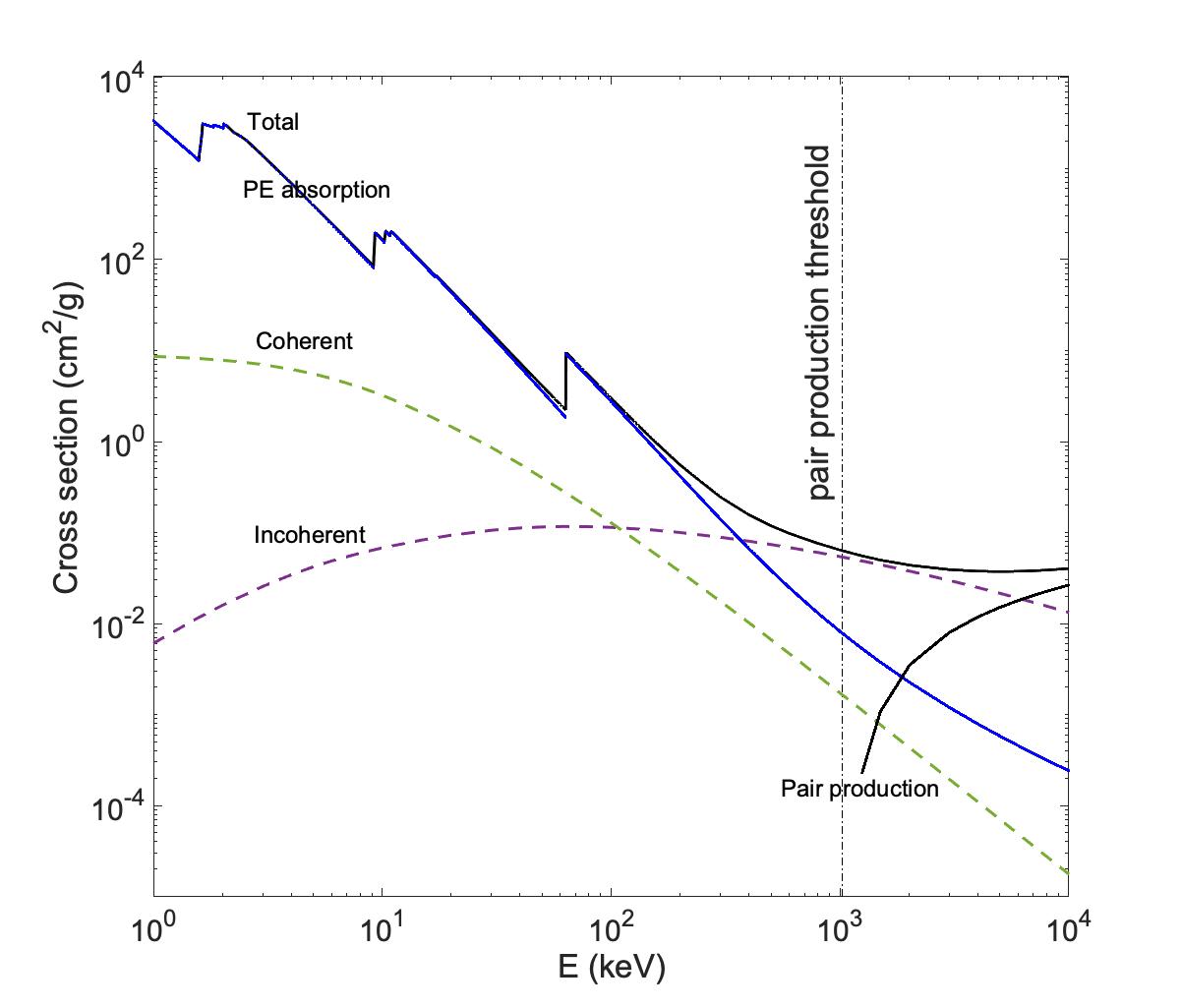}
\caption{Energy-dependent X-ray cross sections in LYSO with an atomic number ratio of Lu:Y:Si:O = 2(1-$x$):2$x$:1:5 and $x$=0.075. The total cross section is a sum of photoelectric (PE) absorption, coherent scattering, incoherent scattering, and electron-positron pair production. The data are from the NIST/XCOM database.}
\label{fig:LYSO}
\end{figure}

In short, X-rays  primarily interact with the electrons in materials except for energies above a few MeV, see Fig.~\ref{fig:LYSO} for an example of energy-dependent X-ray cross sections in lutetium-yttrium oxyorthosilicate (LYSO).  The total X-ray attenuation cross section is a sum of photoelectric (PE) absorption, coherent scattering, incoherent scattering, and electron-positron pair production. 

The above ray-tracing or `particle' model for X-ray attenuation in matter, Eq.~(\ref{eq:att1}), are complicated by X-ray refraction, X-ray diffraction physics or `wave' properties of X-rays~\cite{Warren:1969}. Coherent scattering of x-ray photons leads to X-ray diffraction, and tends to redistribute the X-ray flux and intensities in the forward and backward direction with respect to the X-ray beam propagation. Incoherent scattering can spread X-rays into  the 4$\pi$ solid angle according to the well-known Klein-Nishina formula. Diffraction and scattering therefore can complicate interpretation of absorption-based X-ray radiography through reduced image contrast and signal-to-noise (SNR) ratio~\cite{LJ:1998}. Another limitation in absorption-based X-ray IT methods is the low contrast for low-Z materials such as biological objects when X-rays with energies above 20 keV are used~\cite{Wilkins:1996}.

\begin{figure}[htb!]
\centering\includegraphics[width=1.0\linewidth]{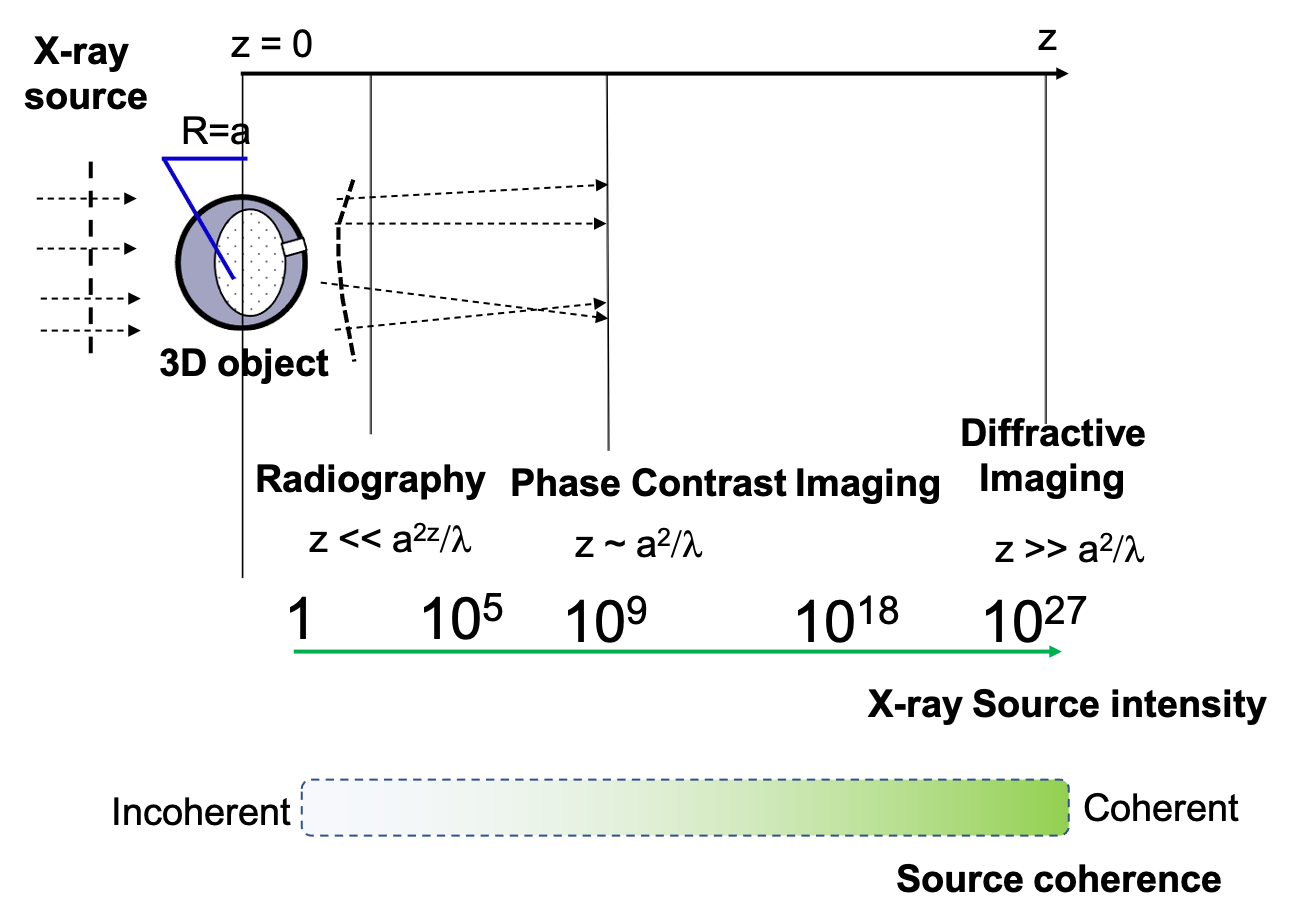}
\caption{Several lensless X-ray radiography and imaging modalities depends on the X-ray source properties (coherent and incoherent), X-ray interactions (absorption, scattering) with the object, post-interaction X-ray propagation (interference and diffraction), and the X-ray detector distance to the object.}
\label{fig:config}
\end{figure}

Meanwhile, X-ray diffraction and interference can also be used to X-ray imaging when a sufficiently high-intensity monochromatic X-ray source such as the third-generation synchrotrons is available~\cite{Snig:1995,TCZ1998}. While the modern X-ray tubes using rotating anodes can deliver 10$^5$ times the X-ray flux available to R\"ontgen, The third-generation synchrotron X-ray sources can deliver 10$^{18}$ times the X-ray flux and keep improving. The high X-ray fluxes and associated X-ray doses are now causing significant scintillator heating and potential reduce the lifetime of scintillator due to radiation damage. X-ray phase contrast imaging (PCI) has seen great success using synchrotrons, see Fig.~\ref{fig:config}. Hard X-ray PCI is also effective for low-Z materials, in part due to the fact that X–ray phase shift cross section can be a thousand times larger than the X–ray absorption cross section for light elements such as hydrogen, carbon, nitrogen and oxygen~\cite{MTI:1996}. In X-ray PCI, the distance between the object and detector satisfies $z \sim \displaystyle{a^2/\lambda}$, which corresponds to a distance $z \sim 2$ m for an object resolution $a \sim 10$ $\mu$m at the X-ray energy of 25 keV. Dynamic X-ray PCI, or movies of X-ray PCI images, are also possible due to the repetitive x-ray pulses at 10s of ns rates. X-ray free electron lasers (FELs) are currently the most intense, coherent laboratory X-ray sources for coherent imaging and related applications~\cite{PS:xx}. X-ray coherent diffractive imaging~\cite{Miao:2015}, X-ray Bragg projection ptychography~\cite{HHM:2012} from XFELs are used to image non-periodic material structures and lattice dynamics with nm-resolution. X-ray computed tomography (CT), introduced as a method to  reconstruct three-dimensional (3D) models from 2D radiographs of many different angles, was introduced in the 1970s. X-ray diffraction computed tomography was introduced based on the coherent scattering~\cite{HKN:1987}. In addition to continuous improvements in resolution, another trend in X-ray IT is to improve temporal resolution and reduces the number of 2D projections towards time-resolved CT.

It should be mentioned that table-top microfocus X-ray sources with a spot size less than 100 $\mu$m, which have high spatial coherence at the object location but not necessarily monochromatic in energies, have also been successfully used for X-ray PCI~\cite{Wilkins:1996}. This is made possible in part by the use of the high-performance detectors including the use of scintillators in conjunction with high-resolution pixelated cameras~\cite{Lumpkin:2018}. In short, further advances in X-ray IT critically depends on advances in scintillators light yield and other metrics, which we elaborate in the next section.

\subsection{RadIT Metrics}

\subsubsection{Spatial resolution}

Spatial resolution measures the ability to differentiate {\it the smallest} spatial variations in density and other physical quantities such as the temperature, velocity, lattice structure or phase of an object~\cite{adj:1997}. Similar to optical imaging, a point-spread function (PSF) can be used to describe the finite resolution or the image blur in X-ray IT and RadIT. A PSF may be interpreted mathematically as a two-dimensional (in imaging) or three-dimensional (in tomography) intensity distribution as the result of the blurring of a point intensity, which is described by a 2D or 3D Dirac delta function~\cite{DS:1974}. PSF is most useful for the analysis of a linear information system like an image, which may be treated mathematically as a linear superposition of intensities, or the convolution of PSF with the unblurred image. 

Image blur occurs due to a number of reasons: the finite X-ray source size or equivalent in particle-based IT, dispersion of the optics or propagation in lensless imaging, diffraction and scattering of the X-rays by the object and the scintillator, motion of the object in a dynamic experiment such as material compression and deformation due to implosion, vibration of the scintillator and the instrument, and different mechanisms of instrumentation broadening such as the isotropic emission of scintillating light in a bulk scintillator or the charge sharing among neighboring pixels in CCD or CMOS cameras. Assuming that each source of blurring is a mutually independent Gaussian process, the overall resolution ($\delta$) may be estimated as a sum of blurring widths ($\delta_k$),
\begin{equation}
\delta = \sqrt{\sum_k \delta_k^2}
\end{equation}
for each of the blurring mechanism $k$.
Micrometer spatial resolution has now been routinely obtained for small ($\sim$ 1 mm$^3$) objects by using synchrotrons and scintillator cameras~\cite{MK:2006}. Resolution also depends on the object size, the wavelength of the X-ray or the energy of the particle in a particle IT such as neutron or proton radiography, and the magnification. Sub-micrometer resolution down to atomic dimension is possible in X-ray microscopy, coherent diffractive imaging for small objects less than 10$^6$ $\mu$m$^3$ in volume. In medical and industrial X-ray CT, non-destructive imaging and tomography of thicker objects ($>$ 10$^3$ cm$^3$ in volume) are achieved with a compromised resolution at  $\delta > 100$ $\mu$m. 

\begin{figure}[ht!]
\centering\includegraphics[width=0.9\linewidth]{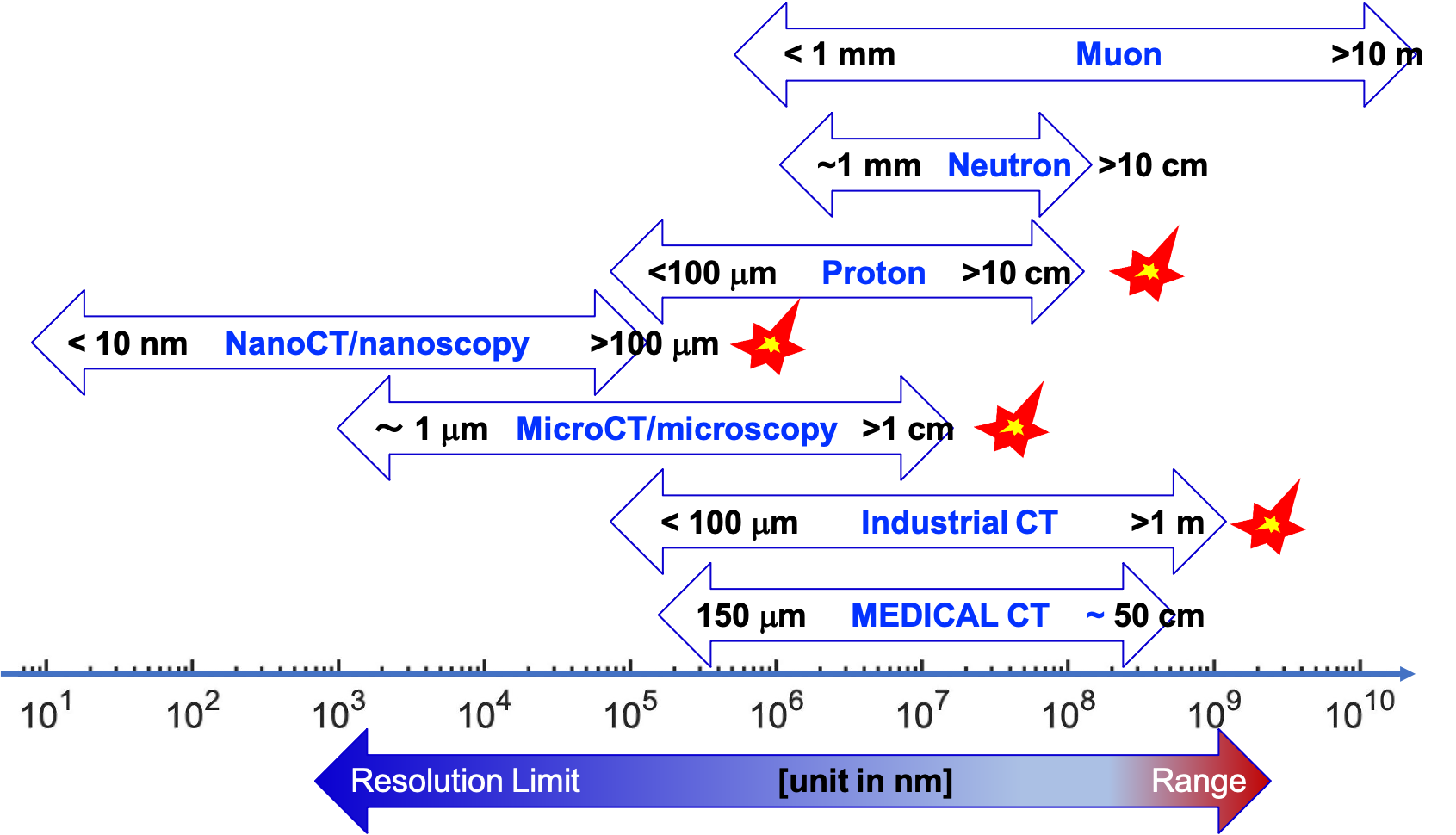}
\caption{A comparison of spatial resolution and range or the size of the field-of-view (FOV) for different RadIT modalities. \mychar symbolizes flash or single-pulse (or a very few pulses) time-resolved radiographic capabilities. }
\label{fig:ResRan}
\end{figure}

A comparative summary of the spatial resolution ($\delta$), or voxel resolution in tomography for different forms of RadIT is given in Fig.~\ref{fig:ResRan}. A previous consideration was given to single photon emission computed tomography (SPECT)~\cite{Meikle:2005}. Depending on the interaction cross sections, the radiation dose $\mathcal{D}$ scales with the spatial resolution $\delta$ as $\mathcal{D} \propto 1/\delta^k$. For X-ray radiation-damage-limited dose, $k$ was found to be 4~\cite{Howells:2009}. Similar consideration may be used for protons, neutrons, muons and others. The X-ray dose has to increase by a factor of 10$^4$ to
maintain a constant signal/noise ratio with a factor of 10 improvement in resolution; for example, imaging at 50 $\mu$m resolution requires 10$^4$ higher
dose than at 500 $\mu$m resolution~\cite{CRA:2012}. Therefore, the dose constraint poses significant limitation in the achievable spatial resolution in practice in X-ray IT, especially in in-vivo and in-situ medical IT. For neutron, proton, and muon IT, as well as time-resolved X-ray IT, the spatial resolution is typically limited by the radiation source intensity.

Scintillators become a significant contributor to the image blur and PSF in high-energy photon IT and neutron IT. In MeV photon radiography, tens of mm thick scintillator such as LSO:Ce are needed for efficient detection of the high-energy photons. In fast neutron imaging, mm and thicker scintillator are also needed. Example calculations of are given in Sec.~\ref{sec:WWH} below for modulation transfer functions (MTF). MTF is a frequency space representation or Fourier transform of the line spread function (LSF).  LSF is related to a 2D PSF through an integration that  reduces the dimension by one~\cite{DS:1974}. Due to the complex interplay of different mechanisms for PSF, Monte Carlo simulations are often used for image analysis including PSF, LSF and MTF~\cite{DS:1974,WGG:1999, WWH:2017}. Some factors such as visible light transport in scintillators, electronics in digital cameras, which are usually ignored in the Monte Carlo (MC) calculations, require experimental data inputs due to the lack of accurate models.

Several methods are used to improve the spatial resolution when thick scintillators are used. Traditionally, columnar scintillators and segmented scintillator are used to confine the optical emissions along the direction of X-ray and particle beams~\cite{Winch:2018}. Alternatively, the multilayer thin-scintillator configuration is also useful to improve the efficiency without compromising the spatial resolution by minimizing the optical photon pathlength before reaching the photodetectors~\cite{Wang:2015}. Some recent work on using micro-meter-thick scintillators are summarized in Sec.~\ref{sec:duj} below. More recently, photonic crystal scintillators, or structured scintillators with features comparable to and less than optical wavelengths are showing promising results to guide optical emissions to the detectors with minimal loss and spatial spread~\cite{Firat:2021}.

\subsubsection{Field of View \& Depth of Field} Centimeter and larger objects are frequently encountered in RadIT which require commensurate large field of view and depth of field. As illustrated in Fig.~\ref{fig:ResRan}, Medical CTs are designed to fit a human body. Industrial CTs are used for quick (seconds to minutes on many occasions) and non-invasive inspection of cargo containers, airport security, non-destructive testing in aerospace industry, automobile industry and lately additive manufacturing.  Cosmic-ray muon tomography was used to look for hidden chambers in the pyramid and recently been used to inspection of damaged nuclear reactors. Compton scattering of X-rays, nuclear scattering of neutrons, and multiple-Coulomb scattering of charged particle beams can spread the primary ionizing beams and secondary particles up to the 4$\pi$ solid angle.  

There are a number of practical limitations to cover the full field of view, which ideally should intercept all the detectable signals in the the 4$\pi$ solid angle. For a 100 cm $\times$ 100 cm field of view, a spatial resolution of $\delta = $ 1 mm would require a pixelated detector with $N_p = $ 10$^6$ pixels ($N_p$ symbolizes the number of pixels of a pixelated sensor such as a CMOS or CCD camera). To resolve the point-spread function, a factor 3 decrease in pixel size would increase the number of pixel by a factor of 9. The spatial dynamic range, which is equivalent to $\sqrt{N_p}$, is traditionally limited by the availability of large-format imaging sensors and image sensor cost. Recent advances in CMOS sensors at lower cost provide opportunities for billion pixel and large format RadIT camera designs~\cite{Wang:2020}. Monolithic scintillators of 1 m in size are rare due to for example the crystal growth cost. Segmented scintillators (see Sec.~\ref{sec:WWH}) and tiled scintillators (see Sec.~\ref{sec:MF}) are therefore often used for large field of view. Due to the refractive index mismatch, scintillator light spread at the tile boundaries can lead to undesirable artifacts.  

\subsubsection{Time or temporal resolution} R\"ontgen's first X-ray radiograph was static. The nature is fundamentally dynamic and in perpetual motion according to Heisenberg's uncertainty principle. Time-resolved RadIT methods have been increasingly used to examine the changes or dynamics of materials since R\"ontgen's pioneer work. In high-speed imaging such as GHz X-ray imaging~\cite{Wang:2012, Wang:2013}, it is known empirically that the time or temporal resolution ($\delta_\tau$) required is proportional to the spatial resolution ($\delta$) with $\delta/\delta_\tau \sim 1 - $100 km/s~\cite{ZT:2010}, limited by the achievable speed in the laboratory.  To image the motion requires a sufficient number of X-rays and other particles (10$^7$ or greater per megapixel image for low noise detectors, see the discussions on {\it Feature detactability and noise} below) for at least two images separated by $\delta_\tau$, which usually require a sufficiently bright source of X-rays or particles, and an efficient scintillator converter and photodetector. When a scintillator is used, the scintillator decay time needs to be a fraction ($\sim$ 1/3) of $\delta_\tau$ for consecutive frames of images, which result in a scintillator light decay by a factor of $e^{-3}$, sufficient to avoid image latency from one image frame to the next as in high-speed synchrotron X-ray imaging~\cite{Campbell:2021}.

\subsubsection{Feature detectability and noise}
One of the central questions in RadIT, similar to other forms of IT such as optical IT, ultrasound IT, MRI, {\it etc.} is what tiny features may be resolved in the ubiquitous presence of noise. This is sometimes known as the {\it detectability problem}. A theoretical framework for feature identification in a noisy environment, which is intrinsically statistical, now exists, following the pioneer work of A. Rose, C. E. Shannon and others~\cite{Rose:1973,DS:1974}.  Many useful concepts such as contrast,  contrast transfer function, contrast threshold~\cite{Burg:1999}, noise-equivalent power, contrast to noise ratio, signal to noise ratio, {\it etc.} are applicable to normal vision, as well as RadIT. For example, contrast ($C$) is intuitively defined as the difference between observed intensity for feature A ($I_A$) and a reference feature B ($I_B$), $C = 2 | I_A-I_B | / (I_A + I_B)$, or equivalently $C^2 = 4 (I_A-I_B)^2 / (I_A + I_B)^2$. In the absence of a reference feature intensity, a `dark field' (with illumination such as the X-ray source off) and a `white field' (with illumination on but without the object) may be taken as references for images with an object of interest. In another example, detective quantum efficiency (DQE) as a function of spatial frequency ($f$, in lp/mm) may be defined as
\begin{equation}
DQE (f) = \frac{SNR_{out}(f)}{SNR_{in}(f)} \propto \frac{MTF(f)^2}{NPS (f)}
\end{equation}
Detective quantum efficiency
(DQE) was introduced by A. Rose as a measure
of ‘useful quantum efficiency’ or noise-equivalent quantum
efficiency of a detector~\cite{Rose:1946}. Noise power spectrum (NPS) measures the noise of the imaging
system as a function of spatial frequency. NPS for image can be estimated by, for example, using a method given in~\cite{Han:1998}. 

Contrast, feature detectability and noise are detector and scintillator dependent. One of the oldest detectors with single visible photon sensitivity is photomultiplier tube. There is now a growing number of photodetector technologies with single photon detection sensitivity and high quantum yield in the visible wavelength regime such as silicon photomultiplier (SiPM), Multi-Pixel Photon Counter (MPPC), CCD cameras, and more recently CMOS pixelated sensor arrays or CMOS cameras~\cite{MMS:2017, Wang:2020}. 
CCD and CMOS cameras ushered in the age of digital and real-time RadIT. 

\begin{figure}[ht!]
\centering\includegraphics[width=0.9\linewidth]{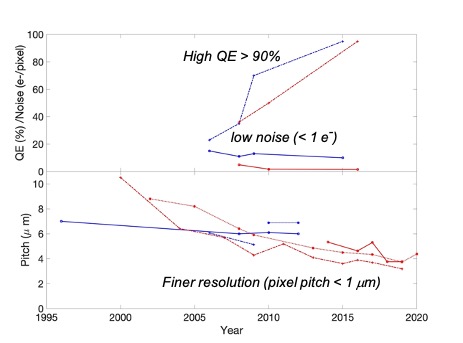}
\caption{ A recent survey of the evolutionary trends of CCD (in
blue) and CMOS sensors (in red) over the last 25 years. The
quantum effciency (QE) for visible photons has now exceeded
90\%. The noise level per pixel continues to decline, reaching
1 electron per pixel per readout cycle or less. Individual pixel
size or pitch is  $<$ 1 to 5 $\mu$m as of 2020. These performance
trends, in combination with continuing decline in cost, allow
flexibility in RadIT designs and applications. Figure credit~\cite{Wang:2020} and reproduced with permission from AIP publishing.}
\label{fig:HTPDFig1}
\end{figure}

\begin{figure}[ht!]
\centering\includegraphics[width=0.95\linewidth]{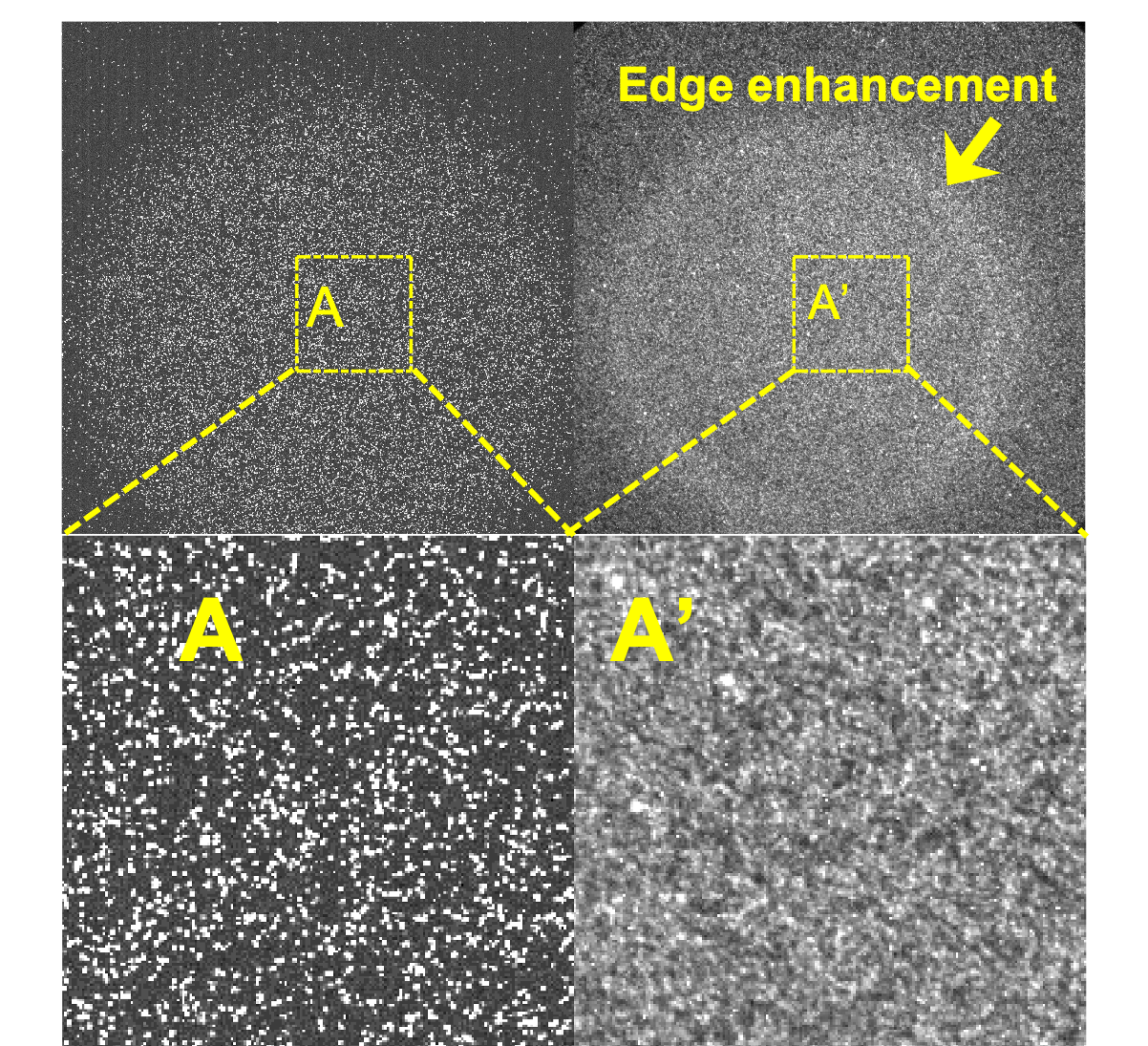}
\caption{A comparative study of X-ray source imaging with (Right) and without (Left) a LYSO scintillator.  An image intensified was used to enhance the sensitivity of scintillator detection. The X-ray source was an Amersham $^{241}$Am $\alpha$-induced K$_\alpha$ and K$_\beta$ emissions from Ba.}
\label{fig:directIndirect}
\end{figure}

The use of low-cost and high-performance CMOS cameras, including scientific CMOS cameras, is now growing in RadIT. Low cost is in part due to the large quantities of CMOS sensors used in both scientific and consumer applications such as in cell phones. High performance is a combination of small pixel pitch ($<$ 1 $\mu$m), high visible-light quantum yield ($>$ 90\%), low electronic noise ($<$ 1 $e^{-}$), see Fig.~\ref{fig:HTPDFig1}, and large format or field-of-view exceeding 10 million pixels. CMOS sensors are directly sensitive to X-rays, charged particles~\cite{PSW:2016}, and neutrons with a layer of neutron absorber such as $^{10}$B deposited on the sensor surface~\cite{Kuk:2021}. An example of direct CMOS detection of X-ray (K$_\alpha$ 32.06 keV and K$_\beta$ 36.55 keV lines of Ba), in comparison with the use of a LYSO converter, is shown in Fig.~\ref{fig:directIndirect}.

Individual X-ray responses show up as bright dots in the zoomed-in region (A) on the left in Fig.~\ref{fig:directIndirect}, while the corresponding scintillator response (A$'$) is blurry as expected.  Another observation of scintillator response is the enhanced edge in the scintillator response, which might be due to edge scintillator light distortion. Another example of edge effects of a scintillator called `tile glow' is given in Fig.~\ref{fig:Mo2} below. Further systematic studies of scintillator-based systems on feature detectability in different noise environment will likely provide necessary theoretical foundation to guide future innovations in scintillators and RadIT.

\subsection{Scintillator Metrics}
\begin{figure}[ht!]
\centering\includegraphics[width=0.95\linewidth]{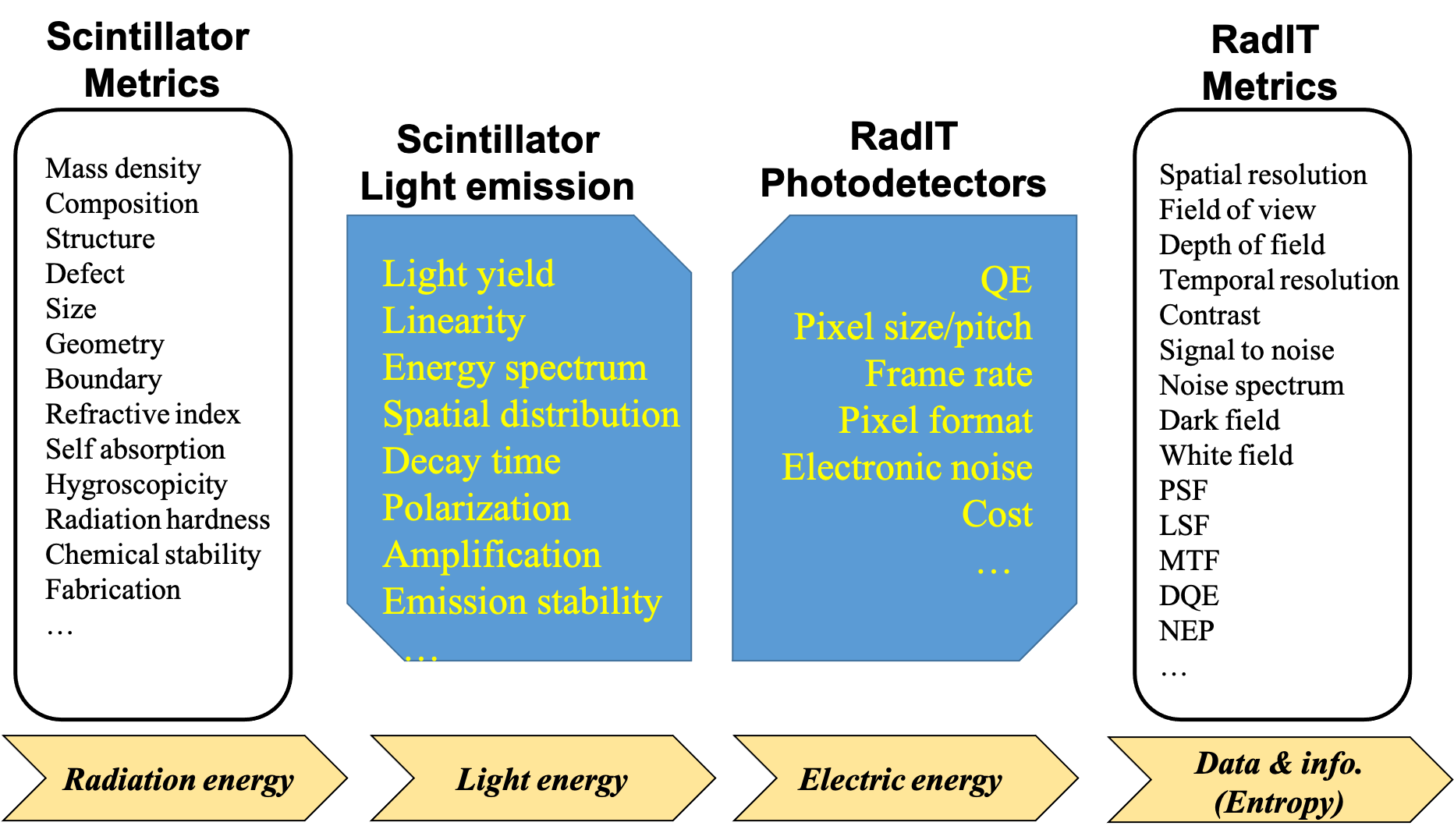}
\caption{RadIT metrics and scintillator metrics are correlated through the energy and data (information) flows. Material properties such as density, composition, {\it etc.} determine how the information encoded in radiation such as X-rays are converted to (visible) light emission, then electrical signals in photodetectors, before being recorded as data such as images, which can be characterized by metrics such as spatial resolution, field of view, DQE, {\it etc.}}
\label{fig:Metric1}
\end{figure}

The complex correlations between RadIT metrics and scintillator metrics are summarized in Fig.~\ref{fig:Metric1} through the energy and data (information) flows. Data and information encoded in ionizing radiation such as X-rays, neutrons  and charged particles determine the luminescence from scintillator, {\it i.e.} radioluminescence. In addition to light yield, other metrics for radioluminescence may include the emission spectrum, spatial distribution of the light, decay time, polarization, amplification (in active scintillator medium, which is relatively rare for now), and emission stability or degradation due to radiation damage of scintillator.  Light emission depends on a number of scintillator metrics, which include scintillator mass density, material composition, material structure, impurity and defects, scintillator size, scintillator geometry, scintillator boundary condition, scintillator chemistry and responses to the environment such as temperature, moisture, coupling to the photodetectors, refractive index, and self absorption. The light emission also depends on type of radiation, as discussed in more details in Sec.~\ref{sec:css} for X-rays of different energies, neutrons, and charged particles. 

Due to the complex interplays of different metrics and material properties, quantitative mappings between scintillator metrics and RadIT metrics usually require multi-physics codes as GEANT4~\cite{web:geant4} and MCNP~\cite{web:mcnp}. First-principles simulations of full RadIT systems are still beyond the current scope of GEANT4 and MCNP due to the lack of sufficient accuracy for data interpretation and object reconstruction from images and tomographic measurements. In addition to improvements in RadIT system specific modeling and simulations, laboratory experiments to characterize scintillators, scintillator-photodetector prototypes (see Sec.~\ref{sec:css}, Sec.~\ref{scin:new}) remain essential in scintillator selection for RadIT applications and system optimization. Recently, data science and use of RadIT such as neutron imaging for in-situ scintillator development and characterization are  highlighted as examples of emerging and trendy topics in Sec.~\ref{sec:TO}. Table.~\ref{scint:tab} in the appendix lists metrics  requirements for specific applications. It is apparently impossible to meet all the scintillator requirements at the same time in most applications. A trade-off between different metrics are often necessary, as briefly summarized here.

\subsubsection{Stopping power}
The stopping power or linear attenuation coefficient depends on density, material composition (effective atomic number), and cross sections. A thicker scintillator corresponds to high stopping power. However, the spatial resolution usually degrades. Therefore a tradeoff between stopping power and resolution is often necessary.

\subsubsection{Light yield}
Light yield can vary from X-rays to charged particles to neutrons for the same amount energy deposited. For X-rays, there is an additional complications due to the different X-ray-electron interactions. All the X-ray energy transfers to an electron in the scintillator in photoelectric absorption. No energy transfer happens in the coherent scattering by the scintillator. Only a fraction of X-ray energy is absorbed by the scintillator for an incoherent scattering event. For the pair-production process, the positron may carry a fraction of energy to a different location before re-absorption through electron-positron annihilation.

Many scintillators emit a relatively broad spectrum of wavelengths, and therefore light yield (LY) is wavelength dependent. Other factors such as linearity or proportionality between the radiation dose and LY, self-absorption, refractive index matching. It is known empirically that many scintillator emissions have multiple decay time constants for different wavelengths. A tradeoff between LY and decay time is often necessary.


\subsubsection{Decay time}

The scintillator decay times are normally determined by the spontaneous emission rate at the luminescence centers, which
is normally an intrinsic property of the scintillator.  Luminescence efficiency can be compromised by the quenching/ionization processes affecting excited state of luminescence centers which affects also its timing characteristics ~\cite{KL:2015}.  Furthermore, the spontaneous emission in a bulk material is isotropic, some of the emitted photons may not reach the photodetector due to, for example, internal reflection at the boundary of the scintillator. Photonic crystal structures have been proposed or recently demonstrated to modify spontaneous emissions so that it is possible to obtain anisotropic emissions~\cite{KSS:2020}, higher efficiency of the scintillator light collection~\cite{CRC:2022}, as well as the reduction in the intrinsic spontaneous emission rate, all in the visible range of wavelengths~\cite{KAL:2008}. The last one is known as the Purcell effect, which was initially proposed for the radio frequencies of the electromagnetic spectrum~\cite{Purcell:1946}. By locally enhancing the electric field,  higher emission rate may be obtained even if the probability of the
electronic transition is weak~\cite{Sil:2019}. 

\subsubsection{Radiation hardness}
Usually is of passive character, under irradiation, generated charge carriers are relocated to material (point) defects and color centres arise, the absorption of which overlaps with luminescence spectrum. Arising reabsorption decreases the externally measured light yield. So called radiation damage can be bleached, even spontaneously, by release of charge carriers from trap so that the radiation induced absorption shows distinct time and temperature dependence Radiation hardness (RH) of scintillators is an important consideration for RadIT as the X-ray and particle sources continue to become more brighter~\cite{KLM:2021}. RH requirement can sometimes rule out scintillators with high LY or fast decay time.

\subsubsection{Size and scalability}
A large volume of scintillators is needed in HEP. Large-size scintillators are needed for RadIT to capture scattered particles. A tradeoff between the size and scintillator cost, and a tradeoff between the volume and fabrication time are often necessary.

\subsubsection{Cost and fabrication}
Raw materials, fabrication process, and crystal quality affect the scintillator cost. Some of the highest LY or fast scintillator may require single crystalline structures with minimal self-absorption. A tradeoff between cost and LY, decay time may be necessary in applications.

\subsubsection{Stability and  mechanical ruggedness}
Some scintillators are hygroscopic. Low temperatures including cryogenic temperatures can reduce the thermal quenching of excited states in a scintillator and therefore increase the LY and reduce the radiative decay time for some fast scintillators. However, the use of cryogens can complicate the measurements. Therefore a tradeoff between the operating scintillator environment and performance may be necessary. 

\section{Current Status of Scintillators in RadIT \label{sec:css}}
While there is a growing number of competing semiconductor technologies, including high-Z semiconductors such as GaAs, CdTe, CZT for X-ray, $\gamma$-ray IT~\cite{He:1999,BTN:2014, DKL:2016, SPG:2020}, scintillators remain a favorite and sometimes the only option in X-ray and RadIT applications, as summarized in Table.~\ref{tab1} and Table.~\ref{scint:tab} in the appendix. 

Some advantages of the scintillator-based detectors include but are not limit to, built upon the metrics discussions above, the large number of scintillators and photodetectors to choose from, the flexibility in different combinations of scintillator with photodetectors, lower cost, and radiation hardness~\cite{HYZ:2022}. Some disadvantages of the scintillator detection scheme may include more complex data interpretation in order to model the scintillator light propagation with high fidelity, lower intrinsic spatial resolution due to the light propagation and spread, and edge effects due to the mismatch in refractive indices of different materials at the boundaries. There is no fundamental reason not to overcome these disadvantages, which motivate efforts in thin-film scintillators, nano-structured scintillators, meta-scintillators, photonic crystal guiding of scintillator light, data science and other exciting development in this growingly interdisciplinary field. 
\begin{table}
\caption{Different forms or modalities of Radiographic Imaging and Tomography (RadIT)}
\label{table}
\setlength{\tabcolsep}{3pt}
\begin{tabular}{|p{95pt}|p{75pt}|p{40pt}|}
\hline
Modality&
Contrast Mechanism&Scintillator examples  \\
\hline
X-ray radiography&
Absorption& LuAG:Ce
\\
MeV  X-ray/$\gamma$-ray radiography&
Incoherent Scattering& BGO, LSO/LYSO, GLO
\\
X-ray PCI &
Coherent scattering, \par interference & LuAG:Ce
 \\
X-ray CT &
Absorption &
 \\
X-ray CDI &
Coherent scattering, \par interference&
 \\
proton radiography&
Electron scattering, \par {\it or} Coulomb scattering & L(Y)SO
 \\
neutron radiography&
elastic scattering &
 \\
relativistic electron \par radiography&
 electron scattering, \par {\it or} Coulomb scattering & CsI
 \\
PET &
e-e$^+$ annihilation & BGO, L(Y)SO
 \\
P2T~\cite{LNS:2022}&
e-e$^+$ annihilation &
Ref.~\cite{LNS:2022} \\
\hline
\multicolumn{3}{p{251pt}}{Vertical lines are optional in tables. Statements that serve as captions for
the entire table do not need footnote letters. }\\
\multicolumn{3}{p{251pt}}{$^{\mathrm{a}}$ PCI $=$ Phase Contrast Imaging. CT = Computed tomography. CDI $=$ coherent diffractive imaging.  PET = Positron emission tomography. }
\end{tabular}
\label{tab1}
\end{table}

\subsection{X-rays below 100 keV}
Due to the compact table-top X-ray sources and continuing improving detector technologies including scintillator-based detectors, this is by far the most accessible form of RadIT. A comprehensive review of this topic alone is beyond the scope of this paper.

The absorption cross section dominates the X-ray-matter interactions in this energy range. To minimize the X-ray dose, contrast agents can be used to enhance the absorption contrast, and allow not only the structural imaging or static imaging but also functional imaging of in-situ biochemistry and disease pathology~\cite{CRA:2012}.

X-ray diffraction-based imaging techniques at modern light sources are powerful tools for non-destructive materials characterization at the mesoscale. Various X-ray diffraction techniques have been developed over the past decades that provide the ability to map three-dimensional (3D) materials microstructure at various spatial and temporal scales. It is now possible to obtain 3D crystallographic information of a polycrystal at a sub-grain (sub $\mu$m) resolution and subsequently zoom in (sub nm) on an individual crystal to further image defects and dislocations present at the atomic scale.

Since the seminal work of Laue and Bragg, X-ray crystallography has been commonly used for structure determination. Among a variety of techniques employing crystallography, high-energy X-ray diffraction microscopy (HEDM) is a 3D microstructure characterization technique~\cite{R1, R2}. It provides information on spatially resolved crystallographic orientation via near field HEDM (nf-HEDM)~\cite{R3} as well as grain resolved elastic strains and corresponding stresses in polycrystalline materials via far-field HEDM (ff-HEDM)~\cite{R4}. The difference between that near-field and far-field HEDM methods is mainly the sample to detector distance. Both of these HEDM techniques take advantage of the X-ray tomography setup to collect diffraction data for obtaining 3D crystallographic information.

Typically, HEDM measurements are performed using incoherent monochromatic X-rays, where the incident beam illuminates a volume of a sample. The image generated on an area detector corresponds to the X-rays scattered off of atomic planes of crystallographic grains in the illuminated volume that satisfy the Bragg condition at a given projection angle. The diffracted beam passes through a scintillator screen where the X-rays are converted to optical photons which are then recorded on a CCD camera. The sample is rotated in a plane normal to the beam axis and multiple such diffraction images are recorded at various projection angles. Combined with tomography, comprehensive materials characterization can be performed with detailed information on crystal orientations and grain morphology (spatially resolved and grain averaged, texture), mechanical response of individual 3D grains (average lattice strains, and stress associated with it), and volumetric density maps (pores and cracks, second phase inclusions). In situ microstructural information from experiments are crucial for informing, instantiating, and validating various physics models~\cite{R5,R6,R7,R8}. 

Since, most diffractive techniques rely on elastic scattering, the number of photons that get diffracted to form an image on the detector is significantly lower than those generated during X-ray radiography. Therefore, the scintillator efficiency over a range of materials (from low to high $Z$) becomes highly important for diffraction experiments. Existing scintillator capabilities are highly limited when performing experiments on high Z materials with high resolution imaging ($\sim$ 1 $\mu$m or better spatial resolution) that require high-energies (e.g., $>$ 80 keV) to penetrate bulk specimens (e.g., $>$ 500 $\mu$m of uranium)~\cite{R9}. Therefore, choosing scintillators at high energies show significant tradeoff between light yield and imaging resolution. 

\begin{figure}[ht!]
\centering\includegraphics[width=.95\linewidth]{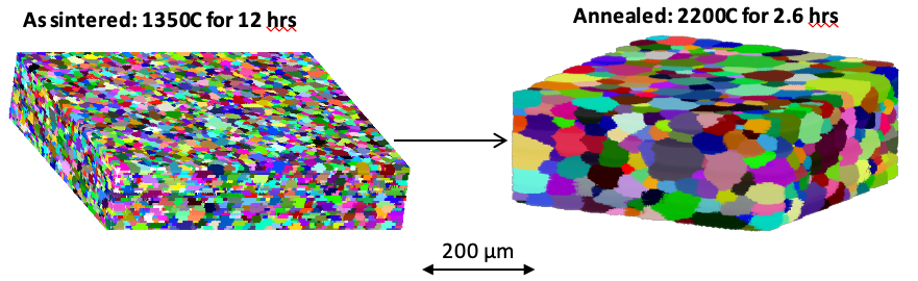}
\caption{ 3D microstructure of UO$_2$ characterized before and after heat-treatment using HEDM.}
\label{fig:Reeju1}
\end{figure}

Additionally, temporal resolution suffers when counting longer to improve signal to noise per image. Fig.~\ref{fig:Reeju1} shows an example of 3D microstructure of UO2 characterized before and after heat-treatment using HEDM with 90 keV X-rays. It is only recently that such measurements have been possible~\cite{R11}, however, the need for long acquisition time to collect high signal to noise diffraction images limited the ability to capture materials dynamics during such experiments. As a result, even though unprecedented before and after 3D data from a high-Z material was acquired, the kinetics of grain growth was not captured. 

\begin{figure}[ht!]
\centering\includegraphics[width=0.95\linewidth]{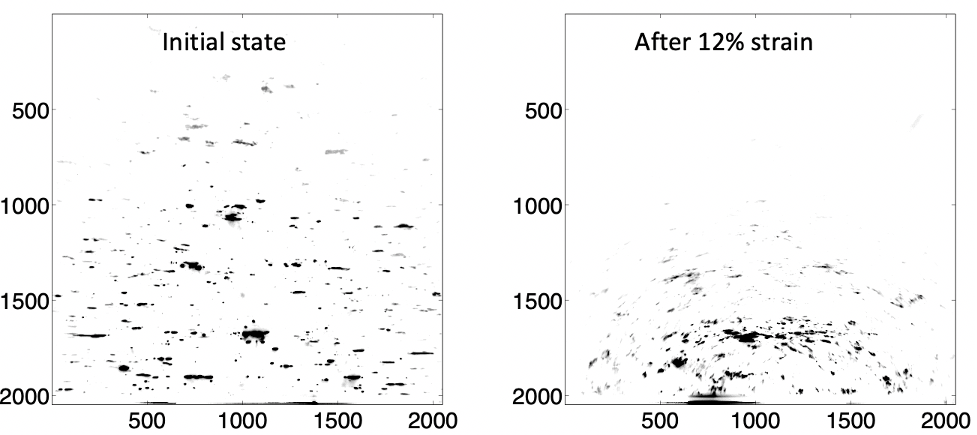}
\caption{2D detector image showing diffraction from multiple grains satisfying the Bragg condition before and after 12\% deformation of polycrystalline copper. }
\label{fig:Reeju2}
\end{figure}

The signal-to-noise ratio also drops when the crystallinity of the as-received sample starts to break by introducing defects and deformation in the material, highly limiting the information extraction capability at large strains~\cite{R10}. Fig.~\ref{fig:Reeju2} shows an example of the diffraction image recorded at the initial, recrystallized state and after 12\% tensile deformation of a polycrystalline copper specimen. The diffraction signal reduces significantly upon deformation. Along with streaking of the diffraction spots, with increasing deformation, the high order scattering intensity is also decreased or lost entirely. This is another example where increased light yield of the scintillators would improve both image quality and information extraction capability. These are only a few examples and there are many more where these advanced microstructural characterization techniques would highly benefit from scintillators with increased absorption efficiency for a wide range of materials and X-ray energies, without compromising spatial resolution.

In addition to static measurements, dynamic measurements are inherently photon starved with photon counts inversely proportional to exposure time, producing comparatively lower quality images and limited to relatively slow dynamics. Dynamic measurements could greatly benefit from higher light yield scintillators. With upcoming upgrades, synchrotron sources such as the APS and next generation FEL-based light sources such as the EuXFEL will produce higher flux and more brilliant beams. New versatile scintillator materials and technologies are needed to take full advantage of upgraded X-ray sources, the benefits they will provide in higher quality images for given exposure times as well as enabling the imaging of faster dynamics cannot be over emphasized. Advances in imaging scintillators have the potential to benefit a large set of problems in various scientific and medical fields employing slow and high-rate radiographic imaging combined with diffraction.

\subsection{X-rays above 100 keV, $\gamma$-rays \label{sec:WWH}}
The large mean-free-path of X-rays at above 100 keV energies and $\gamma$-rays makes them an effective tool to radiograph thick and/or dense objects. Applications include weld inspection, parts inspection (including additive manufacturing), portal monitoring, and hydrotesting. 

The Compton scattering cross section dominates the X-ray/$\gamma$-ray-matter interactions at above 100 keV (the exact energy threshold is material specific). Other interactions such as photoelectric absorption and pair-production can also contribute to the image formation depending on the photon energy. In the example of LYSO scintillator, Fig.~\ref{fig:LYSO}, the Compton or incoherent scattering cross section becomes the dominant process at around 370 keV. Electron-positron pair production becomes dominant at energies above 6.3 MeV. 

The major limiting factors in high energy radiography are the resolution loss due to Compton scattering in the detector and the low DQE of common scintillator and photodetector pairs. High MeV-photon detection efficiency is particularly important for flash radiography applications where fast data acquisitions are required, the source has low flux, or high-fidelity images are required~\cite{W3}. Quantum efficiency can be traded for: time in emergency response applications, patient dose in medical radiography, and the number of views in tomography. Therefore, maximizing the DQE is of paramount importance in these applications.
Some high-density crystals have been developed which are suitable for high energy radiography and tomography, namely Bismuth Germanate (BGO)~\cite{W4}, cerium doped lutetium oxyorthosilicate (LSO) and lutetium yttrium oxyorthosilicate (LYSO)~\cite{W6}. The properties of these scintillators are shown in Table.~\ref{WWH:tb}. LYSO and LSO suffer in some applications due to the intrinsic radioactivity of lutetium~\cite{W9}. These single crystals cannot be grown in large sizes and therefore need to be pixelated or segmented for radiography purposes.

\begin{table*}[htbp]
\caption{Properties of scintillators used for high energy radiography (data from~\cite{W8,W16}).}
\centering
        \begin{tabular}{|l|c|c|c|c|c|}
            \hline
            Scintillator & Formula & Density (g/cm$^3$)  & LY (10$^3$ photons/MeV) & Decay time  (ns) & Scintillator radioctivity \\[3pt]
            \hline
            BGO & Bi$_4$Ge$_3$O$_{12}$ & 7.13 & 9 & 300 & No \\[3pt]
            \hline
            LSO & Lu$_2$SiO$_5$:Ce & 7.4 & 26 & 40 & Yes  \\[3pt]
            \hline
            LYSO & Lu$_{2(1-x)}$Y$_{2x}$SiO$_5$:Ce \textsuperscript{\textit{a}}  & 7.1 & 33 & 36 & Yes \\[3pt]
            \hline
            GLO & Gd$_{3y}$Lu$_{2(1-2y)}$Eu$_y$O$_3$ \textsuperscript{\textit{b}} & 9.1 & 70 & 10$^6$ (1 ms) & Yes  \\[3pt]
            \hline
            \multicolumn{6}{l}{\footnotesize{\textsuperscript{\textit{a}} $x$ $\sim$ 0.2, \textsuperscript{\textit{b}} $y \sim$ 0.1}}
        \end{tabular}
\label{WWH:tb}        
\end{table*}

\begin{figure}[htb!]
\centering\includegraphics[width=0.9\linewidth]{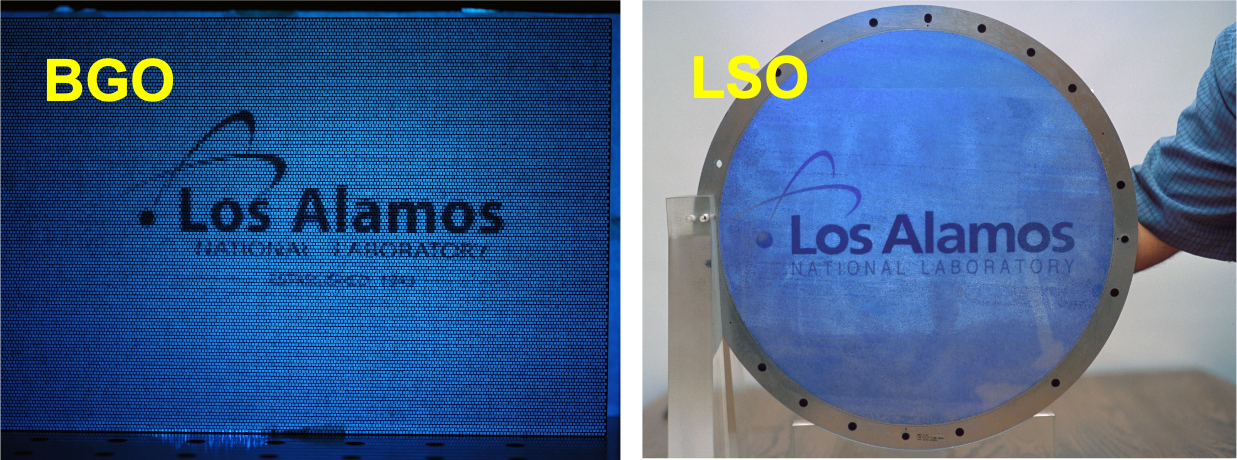}
\caption{Segmented BGO and LYSO ~\cite{W13} scintillators. Both have 1 mm $\times$ 1 mm $\times$ 40 mm scintillator pixels and are separated by a metal septum.}
\label{fig:WWH2}
\end{figure}

Segmented scintillators are composed of individual pixel light pipes, which produce a planar image suitable for fast optics~\cite{W10, WWH:2017, Winch:2018}. The light pipe aspect is extremely important for multiple reasons. First, it allows low-f-number, planar optics to be used, and secondly, it eliminates veiling glare from inclusions, seams, and other defects in the scintillator. Using high density material also reduces the spread of the Compton scattered photons thereby reducing the blur and improving the resolution. Additionally, when the pixels are optically isolated by means of a metal grid or similar the optical scatter is eliminated. Figure~\ref{fig:WWH2} shows two segmented scintillators, a 40 cm diameter LYSO grid with 1 mm $\times$ 1 mm $\times$ 40 mm scintillator pixels ~\cite{W13} and a 10 cm $\times$ 20 cm BGO grid with 1 mm $\times$ 1 mm $\times$ 40 mm pixels. Both are separated by a thin metal septum.

\begin{figure}[htb!]
\centering\includegraphics[width=0.9\linewidth]{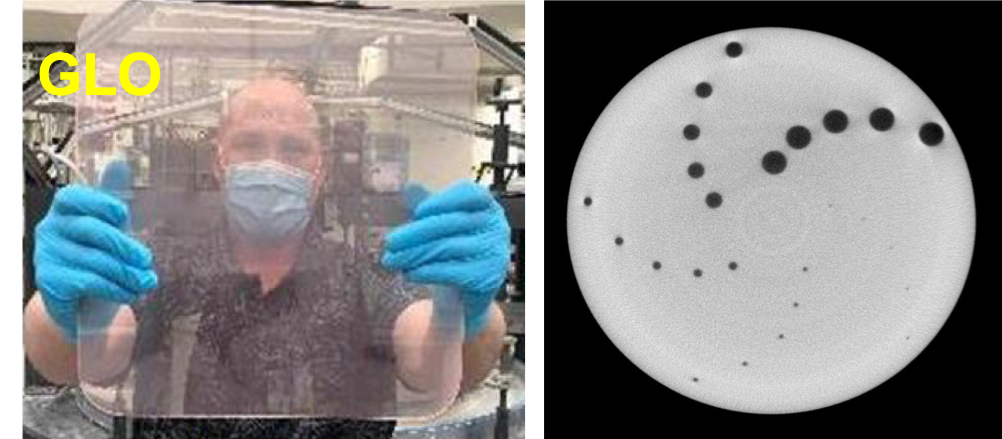}
\caption{Large format (14” $\times$ 14”) GLO scintillator (left, 3.85 mm thick) and radiograph obtained (right) using a 2.5 mm thick GLO plate (reproduced from~\cite{W17}).}
\label{fig:WWH3}
\end{figure}

Recent developments on transparent sintered ceramics have produced a new bixbyite transparent ceramic scintillator, GLO~\cite{W14,W15}. GLO has a very high density and high light yield, but a relatively long decay time (properties shown in Table~\ref{WWH:tb}). The largest scintillator fabricated thus far is roughly 30 cm by 30 cm in dimensions~\cite{W17}, and is shown in Figure~\ref{fig:WWH3}.

Another technique available to take advantage of the large number of commercially available powdered or needle scintillators (of NaI, CsI or Gadox (Gd$_2$O$_2$S)) is to use intensifying screens composed of high Z metals~\cite{W19,W20,W21,W22}. These screens convert the gamma/x-ray photons into Compton scattered electrons which are more readily captured by the scintillator. However, because of the absorption of their own scintillation light, the quantum efficiency of these is still very low with a maximum of 1\%. Intensifying screens and imaging plates can be stacked (at ~1\% per layer) with as many as 40 been demonstrated to obtain ~20\% DQE~\cite{W25}, the downside being each plate must be read out individually and the radiographs aligned and averaged (suited more for storage phosphor imaging plates). 

\begin{figure}[htb!]
\centering\includegraphics[width=0.9\linewidth]{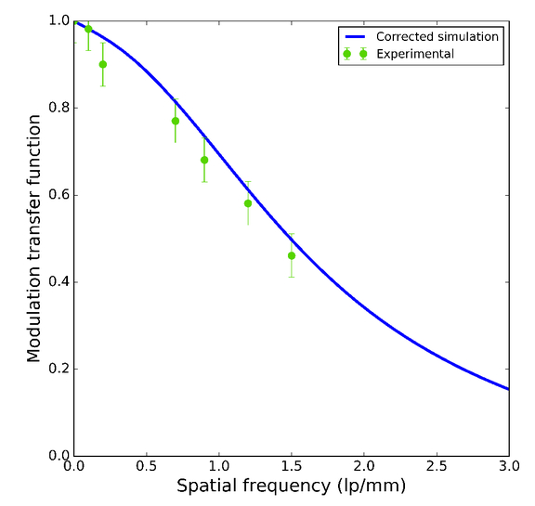}
\caption{Comparison between experimentally determined MTF with simulated MTF for 1 mm thick LSO with a 1.25 MeV source (reproduced from~\cite{WWH:2017}).}
\label{fig:WWH4}
\end{figure}

\begin{figure}[htb!]
\centering\includegraphics[width=1.0\linewidth]{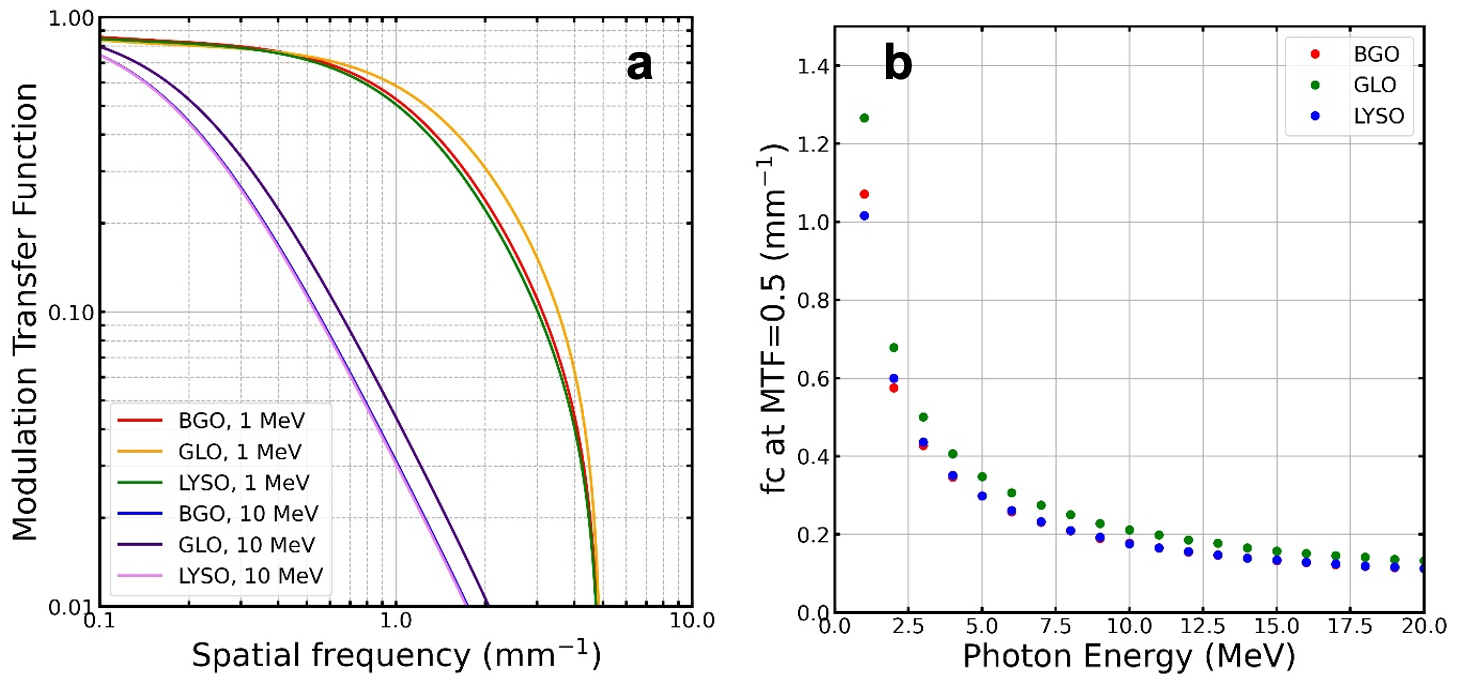}
\caption{(a) Simulated MTFs for 1 MeV and 10 MeV photons and (b) frequency at MTF = 0.5 (fc) as a function of incident photon energy for 10 mm thick BGO, GLO and LYSO scintillators.}
\label{fig:WWH5}
\end{figure}

The MTF curves are easily simulated (see for example Winch et. al.~\cite{WWH:2017}), and are more accurate than measurements which can be sampled and aliased, particularly for segmented scintillators. Fig.~\ref{fig:WWH4} shows an example of the good agreement between simulated and measured MTFs for a 1 mm thick LSO scintillator at a photon energy of 1.25 MeV. Fig.~\ref{fig:WWH5} shows simulated MTFs for 10 mm thick BGO, GLO and LYSO scintillators at 1 and 10 MeV incident photon energy, and the frequency value at MTF = 0.5 (fc) for the same scintillators as a function of photon energy. The MTF is a strong function of the density of the material, as shown by the GLO scintillator having the highest resolution, and both the LYSO and BGO having very similar curves. As the incident photon energy is increased the resolution decreases.

\begin{figure}[htb!]
\centering\includegraphics[width=1.0\linewidth]{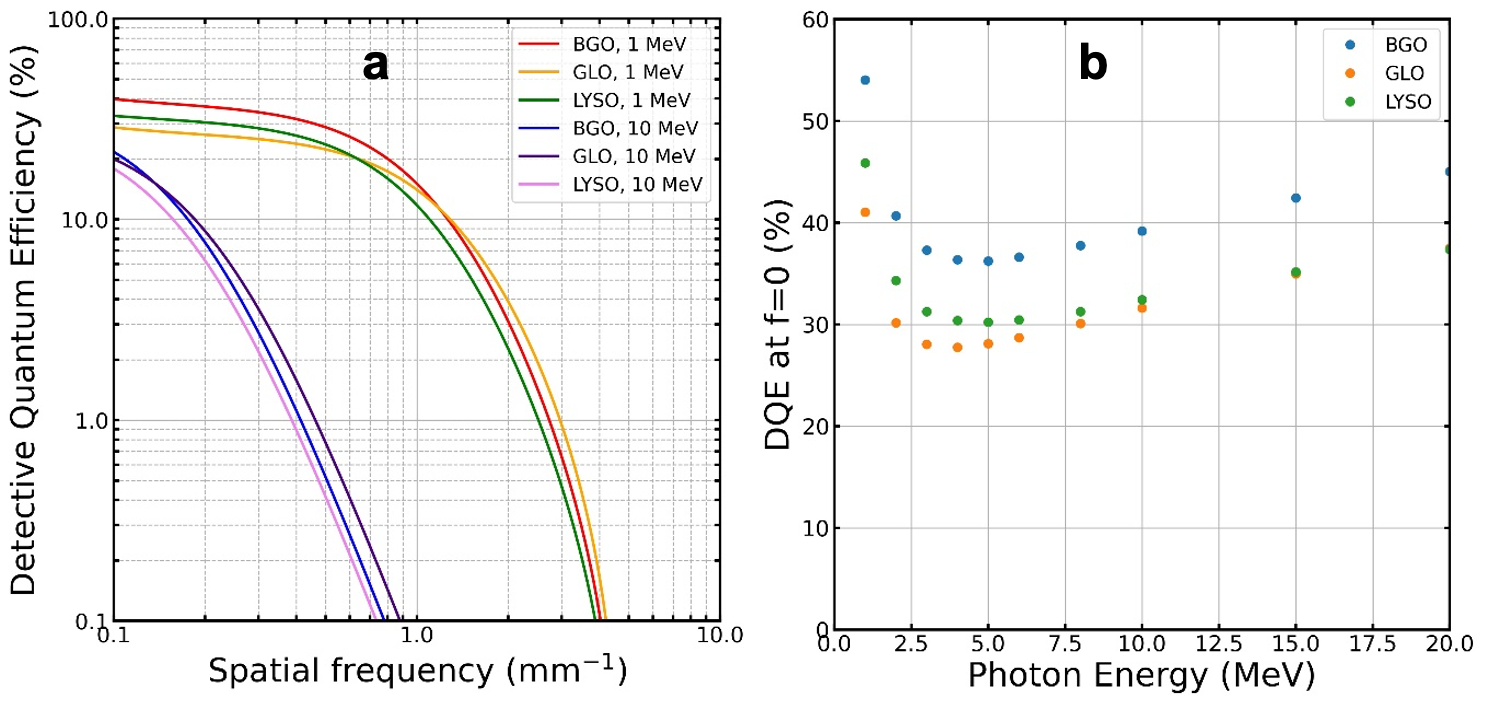}
\caption{(a) Theoretical DQE as a function of spatial frequency using simulated MTFs from Fig.~\ref{fig:WWH5} and (b) zero frequency DQE as a function of incident photon energy for 10 mm thick BGO, GLO and LYSO scintillators.}
\label{fig:WWH6}
\end{figure}

Theoretical DQEs can be determined using the simulated MTFs and the total attenuation of the given photon energy in the scintillator, i.e.
\begin{equation}
DQE(f) \propto MTF^2 (f)\left[ 1 - \exp (-\frac{\mu}{\rho} \rho L) \right],
\end{equation}
where $\rho$ is the scintillator density, $L$ the scintillator thickness and $\mu/\rho$ the total mass attenuation coefficient [25]. Figure ~\ref{fig:WWH6} shows the theoretical DQEs for the $L =$ 10  mm thick BGO, GLO and LYSO scintillators calculated using the MTFs in Figure~\ref{fig:WWH5}a. The DQE is higher for the 1 MeV photon energy, and the BGO shows highest DQE. Figure~\ref{fig:WWH5}b shows the DQE at zero frequency as a function of the incident photon energy. Here the DQE initially decreases with increasing photon energy, and then increases again at a photon energy of around 5 MeV. This follows the trend of the high Z materials mass attenuation coefficients (e.g. Bismuth, Lanthanum, Gadolinium) which make up the scintillators composition.

\begin{figure}[htb!]
\centering\includegraphics[width=0.9\linewidth]{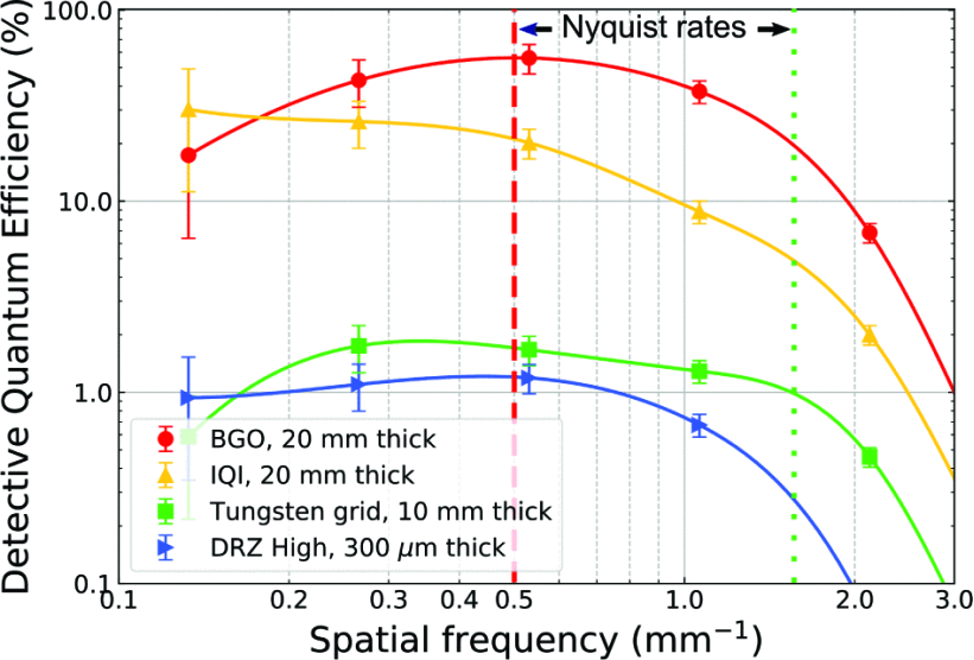}
\caption{Measured DQE for segmented BGO, IQI fiber optic glass plate, tungsten and glass pixelated array and DRZ High (Gadox). The vertical dashed lines show the Nyquist rates for the segmented scintillators. Reproduced from~\cite{Winch:2018} with permission.}
\label{fig:WWH7}
\end{figure}

Figure~\ref{fig:WWH7} shows an example of experimental DQE for various segmented scintillators, reproduced from Winch et. al~\cite{Winch:2018}. These results demonstrated for the first time a 50\% DQE system with a segmented BGO scintillator and a room temperature camera. The glass fiber optic faceplate was shown to have a DQE of 30\%, and is a good compromise between the expensive segmented scintillators and commercial powdered or needle scintillators.

Fundamental scintillator research into high density scintillators is ongoing, and breakthrough scintillators are always of interest. However, one of the key challenges for MeV scintillators is to find a way to manufacture existing scintillators into a myriad of forms, with reasonable cost, large format and shorter timeline. Pixelated or segmented arrays with various thicknesses, dimensions, and even curvatures are one example of taking existing scintillators and modifying them for high energy photon radiography and CT. Another example would be to join transparent crystals like BGO or LSO into large sheets.

\subsection{Neutrons}
Neutron IT as a non-destructive testing tool is relatively new when compared to X-ray IT. The international society for neutron radiology (ISNR) was created in 1996~\cite{LR:2015} to bring the neutron radiography and neutron imaging international community together. Neutrons, mostly from nuclear fission reactors and spallation sources, have now been successfully used for in-situ imaging and 3D tomography of hydrogen fuel cells, diesel particulate filters, nuclear fuel rods, and fossils. Neutron IT have been extensively explored in the neutron energy range from sub-thermal to hundreds of MeV using the LANSCE 800 MeV accelerator, as summarized in a recent review~\cite{NVH:2018}. The latest breakthrough in laser-driven inertial confinement fusion may open up new avenues for neutron IT by providing a prompt intense neutron source.

\begin{figure}[htb!]
\centering\includegraphics[width=0.9\linewidth]{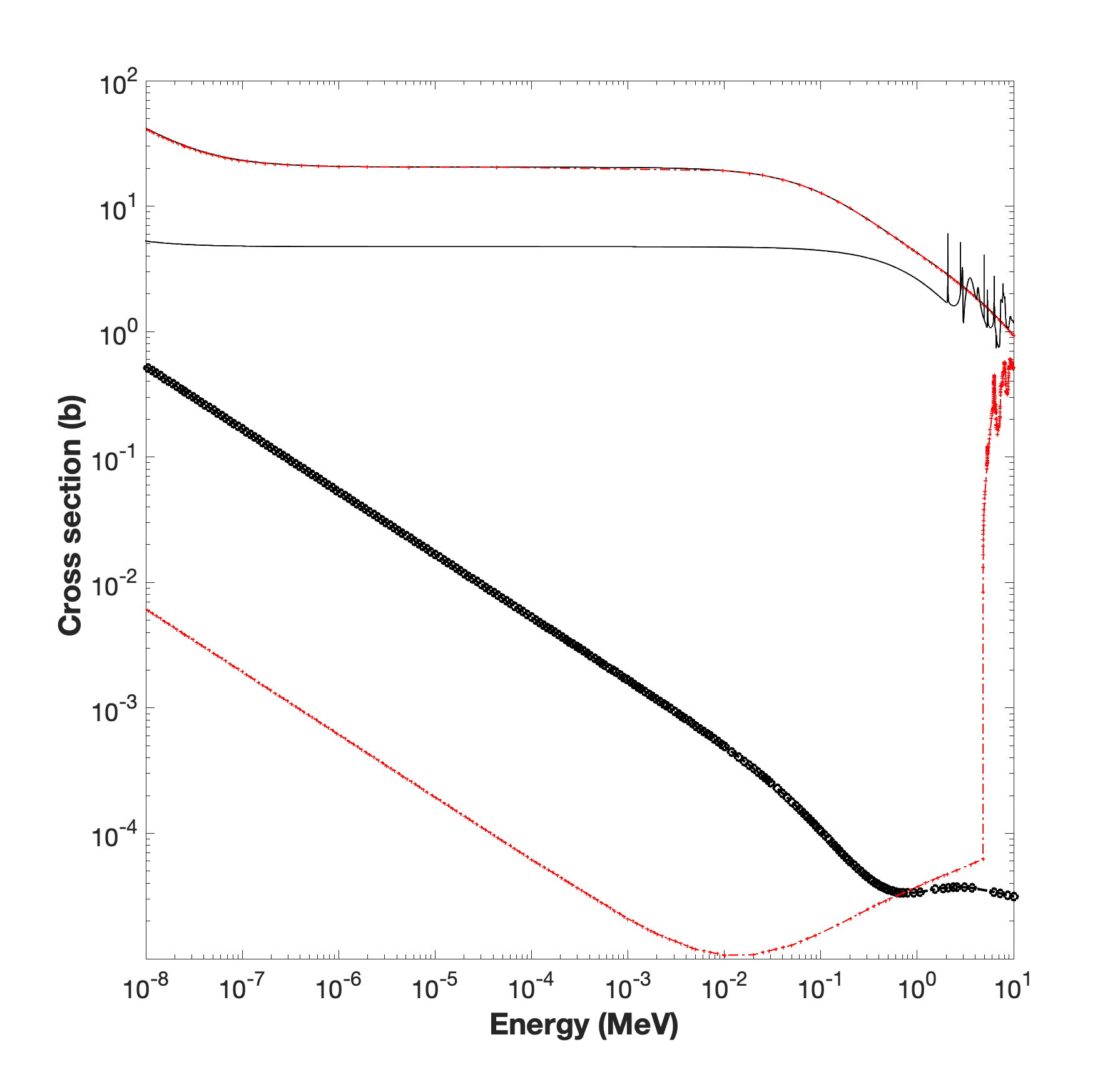}
\caption{Energy-dependent neutron cross sections in a vinyltoluene (C\textsubscript{10}H\textsubscript{9}) based plastic scintillator. The density is assumed to be 1.032 g/cm$^3$. The data are from the ENDF database. Elastic collision cross section dominated over others. {\it https://www.nndc.bnl.gov/endf/}.}
\label{fig:nPlastic}
\end{figure}

Neutrons complement X-rays as a unique material probe due to its relatively weaker interactions with electrons than nuclear interactions.  An example is given in Fig.~\ref{fig:nPlastic} for energy-dependent neutron attenuation cross sections in vinyltoluene plastic scintillator. The transmission of neutrons through a material obeys the same equation as Eq.~(\ref{eq:att1}) with the corresponding neutron cross sections. Similar to X-rays, neutron nuclear interaction cross section is a sum of absorption, coherent and incoherent scattering in the non-relativistic regime. Neutron cross sections on the other hand are highly isotopically sensitive, which make neutron more sensitive to $^1$H than $^2$H for image contrast, for example. The total $^1$H thermal neutron scattering cross section is more than 10 times that of $^2$H. The thermal neutron absorption cross section of $^6$Li is 940 barn, which is orders of magnitude lager than that of $^7$Li, and makes $^6$Li a popular element in scintillators for neutron detectors.

Synergies between neutron scintillator detectors and X-ray scintillator detectors have been recognized~\cite{LR:2015}. An X-ray camera can turn into a neutron camera by switching the scintillators. From the Table.~\ref{scint:tab} in the appendix, it is clear that in addition to neutron-specific scintillators that contains $^6$Li, Gd ($^{157}$Gd and $^{155}$Gd in particular),  composite scintillators that combine materials with differential neutron sensitivities may also be considered. Further discussions on composite scintillators for fast neutrons are given in Sec.~\ref{sec:bren} below. Background reduction in neutron scintillators, esp. $\gamma$-ray background reduction, remains an important consideration and motivate new scintillator innovations. An example of using neutron imaging for in-situ scintillator characterization, Sec.~\ref{sec:AT}, shows that RadIT can not only benefit significantly from scintillator development, but also become a useful tool for scintillator development.

\subsection{Protons and heavy ions \label{sec:MF}}

Proton radiography was first demonstrated in 1968, exploiting the relatively broad range straggling of monoenergetic protons~\cite{f1}. Several years later, scattering-based proton radiography was demonstrated with low-energy protons~\cite{f2}. These two techniques could be used for objects of uniform thickness, or very thin objects, respectively. Lens focused proton radiography uses an energetic beam to image dynamic experiments by focusing transmitted protons using magnetic fields. In the high energy regime, the beam is attenuated by nuclear scattering of the protons~\cite{Mo:1}, and at low energies, Coulomb multiple scattering with a collimator at a Fourier point in the lens system is used to enhance the nuclear attenuation~\cite{Mo:2}. In both cases an image is produced by the transmitted protons in the lens image plane. Using the 800-MeV proton accelerator at the Los Alamos Neutron Science Center (LANSCE), and a magnetic lens system, proton radiography demonstrated fast ($<$ 200 ns), large scale (up to 50 g cm$^{-2}$) imaging of dynamic events. The addition of a magnetic lens allowed for a detector to be safely placed 10 m from an explosion, and for the correction of chromatic aberrations, to first order, that provided 200-$\mu$m spatial resolution~\cite{f5}. The spatial resolution was further enhanced by the implementation of a $\times$3 magnetic lens, that effectively shrinks the field of view and magnifies it to the imaging plane, increasing spatial resolution to 65 $\mu$m~\cite{f6}. A $\times$7 proton microscope further increased the spatial resolution to 25 $\mu$m~\cite{f7}. To increase the areal density sensitivity, special collimation schemes have recently enabled the dark field proton radiography concept~\cite{f8, f9}. Higher energy proton radiography at the GSI accelerator facility in Darmstadt, Germany has dramatically increased the object thickness and spatial resolution possible with proton radiography, using 3-4 GeV protons and the PRoton mIcroscOpe for FAIR (PRIOR). The synchrotron available currently at GSI can deliver 10$^{11}$ protons per bunch, and 4 bunches per pulse, enabling fast, dynamic radiography. A new lens, PRIOR-II, has been developed for use with the enhanced Facility for Antiproton and Ion Research (FAIR) that is coming online within a few years as an upgrade to GSI. FAIR will increase the available proton energy, with more protons per bunch, allowing for more flexibility for dynamic imaging, in conjunction with deliberate partial-bunch extraction from the synchrotron ring. The next frontier in proton radiography will be higher proton energies, for higher spatial resolutions and thicker object imaging, and increased statistics, through upgraded accelerator design or the implementation of synchrotron ring~\cite{f13}.

\begin{figure}[htb!]
\centering\includegraphics[width=0.9\linewidth]{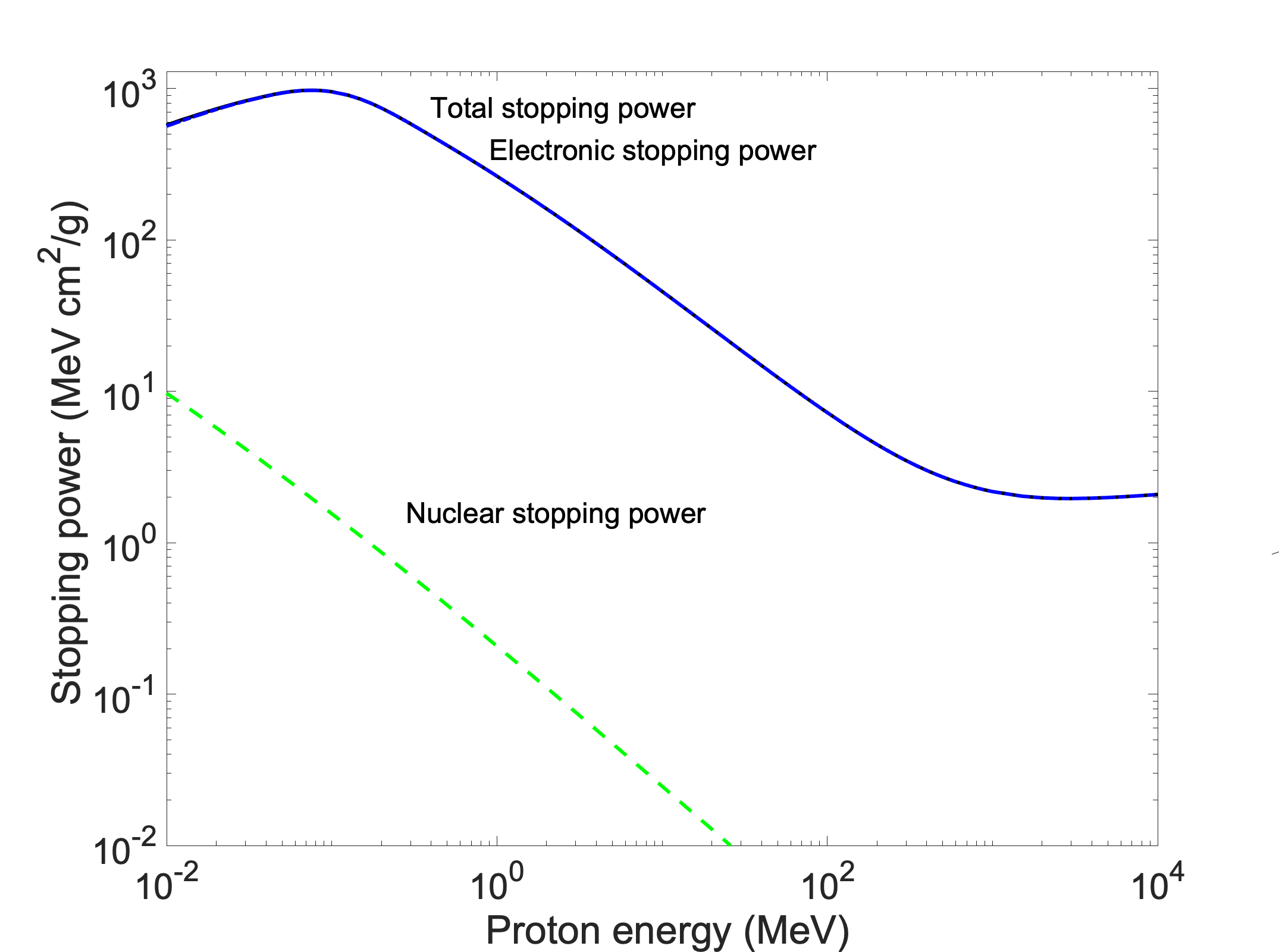}
\caption{Energy-dependent proton stopping power in a vinyltoluene (C\textsubscript{10}H\textsubscript{9}) based plastic scintillator. The density is assumed to be 1.032 g/cm$^3$. The data are from the NIST/PSTAR database.}
\label{fig:pPlastic}
\end{figure}

Energetic charged particles of hadrons such as protons and alpha particles lose energy due to the long-range Coulomb interactions and collisions.  Fig.~\ref{fig:pPlastic} gives an example of energy-dependent proton stopping power as a function of energy in a vinyltoluene-based [CH\textsubscript{2}CH(C\textsubscript{6}H\textsubscript{4}CH\textsubscript{3}) or C\textsubscript{10}H\textsubscript{9}] plastic scintillator. Electronic energy loss is the predominant process in the energy range shown. Nuclear collisions and the corresponding energy loss may be neglected. 

Dynamic experiments require fast proton imaging. This is attained by focusing the light from a scintillator onto an imaging system with a multi-frame multiplexed CMOS camera setup~\cite{Mo:2, Mo:3}. The scintillator currently used at Los Alamos is a 3$\times$2 tiled array of monolithic LSO crystals 1.9 mm thick~\cite{Mo:3}. This system suffers from nonlinearities at the tile edges that produce artifacts in the images, and from background due to totally internally reflected light that escapes due to defects in the crystals. The use of granular screens may address these problems. LYSO screens produce as much as 6 (black backing) or 12 (reflective backing) times more light per areal density thickness than the current tiled scintillator. 

\begin{figure}[htb!]
\centering\includegraphics[width=0.9\linewidth]{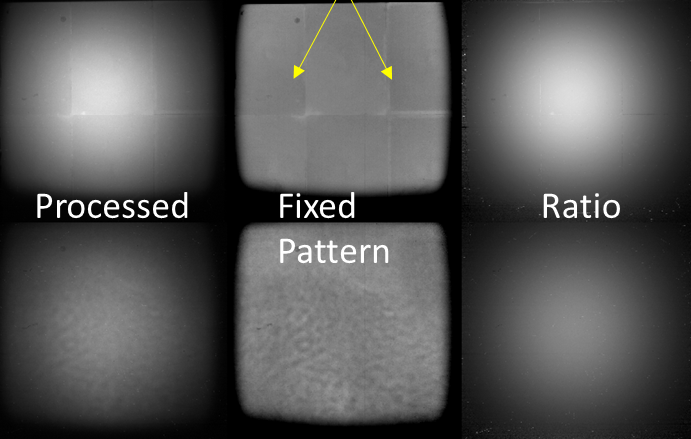}
\caption{Images of a proton beam made with the standard tiled monolithic LSO scintillator crystals (top row) and a LYSO granular screen (bottom row). The arrows point to the crystal scintillator tile boundaries. The fixed pattern noise from both detectors is observed to cancel in the ratio of a single image to a fixed pattern correction.  However, the data from within $\sim$1mm of the tile boundary is not reliable.}
\label{fig:Mo1}
\end{figure}

Figure~\ref{fig:Mo1} shows images obtained using a tiled scintillator along with a large gain (38-76 $\mu$m grain size) screen. The center panel at the top shows clearly visible tile boundaries. Although the tile boundaries cancel when fixed pattern maps are used to correct the image, the data from within 1 mm of the boundary is not reliable because of reflections from the edge of the tile. The screen has a fixed pattern noise that cancels in the ratio.

\begin{figure}[htb!]
\centering\includegraphics[width=0.7\linewidth]{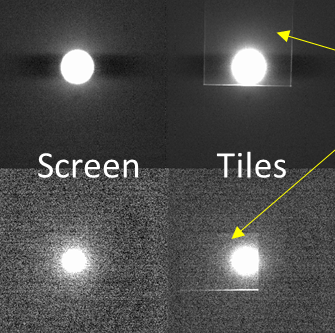}
\caption{Images of a 1 cm diameter proton beam made with a granular LYSO scintillator screen (left) and the standard tiled monolithic LSO scintillator crystals. The bottom row shows the data on an expanded intensity scale. Similar to Fig.~\ref{fig:Mo1}, the ``tile glow” (marked by the arrows) produced a notable artifact in the image that is absent in the screen image.}
\label{fig:Mo2}
\end{figure}

\begin{figure}[htb!]
\centering\includegraphics[width=0.8\linewidth]{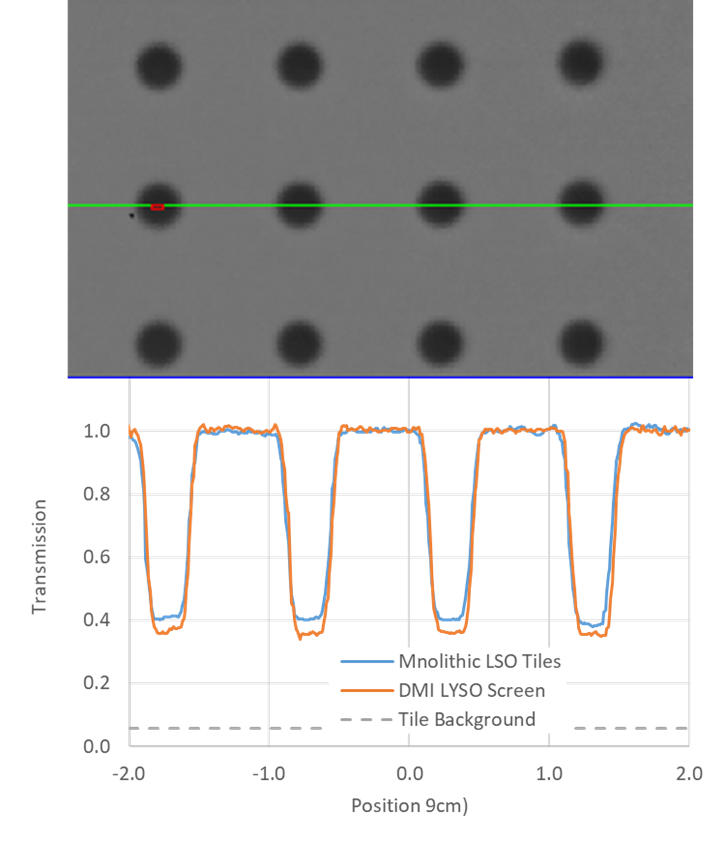}
\caption{Image (top) and lineout plots from the tiled crystal scintillator and the large grain screen (bottom). The screen is observed to have higher contrast.}
\label{fig:Mo3}
\end{figure}

Images made with a pencil beam (Figure~\ref{fig:Mo2}) illustrate another problem. The totally internally trapped light in the tile can scatter and escape resulting in general glow from the tile. This is localized to a given tile by blacking the edge of the tile with a shape pen. This background of about 10\% recuses the image contrast.
Figure~\ref{fig:Mo3} shows a comparison of the contrast produced by a set of 3 mm thick plugs with the tile and screen detectors. The reduced contrast with the tiles is apparent in the lineouts shown in the bottom panel.  The 10\% background produced by the trapped light reduced the dynamic range of proton radiography.
This work has shown the benefits of grainular screen scintillators for proton radiography, which include high specific light output, the absence of tile boundaries, lower backgrounds, and the ability to construct detectors from a wider rage of material that may not be available in suitable size

In addition to using differential proton intensities for imaging contrast, precise measurement of proton energy loss has been
recognized as a potentially powerful approach to radiographic
imaging of thick (areal density above 10 g/cm$^2$) objects~\cite{Mo:4}. The basic idea is to measure proton energy-loss precisely through for example measurement of proton time of flight (ToF). At an energy resolution of $\Delta E/E$ from 10$^{-4}$ to 10$^{-5}$ for 800 MeV protons, or an absolute energy resolution of 8 to 80 keV, the areal density resolution can reach 5 to 50 mg/cm$^2$. Here the energy loss per cm is assumed to be 1.6 MeV cm$^2$/g, which corresponds to the mean energy loss in copper for 800 MeV protons. The requirement for picosecond timing resolution in ToF is due to the fact that in a real experiment, such as the LANSCE, the flight path ($L_0$) is limited to about 20 m.  The required timing resolution ($\delta \tau$) is related to energy resolution through $\delta \tau/\tau = 0.41 (\delta E/ E)$ for 800 MeV protons. $\delta \tau$ is about 3 ps for $\Delta E/E$ of 10$^{-4}$ and the flight time of 79 ns (20 m proton flight path). Sub-ps timing resolution is desirable but difficult in order to achieve energy resolution of 10$^{-5}$ for the same proton energy and real estate. The flight path of 20 m, or the relative positions of the timing detectors, needs to be fixed within 1-mm accuracy for 10$^{-4}$ energy resolution. 

One possible ultrafast timing option for proton ToF is to detect proton-induced Cerenkov light. The amount of light is crucial for precision timing. The number of visible photons (400-700 nm wavelength range) as a function of refractive index ($n$) and proton energy has been estimated for about a dozen materials with $n$ ranging from 1.33 (water) to 4.1 (germanium). The medium thickness is assumed to be 1 mm. For sufficiently large $n$, the number of photons may exceed 10 per proton, which remains a relatively small number. Similar to the 10-ps challenge for PET, see Sec.~\ref{sec:positron}, bright scintillators with very short decay time would be enabling for energy-loss or ToF proton radiography.

\subsection{Electrons}

Electrons in the energy range of 6 to 20 MeV have been used
in treatment of cancers of less than 5 cm depth for many
years~\cite{Addido:2017}. A portable electron radiography setup at the electron energy of 30 MeV has been reported~\cite{MHH:2017}.  Permanent magnet quadrupoles were used to focus electrons to form radiographic images of thin static and dynamic objects at about 2 m away.  The objects had a nominal areal density sensitivity range of 10 - 1000 mg cm$^{-2}$. The spatial resolution was found to be about 100 $\mu$m. Electron radiography was recently extended to 14 GeV at the Linac Coherent Light Source (LCLS)~\cite{MSFe3}, and also called transmission high energy electron microscopy (THEEM). In addition to the highly relativistic  electrons, an additional feature of the THEEM was the very short electron bunch duration down to 1 ps, which offers very high resolution for IT of dynamic processes. A 400-$\mu$m thick columnar CsI scintilator was used in conjunction with a CCD camera to collect the focused electrons with a spatial resolution below 10 $\mu$m. Recently, the development of higher-charge Laser Plasma Accelerator-driven electron production~\cite{MSFe4} has enabled ultrafast (sub-ps) imaging using electrons~\cite{MSFe5}. This source generates a broad spectrum of electrons, with a peak energy  above 200 MeV, and maximum in flux around 20 MeV. Using the OMEGA EP lasers as a source, target-on-detector and projection radiography has been demonstrated on Inertial Confinement Fusion (ICF) scale targets~\cite{MSFe6}. Work is underway now to implement a similar lens-based system within the confines of OMEGA EP to project high quality electron radiography onto a fast detector system, enabling flash electron radiography for ICF experiments.

\begin{figure}[htb!]
\centering\includegraphics[width=0.9\linewidth]{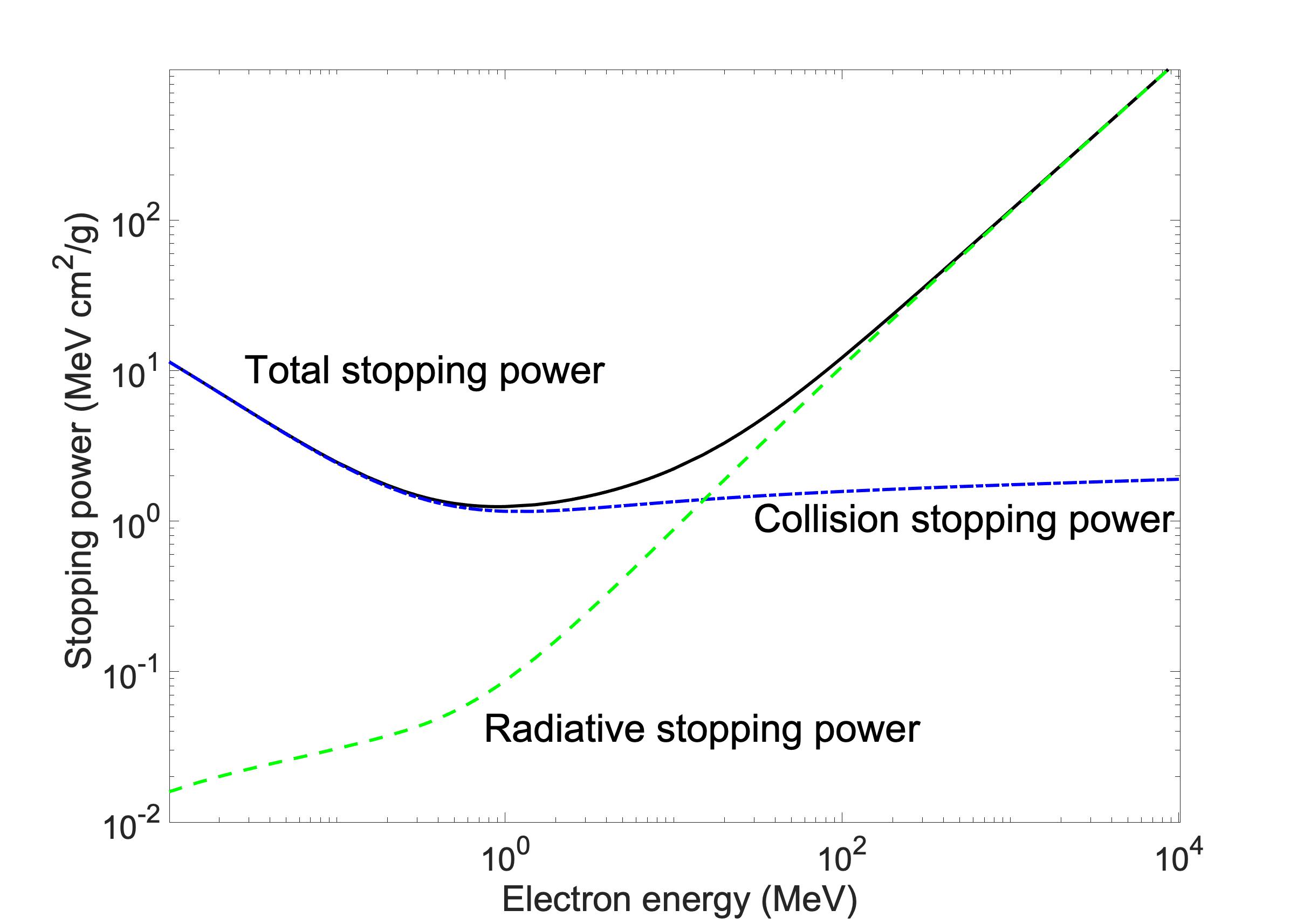}
\caption{Energy-dependent electron stopping power in LYSO with an atomic number ratio of Lu:Y:Si:O = 2(1-$x$):2$x$:1:5 and $x$=0.075. the density is assumed to be 7.2 g/cm$^3$. The data are from the NIST/ESTAR database.}
\label{fig:eLYSO}
\end{figure}

Energetic electrons interact with other electrons and nuclei through collisions and long-range Coulomb force, and lose energy through radiation of photons and direct energy transfers to other electrons in materials. Fig.~\ref{fig:eLYSO} shows the energy-dependent stopping power of energetic electrons in LYSO at energies up to 10 GeV. Collisional and Coulombic scattering energy loss dominate over the radiative energy loss  at low energies up to 15.2 MeV. 

One attractive potential of a THEEM as in the LCLS demonstration~\cite{MSFe3} is that the colocation of the electrons and the XFELs using a single (electron) accelerator would allow dual-probe radiography of electrons and X-rays. Such a dual-probe of electron and X-ray could be simpler than a dual-probe radiography based on protons and photons, which may require two accelerators, one for electrons, the other for protons~\cite{Barnes:2018}. Alternatively, due to the strong bremsstrahlung radiation from a relativistic electron beam, an electron radiography is intrinsically a multiple probe technique by itself if the bremsstrahlung information can be decoded. Different radiation damage mechanisms. Multi-GeV electron radiography or THEEM can potentially deliver 1 micrometer spatial resolution through objects of 1 mm thickness, if scintillator detectors can be optimized together with the beam optics and the electron source.

\subsection{Positrons \label{sec:positron}}
Inorganic scintillators are used in positron emission tomography~\cite{vE:2008}. Recent trends in Positron Emission Tomography (PET) make use of the time-of-flight (TOF) information to increase the signal-to-noise ratio (SNR) in the reconstructed image and improve the location of the annihilation event. The emergence of new technologies in the domain of nanophotonics, microelectronics, artificial intelligence, etc. open new perspectives for PET scanners to break present TOF performance limits. An ultimate goal of 10ps has been set in the frame of the 10 ps TOFPET challenge~\cite{b3}: https://the10ps-challenge.org

Achieving this ambitious goal would allow improving the PET effective sensitivity by a factor of 20 as compared to the best performing TOFPET today, the Biograph Vision from Siemens, opening the way to a reduction of the radiation dose (currently 5-25 mSv for whole body PET/CT), scan time (currently $>$ 10 minutes), and costs per patient (currently $>$ 1000 € per scan), all by an order of magnitude.
One of the most important components in TOF-PET instrumentation is the scintillation crystal. In spite of many efforts, in particular using co-doping strategies,  to reduce the delay between the creation of the hot electron-hole pairs and the capture of the resulting slow charge carriers by the luminescent centres after their multiplication and relaxation in the medium~\cite{L2, L3, L4},  standard scintillation mechanisms in inorganic scintillators are unlikely to produce a scintillation photon rate large enough to break the present barrier of about 200ps coincidence time resolution (CTR) in a realistic PET scanner. Two approaches are presently investigated to boost the timing resolution of scintillator-based X-ray and $\gamma$-ray detectors. The first one consists of exploiting the few Cerenkov photons produced by the recoil electron from the photoelectric $\gamma$-ray interaction in the medium. The other one is based on the concept of metascintillators 
where the recoil electron is sampled in thin layers of fast organic scintillators or ultrafast nanoscintillators.

A third, longer term possibility, is to boost the coupling of the electromagnetic wave associated to particles traversing a medium with the optical states in the medium, increasing therefore the scintillation quantum efficiency and exciton radiative recombination rate.

\subsubsection{Improving Time-of-Flight with Cerenkov light}
Due to their generally high densities and refractive indices, the majority of crystals used in PET scanners have a relatively low Cerenkov threshold, of the order of 100 keV. As the recoil electron following a photoelectric interaction has an energy of 511 keV minus the binding energy of this electron in the deep core level of the constituting heavy ions (91 keV in BGO,  63 keV in L(Y)SO) it is emitted with an initial energy of 420 keV and 448 keV in BGO and L(Y)SO respectively, i.e. much above the Cerenkov threshold. A number of Cerenkov photons (17 on average for BGO, and 12 for LYSO) are therefore produced and can potentially be used to time tag the $\gamma$ conversion events. This bunch of Cerenkov photons increases the photon rate in the leading edge of the scintillation pulse, as shown in~\cite{L5}.  

Considering the energy spectrum of the Cerenkov emission, with the highest number of photons emitted near the optical band edge of the crystal, where the probability of self-absorption is the highest, due to non-radiative impurities and defects generally concentrated in this region, as well as the non-spectral-matching quantum efficiency of the photodetectors, the number of detected photoelectrons is generally very small, not more than 5 on average per event, with large fluctuations from event-to-event. This poses severe constraints on the electronics and results on a non-negligible number of events, where zero Cerenkov photons are detected in at least one (if not both) crystals in coincidence. 
However, a clever sorting of all the events in a data set in several classes associated to different amounts of Cerenkov photons detected in BGO crystals in coincidence, has led to a significant improvement in the CTR for a reasonable number of events, which can provide useful information for improving the signal-to-noise ratio of the reconstructed image ~\cite{L6}. The value of about 200ps obtained for 20mm long BGO pixels is interesting as it is similar to the state-of-the-art with LYSO crystals in the Biograph Vision PET scanner from Siemens, but with the 3 times cheaper BGO than LYSO. 

\begin{figure}[htb!]
    \centering
    \begin{subfigure}{0.4 \textwidth}
    \includegraphics[width= \textwidth]{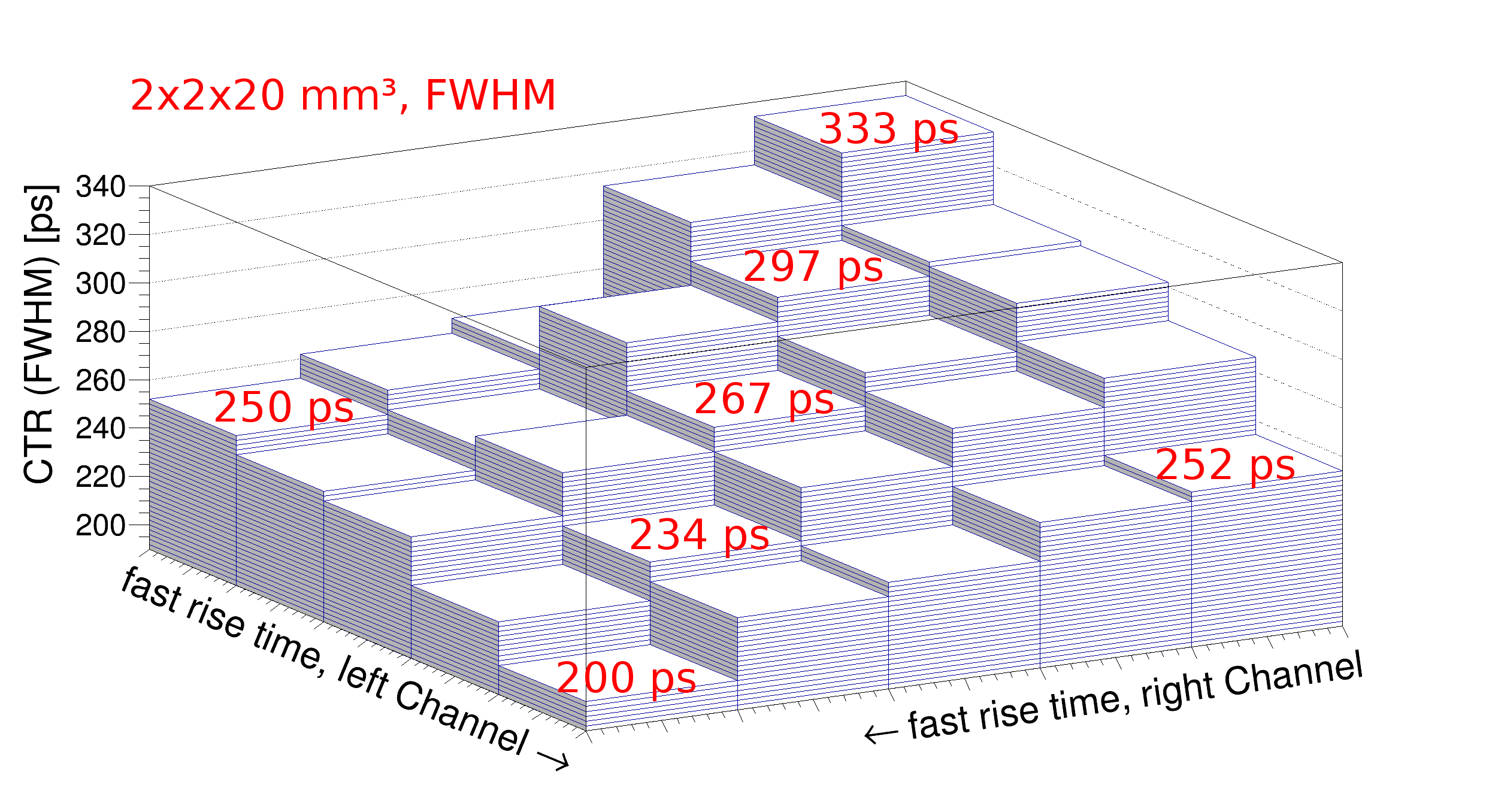}
    \caption{}
    \label{}
    \end{subfigure}
    \hfill
    \begin{subfigure}{0.4 \textwidth}
    \includegraphics[width= \textwidth]{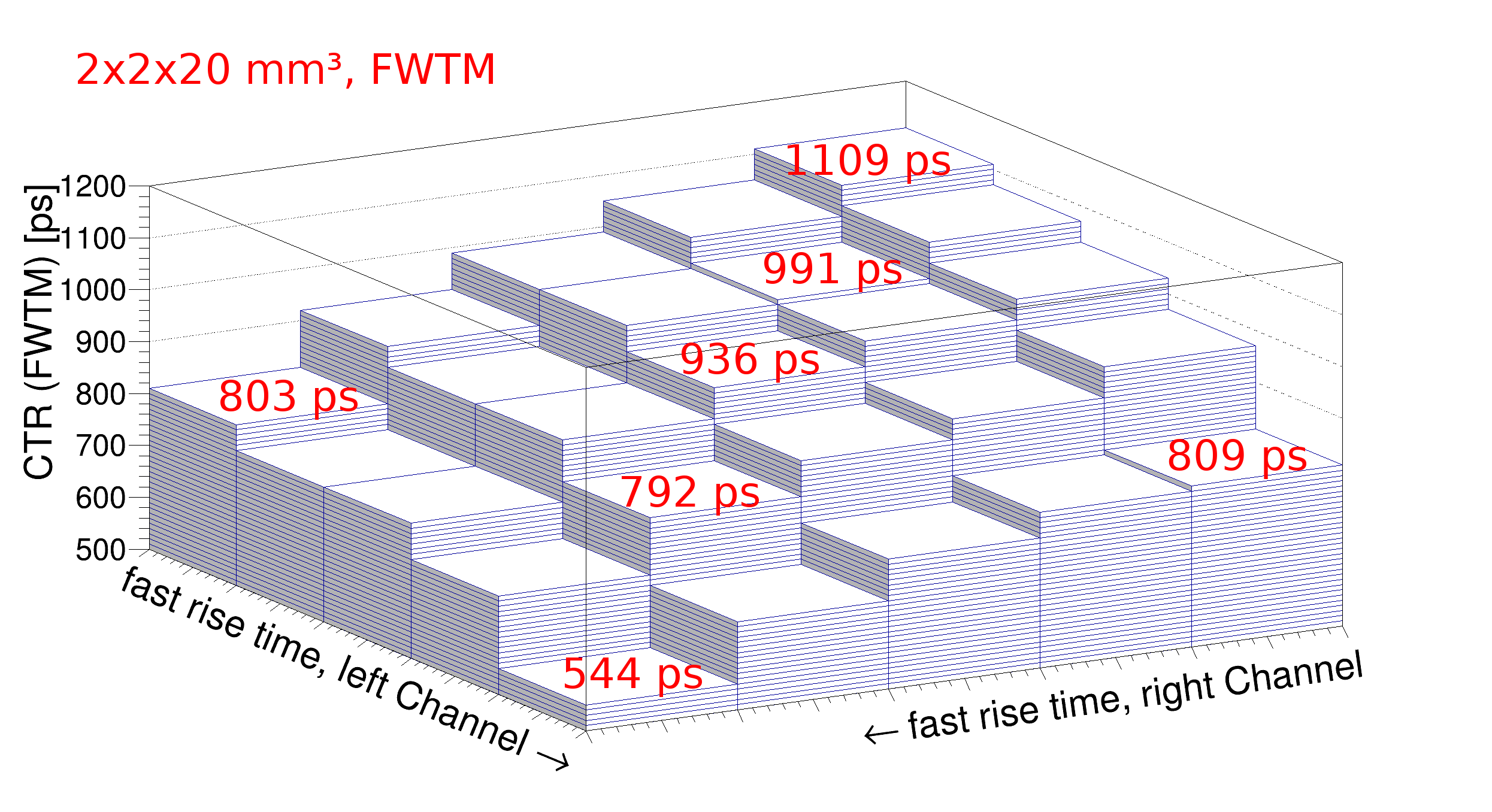}
    \caption{}
    \label{}
    \end{subfigure}
\caption{CTR FWHM (top) and FWTM (bottom) from the 25 coincidence categories between 5 classes of events per detector (20\% of events each) sorted by their individual timing resolution in 20mm long BGO pixels. From Ref.~\cite{L6} with permission.}
\label{fig:Le1}
\end{figure}

Another interesting approach, based on Cerenkov Charge Induction (CCI) detectors, is to use a non-scintillating semi-conductor Cerenkov radiator, such as TlCl or TlBr. The principle is to take benefit from the excellent energy resolution of these semi-conductor crystals by the direct electrons and holes collection, while using the clean fast Cerenkov light signal to time tag the event at the level of 320ps ~\cite{L7, L8}. Moreover, the high density and photoelectric cross section of TlCl and TlBr @511 keV are a clear advantage for a good PET sensitivity.
It remains to be seen if the low noise and high frequency electronics used in these different attempts to exploit the Cerenkov radiation can be scaled-up at a reasonable cost for a PET scanner. On the other hand, this can open the way to a new life for BGO, 3 times cheaper than LYSO, for a final objective of 300 to 500ps CTR at a system level, which can offer an attractive solution for a performant cost-effective Total Body PET. 

\subsubsection{Metascintillators~\label{sec:meta}}
\begin{figure}[htbp]
    \centerline{\includegraphics[width=0.7\columnwidth]{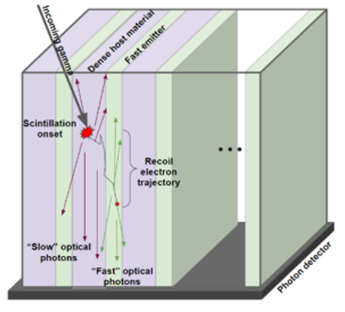}}
    \caption{The metascintillator concept. From Ref.~\cite{L13} with permission.}
    \label{fig:Le2}
\end{figure}

The metascintillator concept, introduced in 2008~\cite{L9} and first tested in 2017~\cite{L10}, is based on composite scintillator topologies allowing the sampling of the recoil electron produced by the $\gamma$-ray conversion in dense scintillator regions in much faster scintillators, such as organic, cross-luminescent or nano-scintillators (Fig. ~\ref{fig:Le2}).

A first generation of metascintillators is now extensively developed by several groups, combining the high stopping power of BGO or LYSO crystals with the fast emission of plastic scintillators~\cite{L11} and/or cross-luminescent crystals, such as BaF\textsubscript{2}~\cite{L12, L13}.
The stochastic nature of the energy sharing between the metascintillator components introduce some challenges on the energy resolution and event selection if the different components have a different light yield, as well as on the timing resolution which is different from event-to-event, depending on the amount of energy leakage to the fast emitting regions. However, different surrogates, based on the light pulse analysis, allow to estimate the amount of energy sharing on an event-to-event basis and to apply the necessary corrections~\cite{L14}.

The results obtained so far for this first generation of metascintillators confirm that a CTR of 200 ps at the system level can be reached with BGO-based metascintillators, i.e. equivalent to the state-of-the-art with bulk LYSO crystals, but with scintillators (BGO, plastic, BaF\textsubscript{2}) much cheaper than LYSO. On the other hand, LYSO-based metascintillators allow reaching 100 ps CTR, i.e. twice better than the state-of-the-art.
It is worth noticing that the concept of metascintillator has been recently extended to a semi-monolithic block geometry~\cite{L12}, benefiting from these excellent timing properties, while allowing the determination of the depth of interaction (DOI) of the $\gamma$-ray interaction at a 2 to 3 mm precision. From the knowledge of the DOI, a correction on the timing can be applied, allowing a further improvement of 20 to 30 ps to the CTR.  

A second generation of metascintillators is being prepared, where plastic and BaF\textsubscript{2} will be replaced by nanoscintillators. Indeed, the quantum confinement of the excitons in nanostructures (quantum dots, nanorods, nanoplatelets, etc.) considerably increases the quantum efficiency and the rate of radiative recombination of the electron-hole pairs. Decay times of 100 to 500 ps have been observed in ZnO:Ga quantum dots~\cite{L15}  or CdSe nanoplatelets (NPL)~\cite{L16}.

\begin{figure}[htbp]
    \centerline{\includegraphics[width=\columnwidth]{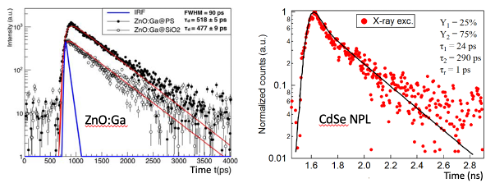}}
    \caption{X-ray excited scintillation decay time of ZnO:Ga nanopowder deposited on polystyrene or SiO2 (left) and of CdSe nanoplatelets (right). Adapted from Ref.~\cite{L15} and ~\cite{L16} with permission.}
    \label{fig:Le3}
\end{figure}

This has triggered extensive research by several groups all over the world, with a particular focus on CdSe NPLs~\cite{L17}, ZnO:Ga nanopowder~\cite{L18}, but also perovskites~\cite{L19}  and in particular mixed inorganic-organic perovskites such as CsPbBr3 with methylammonium as well as GaN/InGaN multiple quantum wells~\cite{L20}.

An interesting attempt has been made to deposit layers of CsPbBr\textsubscript{3} nanocrystals on GAGG:Ce plates, resulting in a clearly identified bunch of prompt photons on top of the GAGG scintillation pulse~\cite{L19}. 
A number of problems remain to be solved for benefiting from the excellent scintillating properties, and in particular the ultrafast emission of nanoscintillators. These problems are related to the important self-absorption in several of them and the long-term stability of perovskites. We can fortunately rely on the huge R\&D and industrial effort on perovskites for photovoltaic applications and we can expect progress on the large scale and cost-effective production of excellent quality and stable perovskites.  Different solutions are being investigated to mitigate the problem of self-absorption.  One of them consists of embedding them in transparent organic materials such as polyethylene or polystyrene~\cite{L22}, with different strategies to transfer the excitation from the nanocrystals to or from the organic host~\cite{L23}.

\begin{figure}[htbp]
    \centerline{\includegraphics[width=0.9 \columnwidth]{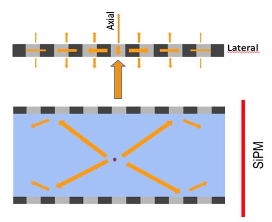}}
    \caption{PhC change of propagation mode from axial to lateral. Courtesy of L. Zhang, Multiwave Metacrystal SA.}
    \label{fig:Le4}
\end{figure}

Another approach is to use the high potential of photonic crystals (PhC) to control the behavior of optical photons. In a perpendicular arrangement relative to the photodetector entrance face, PhCs can reorient the propagation modes, generally isotropic at the point of light emission into lateral propagation modes in the direction of the photodetector, reducing therefore the self-absorption in the material and the time jitter at the reception of these photons (Fig.~\ref{fig:Le4}).

Different PhC designs have been simulated, using several approaches such as Rigorous Coupled Wave Analysis, Plane Wave Expansion, Guided Mode Expansion and Finite-difference time-domain methods, implemented in proprietary and open-source software. All these optical simulations demonstrate the ability of designed PhC slabs to bend light towards the metascintillator extraction surface, reducing the propagation modes and associated time jitter of photons within the scintillator.

\subsubsection{Toward a 10-ps TOFPET \label{sec:psch}}

A number of nanophotonic features would allow to considerably increase the coupling of the electromagnetic wave associated to a particle traversing a medium to the optical states in this medium. Hyperbolic metamaterials have already been proposed to boost the photon conversion efficiency in well-defined regions of SiPM photodetectors, an additional key to improve the timing resolution of scintillators-based ionization radiation detectors ~\cite{L24}. Another possibility is to take advantage of the Purcell effect in polaronic structures engineered in the crystal to create local electromagnetic cavities and increasing the local density of optical states in the material~\cite{L25}.  

An ultimate approach for improving the time resolution is to exploit sub-ps transient phenomena taking place when ionization radiation interacts with a medium. In transparent scintillators the polarization of the environment near the particle trajectory produces hot polaronic states, which locally modify the refractive index and polarization of the medium ~\cite{L26}. An intermediate goal towards this 1 ps target, which cannot presently be achieved with scintillating detectors, is to develop faster alternatives to probe these transient phenomena with high frequency external signal, following the principle of a heterodyne detection ~\cite{L27}. Only such a paradigm-change of detecting optical modulation of a laser probe by ionizing radiation would allow to go faster than scintillation with a femtosecond temporal resolution ~\cite{L28}. This novel approach would require novel detector configuration to achieve high sensitivity to small optical modulations.

\subsection{Dosimetry in radiation therapy}
More than 1 million new cases are diagnosed, and more than 600,000 people die from cancer in the US every year. Besides surgery and chemotherapy (now immunotherapy as well), radiotherapy (RT) is a standard and effective treatment option used for nearly 50-70\% of cancer patients. RT is more than 120 years old, and it all started with two Nobel prize laureates: Marie Curie-Sklodowska (1867-1934) and Wilhelm R\"ontgen (1845-1923). 

There are several means of delivery in RT: (1) brachytherapy, in which relatively weak radioactive sources are surgically implanted into the patient to kill tumors; (2) early external beam RT (obsolete in the US already), in which strong radioactive sources (such as Colbalt 60) are used at a distance of $\sim$80 cm from the patient with collimators to kill the tumors; (3) contemporary external beam RT, in which electron accelerators (i.e., linear accelerators) produce photon beams through bremsstrahlung to kill the tumors; and (4) particle therapy, in which protons and heavier particles are used to kill the tumors.    

\begin{figure}[htb!]
\centering\includegraphics[width=0.9\linewidth]{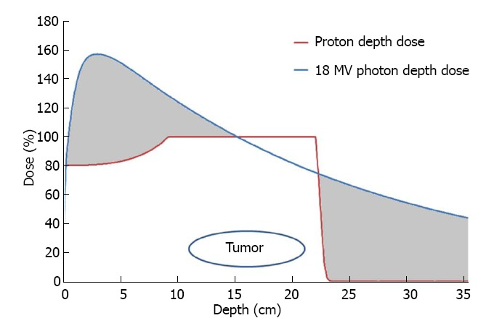}
\caption{Depth-dose comparison of a photon beam and a proton beam. Figure courtesy of Schild et al.~\cite{M1}, used with permission.}
\label{fig:Mayo1}
\end{figure}

\begin{figure}[htb!]
\centering\includegraphics[width=0.9\linewidth]{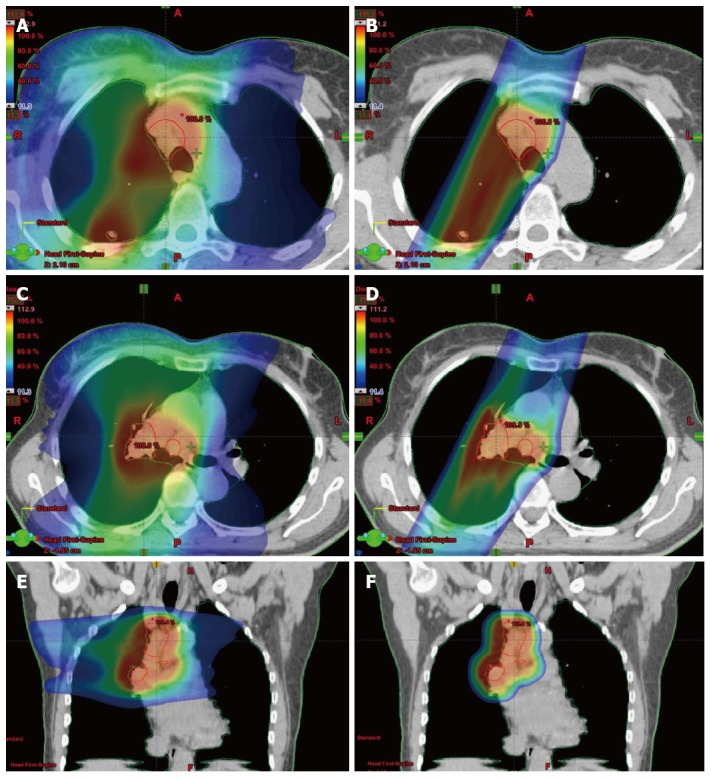}
\caption{Dosimetric comparisons between photon [left (A, C, E) and proton right (B, D, F)] plans for a stage III lung cancer patient. Radiation doses shown as color washes warmer colors (red) represent higher doses compared to cooler colors (blue). The first pair of axial slices (A and B) are at a level superior to axial cuts C and D; It is apparent that more normal lung is irradiated on the left than the right; E and F are coronal cuts with dosimetry (Figure courtesy of Schild et al.~\cite{M1}).}
\label{fig:Mayo2}
\end{figure}

To improve patient outcomes, it is important to deliver the tumoricidal dose to tumors, while minimizing the possible adverse events (AEs) in organs at risk (OARs). Proton therapy (PT) helps in this regard~\cite{M1,M2,M3,M4,M5}. Since proton beam has a finite range (i.e., the Bragg Peak), it can be designed to stop right after the tumor site, and thus no dose is delivered beyond the tumor (Fig.~\ref{fig:Mayo1})~\cite{M1}. These features provide highly conformal tumor (target) coverage, while sparing adjacent OARs. Spot scanning proton therapy (SSPT), the most advanced generation of PT, delivers dose as a pencil-sized spot (beamlet) in the patient’s anatomy for 3D dose painting, allowing an even more precise and conformal radiation dose delivery (Fig.~\ref{fig:Mayo2})~\cite{M1}. Therefore, SSPT achieves better tumor control for cancer patients, while better protecting nearby OARs. 

Scintillators can be applied as radiation dosimeters for machine characterization and patient dose measurement in radiation therapy. One common detector format is the miniature plastic scintillation dosimeter (PSD), where a small volume of plastic scintillator is attached to a fiber-optic light guide leading to readout electronics.  The intensity of the scintillator light output can be converted to an absolute radiation dose reading.  
Table.~\ref{Mayo:tb} compares scintillator materials that have been used or considered for radiation dosimetry applications.  Organic scintillators have many good physics characteristics for radiation dosimetry, including water equivalence in the clinical photon and electron beam energy range with minimal beam perturbation and energy-independency above 125keV, reproducibility, linearity, and dose rate independence~\cite{M6, M7, M8}. Due to these unique physical characteristics, scintillators usually provide similar dosimetry results to ionization chambers (the standard radiation detectors for most radiation therapy applications) in both photon and electron beams, while being much smaller and much more flexible~\cite{M7,M9}. Certainly the temperature dependence of scintillator light output needs to be considered to get correct results, especially in cases of in-vivo dosimetry where the scintillator will be operated at body temperature but may be calibrated at room temperature~\cite{M10}.  It is also important to correctly account for Cerenkov light produced in the optical fiber~\cite{M6,M11}.

\begin{table*}[htbp]
\caption{Comparisons of scintillators used in radiation dosimetry applications.}
\centering
        \begin{tabular}{|l|c|c|c|c|c|c|c|}
            \hline
            Scintillator & Type & $\lambda_{\max}$ (nm) & Brightness (\% Anthracene) & Density (g/cm\textsuperscript{3}) & $n$ & Z\textit{\textsubscript{eff}} & Hygroscopic \\[3pt]
            \hline
            BC-400~\cite{M12}& Organic plastic & 423 & 65\% & 1.032 & 1.581 & 5.7 & No \\[3pt]
            \hline
            EJ-260~\cite{M13} & Organic plastic & 490 & 60\% & 1.023 & 1.58 & 5.7 & No \\[3pt]
            \hline
            BC-531~\cite{M14} & Organic liquid & 425 & 59\% & 0.87 & 1.47 & $\sim$6 & No \\[3pt]
            \hline
            Scintacor GS1~\cite{M15} & Ce-doped glass & 395 & $\sim$17\% & 2.64 & 1.58 & $\sim$14 & No \\[3pt]
            \hline
            ZnSe:O~\cite{M15} & Inorganic crystal & 595 & 418\% & 5.27 & 2.66 & 32 & No \\[3pt]
            \hline
            CsI:Tl~\cite{M15}& Inorganic crystal & 530 & 327\% & 4.51 & 1.79 & 54 & Slightly \\[3pt]
            \hline
        \end{tabular}
\label{Mayo:tb}        
\end{table*}

Inorganic scintillators, while lacking radiological water-equivalence, typically provide a higher light output and are denser, which can be advantageous in some applications~\cite{M15}  However, the high cost and the size limitations imposed by the need to grow single crystal ingots limit the application of inorganic scintillators for some applications.  One example is 3D scintillation dosimetry, which has used organic liquid~\cite{M16} and plastic~\cite{M13} scintillators due to their low cost and the ability to make large scintillator volumes in almost any shape or size.

Scintillators have been widely used in radiation therapy, especially in X-Ray based photon therapy. The major clinical applications include: (1) small field dosimetry and (2) in-vivo dosimetry. With the increased use of stereotactic body radiation therapy (SBRT) and stereotactic radiation surgery (SRS) (for example, GammaKnife and CyberKnife) to treat small tumors, small field dosimetry has become more important in radiation therapy. Unfortunately, it is very challenging to get accurate results in small field dosimetry due to (1) loss of charged particle equilibrium, (2) energy dependence, and (3) partial volume effects if the detector is too large (e.g., ion chambers). Scintillators are an ideal solution in these scenarios. Due to the radiological water-equivalence of plastic scintillator materials, charged particle equilibrium is maintained and the detectors’ energy dependence mimics that of water. The high light output of many scintillators allows PSD’s to be made very small, avoiding partial volume effects. Underwood et al. and Morin et al. have shown that given a field size smaller than 10 mm, the measurement results from scintillators were closer to the Monte Carlo simulation results compared to diode, microdiamond, and microLion chambers~\cite{M17,M18}, which are commonly used in the small-field dosimetry. Some commercial products have become available in the market, including the Exradin W1 and W2 Scintillator detectors (Standard Imaging, Middleton, WI), which have become popular for small-field dosimetry~\cite{M8}.  

Another important application of scintillator dosimetry in X-Ray-based radiation therapy is in-vivo dosimetry, due to their radiological water-equivalence, small size, freedom from beam perturbation, and their ability to provide accurate dose measurements without the use of a high bias voltage.  The flexibility of PSD’s is well suited, for example, to rectal balloon mounted detectors for prostate dose verification such as the OARtrac Dose Monitoring System (RadiaDyne, Houston, TX) and detection of radioactive source position in needles or catheters for high dose rate (HDR) brachytherapy~\cite{M19}.   

\begin{figure}[htb!]
\centering\includegraphics[width=0.9\linewidth]{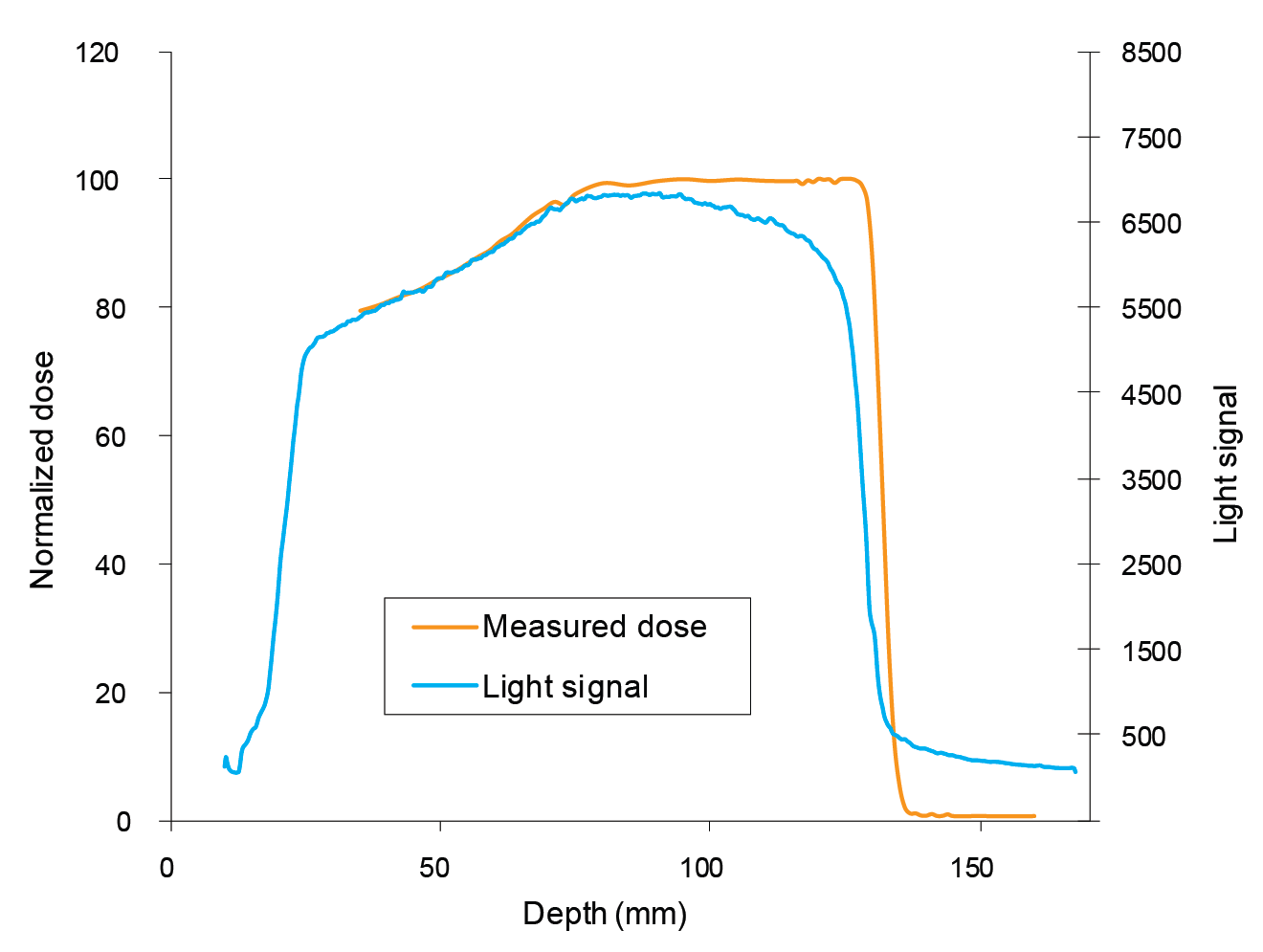}
\caption{Depth-dose curves of a proton spread-out Bragg peak measured with an ionization chamber (orange) and a liquid scintillator detector (blue). The scintillator signal is depressed near the end of the beam range due to ionization quenching.}
\label{fig:Mayo3}
\end{figure}

Scintillators also have applications in machine quality assurance (QA) for photon external beam radiation therapy.  Scintillator detectors have been developed for machine testing of X-Ray radiosurgery devices~\cite{M20} and to measure patient treatment plans for X-Ray-based radiation therapy~\cite{M21}.  
Scintillator detectors have also found extensive use in proton beam therapy applications for beam performance quality assurance testing.  Errors in the proton beam range might result in missing the target or overdosing nearby critical structures, both leading to unfavorable patient outcomes. Therefore, it is critically important to check proton beam range routinely.  American Association of Physicists in Medicine (AAPM) Task Group (TG) 224~\cite{M22} recommends a monthly consistency check of proton beam range. However, the conventional method of using a multiple layer ion chamber (MLIC) can be very time consuming and provides limited spatial resolution. Scintillator detectors have been developed specifically for efficiently measuring proton beam range, which provide fast, accurate, and high-resolution beam range measurements~\cite{M23,M24}. Scintillator-based detectors are also used for other proton machine QA tests, including beam isocentricity~\cite{M25}, pencil beam profile and positioning testing~\cite{M26}, and daily comprehensive beam testing~\cite{M27}.  One limitation of scintillator detectors for proton therapy is the ionization quenching of most scintillator materials.  As the protons’ linear energy transfer (LET) increases towards the end of their range~\cite{M28}, the linear dose-to-light relationship breaks down, leading to an under-response of the scintillator, as shown in Fig.~\ref{fig:Mayo3}.  This relationship is described by the well-known Birks relationship~\cite{M29}, which can be used to provide quenching correction factors, presuming the LET can be determined~\cite{M14,M30}. 

In summary, radiation therapy is an important tool in the treatment of cancer, including brachytherapy, external beam radiation therapy, and particle therapy. Scintillator dosimetry plays an important role in radiation therapy; however, its potential has not been fully exploited. Additional R\&D and collaboration, especially from outside the medical physics community, is needed to use scintillator dosimetry more to improve the therapeutic efficacy of radiation therapy.

\section{Recent scintillators and concepts \label{scin:new}}
In addition to progress in new inorganic scintillators, Sec.~\ref{sec:utk}, we shall highlight results and progress in structured scintillators presented during the SCINT22 conference, ranging from nanostructures, Sec.~\ref{sec:nano}, micrometer-thick thin films, Sec.~\ref{sec:duj}, to bulk composite structures, Sec.~\ref{sec:bren}. As presented, each topic emphasizes a particular application: Sec.~\ref{sec:utk} on new organic scintillators for fast timing, Sec.~\ref{sec:nano} on nanostructures for light guiding and higher light yield, Sec.~\ref{sec:duj} on thin film for high X-ray imaging resolution, Sec.~\ref{sec:bren} on bulk composites for fast neutrons. Apparently, many other new applications in RadIT may be found for each of the scintillators and novel structures.

\subsection{Novel Ultrafast Inorganic Scintillators \label{sec:utk}}

The demands in recent years for scintillators with faster timing capabilities have motivated research efforts toward development of new ultrafast materials. Much of this work is aimed toward satisfying the requirements for high energy physics (HEP) experiments and time-of-flight positron emission tomography (TOF-PET) with an emphasis on pushing the limits of timing resolution, as explained in Sec.~\ref{sec:positron} above. One of the main factors limiting detector time resolution is the photon time density of the scintillation pulse, or the ratio of light output to decay kinetics of the emitted light. The best coincidence time resolution (CTR) can be obtained with a combination of high light yield (LY) and short decay ($\tau$\textsubscript{d}) and rise ($\tau$\textsubscript{r}) times as described by the relationship \textit{CTR} $\propto \sqrt{\tau \textsubscript{r}\tau \textsubscript{d}/\textit{LY}}$.

\begin{figure}[htbp]
    \centerline{\includegraphics[width=\columnwidth]{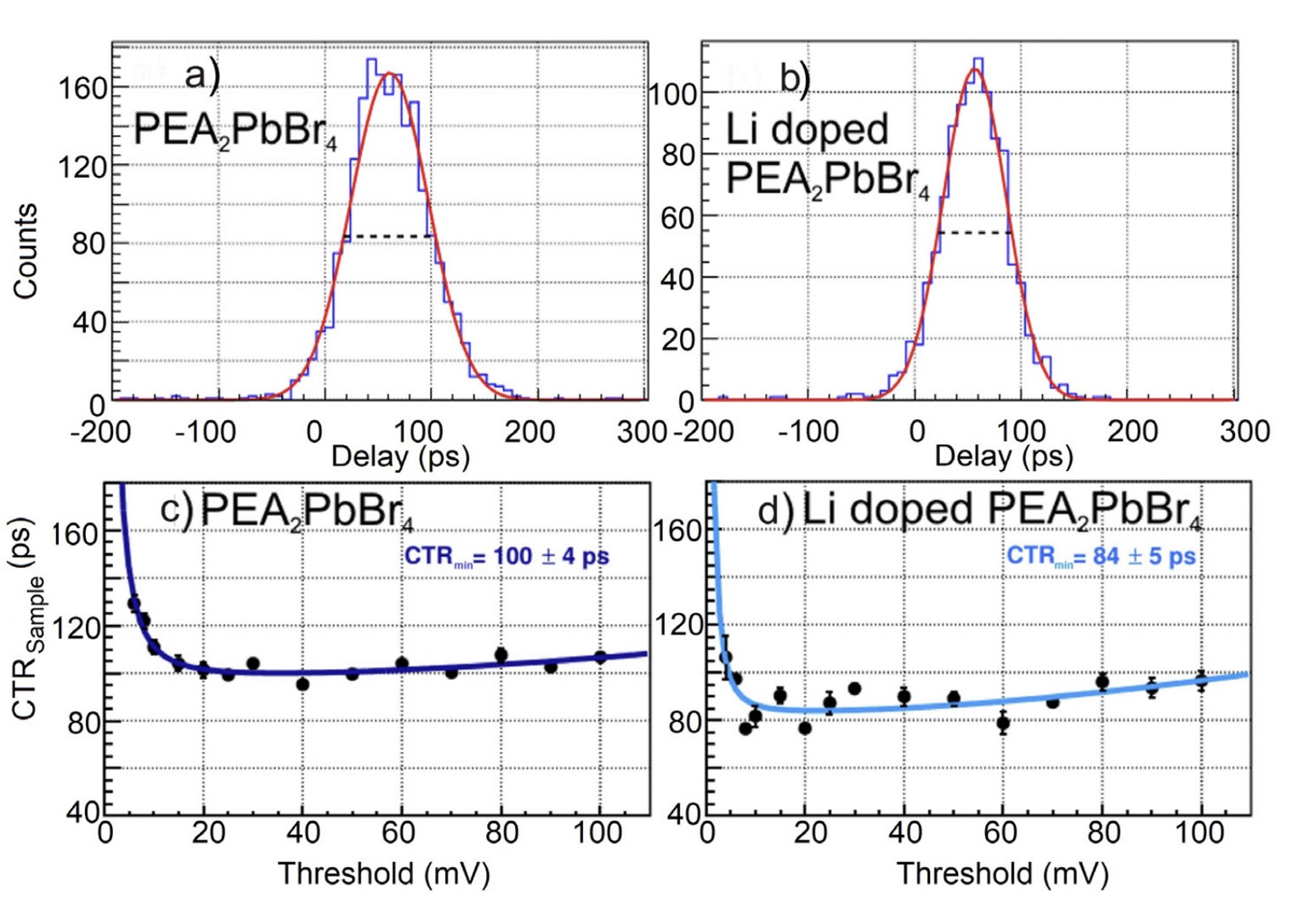}}
    \caption{Delay distributions at 20 mV of (a) undoped and (b) Li-doped PEA\textsubscript{2}PbBr\textsubscript{4} showing the FWHM (dotted lines). CTR\textsubscript{sample} vs leading edge threshold for (c) undoped and (d) Li-doped PEA\textsubscript{2}PbBr\textsubscript{4} crystals showing a CTR\textsubscript{min} below 100 ps when Li-doped. Reprinted from \cite{b8}, with the permission of AIP Publishing.}
    \label{figUTK1}
\end{figure}

Other review papers \cite{DAB:2018,b3,b4} have already addressed the specific needs for these applications and discussed processes for achieving ultrafast (sub-nanosecond up to a few nanoseconds decay times) emission with inorganic materials – Cerenkov, hot intra-band luminescence, and core-valence luminescence. Sub-ns decay times resulting from quantum confinement effects are also possible in nanocrystals and nanocomposites \cite{b5,L22,b7}. For example, a prompt component of 0.79 ns (17\% abundance) and effective decay time of about 1.9 ns is observed for lead halide perovskite CsPbBr\textsubscript{3} nanocrystals embedded in a polystyrene matrix, which has been measured to have time resolution twice as good as LYSO (used in commercial TOF-PET scanners) at low energies of 10 keV~\cite{L22}. Lead halide perovskites used in bulk crystalline form have also shown promising timing characteristics. Sub-100 ps CTR was recently demonstrated with a two-dimensional hybrid organic-inorganic halide perovskite scintillator (Fig.~\ref{figUTK1}), Li-doped PEA\textsubscript{2}PbBr\textsubscript{4} (CTR of 84 ps FWHM) \cite{b8}. Semiconductor scintillators, such as ZnO:Ga \cite{b9,b10}, are another class of material being considered for fast timing applications due to their ultrafast decay times.

Besides exploiting fast emission processes in inorganic materials, an alternative strategy for designing fast timing detectors currently receiving a great deal of attention is the concept of heterostructures or metascintillators \cite{L11,L12,L10}. Since a high detection efficiency is required for applications such as TOF-PET, plastic scintillators alone are not a suitable choice (due to low Z\textit{\textsubscript{eff}} and density) despite having ultrafast timing properties (about 700 ps decay times for EJ-232Q and BC-422Q). By alternating between layers of a dense scintillator, like BGO, and a fast plastic scintillator, the desired properties from each can be harnessed to achieve better timing resolution without sacrificing detection efficiency. CTR as good as 55 ps full width at half maximum (FWHM) has been measured with LYSO + BC-422 \cite{L10}. 

Of the various potential avenues for achieving faster timing performance, core-valence luminescence (CVL) is of particular interest due to overall well-balanced set of properties that can be obtained – sub-ns decay time, moderate density (3 to 6 g/cm\textsuperscript{3}), good chemical stability, and relatively bright emission, for example. Unlike Cerenkov emission and hot-intraband luminescence, which produce very few photons per gamma interaction ($\sim$17 photons per 511 keV gamma for BGO \cite{b14}), CVL scintillators typically have light yields in the range of 1,000 to 2,000 ph/MeV (at 662 keV), making them more practical for use in a wide range of applications. Likewise, the ability for these materials to be used in bulk form without significant effects from self-absorption provides an advantage over semiconductors. The generally higher density and Z\textit{\textsubscript{eff}} of fully inorganic CVL scintillators compared to halide perovskite nanocomposites and hybrid organic-inorganic crystals provides an advantage over these materials. With these considerations in mind, highlights from recent developments on CVL scintillators will be presented next, and areas in which future efforts should focus will be identified.

Early investigations into CVL in the 1980’s and 1990’s focused mainly on BaF\textsubscript{2}, CsF, CsCl, and CsMCl\textsubscript{3}-type (M = Mg, Ca, Sr) compounds. By far the most successful material to emerge from this group has been BaF\textsubscript{2}, a commercially available scintillator with a 600 ps decay time and relatively high CVL light yield of 1,400 ph/MeV (at 662 keV) that is still being considered for use in fast timing applications \cite{b15}. The drawbacks, however, lie in its VUV emission that is not well-matched with the spectral sensitivity of common photodetectors, as well as the presence of a dominant slow decay component ($\sim$630 ns). Advances in photodetector technology (e.g. VUV sensitive silicon photomultipliers (SiPMs)) \cite{b16} and the use of rare-earth ion doping (BaF\textsubscript{2}:Y) \cite{b10} to suppress the slow emission have allowed the possibility for better performance with BaF\textsubscript{2} (CTR of 51 ps FWHM) \cite{b17}, however, there is still interest in finding alternative materials that bypass these issues completely.

A brief description of the core-valence luminescence mechanism, also called cross luminescence (the term used in Sec.~\ref{sec:meta}), in halide scintillators is warranted in order to establish the relevant compositional space in which new materials may be discovered. For a detailed description of the process, refer to \cite{b18}. To summarize, an electron from the valence band (comprised mainly of the halogen p orbital) may recombine with a hole generated in the outermost core level (comprised mainly of the metal cation p orbital) leading to the emission of light if the condition E\textsubscript{vc} $<$ E\textsubscript{g} is satisfied, where E\textsubscript{vc} is the energy difference between the top of the valence band and top of the outmost core level and E\textsubscript{g} is the bandgap energy. It is apparent based on this criterion that wide bandgap halides (i.e. fluorides and chlorides) are most favorable. The compositional space of interest can further be narrowed down to compounds containing Ba, Cs, Rb, and K \cite{b18}. Impurity-induced CVL has also been demonstrated in several Cs- and Rb-doped materials \cite{b19,b20}.

Novel CVL scintillators that have been discovered in the last decade include Rb\textsubscript{2}ZnCl\textsubscript{4} \cite{b21,b22}, Cs\textsubscript{2}BaCl\textsubscript{4} \cite{b23}, Cs\textsubscript{3}ZnCl\textsubscript{5} \cite{b24}, K\textsubscript{2}BaCl\textsubscript{4} \cite{b25}, Rb\textsubscript{2}BaCl\textsubscript{4} \cite{b25}, and BaGeF\textsubscript{6} \cite{b26}. In addition to these novel materials, there has been renewed interest in the more traditional CVL scintillators CsSrCl\textsubscript{3}, CsMgCl\textsubscript{3}, and CsCaCl\textsubscript{3} due to the advancements made to photodetector technology and signal processing methods since they originally drew interest. The most promising of these seems to be CsCaCl\textsubscript{3} due to its high light yield of 1,371 ph/MeV and fast decay time of 2.47 ns. The CTR has recently been reported to be 148 ps FWHM for a 2 × 2 × 3 mm\textsuperscript{3} CsCaCl\textsubscript{3} pixel measured with a VUV SiPM (Hamamatsu, S13370–3075CN), which is superior to that of BaF\textsubscript{2} (CTR of 164 ps FWHM) measured with the same setup \cite{b25}.

\begin{figure}[htbp]
    \centerline{\includegraphics[width=\columnwidth]{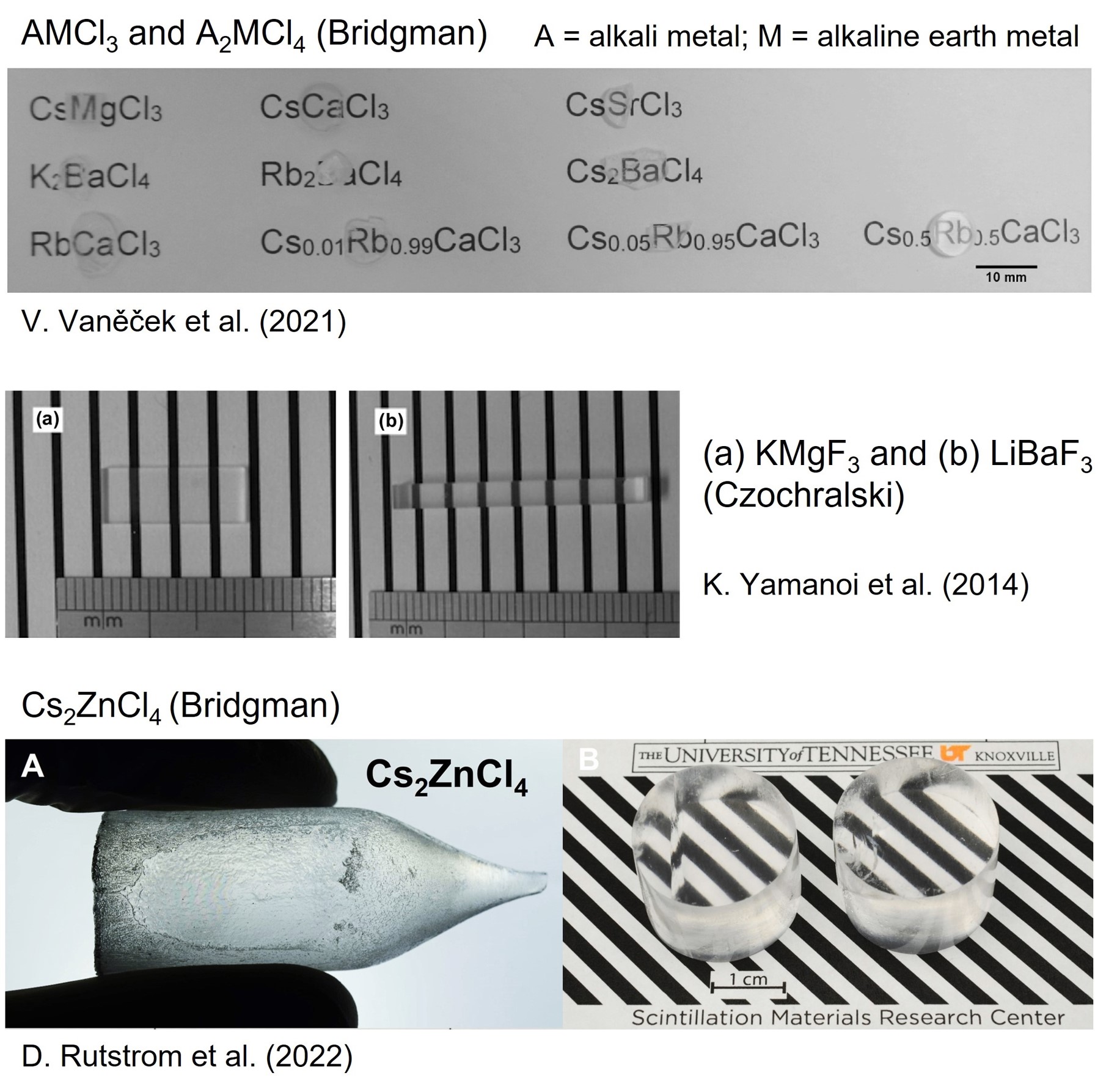}}
    \caption{Examples of some CVL crystals grown in the last decade. The growth methods are noted in parentheses next to each composition. Reprinted from \cite{b25}, \cite{b27}, and \cite{b28}, with permission from Elsevier.}
    \label{figUTK2}
\end{figure}

Cs\textsubscript{2}BaCl\textsubscript{4} is one of the fastest and brightest (1.68 ns decay time and 1,369 ph/MeV light yield) materials recently studied in \cite{b25}. In a separate study, an even shorter decay constant of 1.2 ns and higher light yield of 1,700 ph/MeV (for the fast component) are reported \cite{b23}. Unfortunately, the instability of this compound at room temperature may hinder its usage, as Cs\textsubscript{2}BaCl\textsubscript{4} reportedly decomposes upon cooling \cite{b25,b29}. This means growth from the melt will present substantial challenges.

Similar to the resurgence of AMCl\textsubscript{3}-type compounds, Cs\textsubscript{2}ZnCl\textsubscript{4} has resurfaced as a promising new ultrafast scintillator despite being discovered almost 20 years ago. Between 2003 (when it was initially discovered) and 2019 there were only 3 publications that reported scintillation properties of Cs\textsubscript{2}ZnCl\textsubscript{4} \cite{b30,b31,b32}. In the past few years alone, that number has now doubled \cite{b24,b28,b33}. This is partly due to improvements that have been made to crystal growth techniques that have allowed for better quality and larger volume crystals to be fabricated. Fig.~\ref{figUTK2} shows some examples of different CVL crystals grown in recent years.

\begin{figure}[htbp]
    \centerline{\includegraphics[width=\columnwidth]{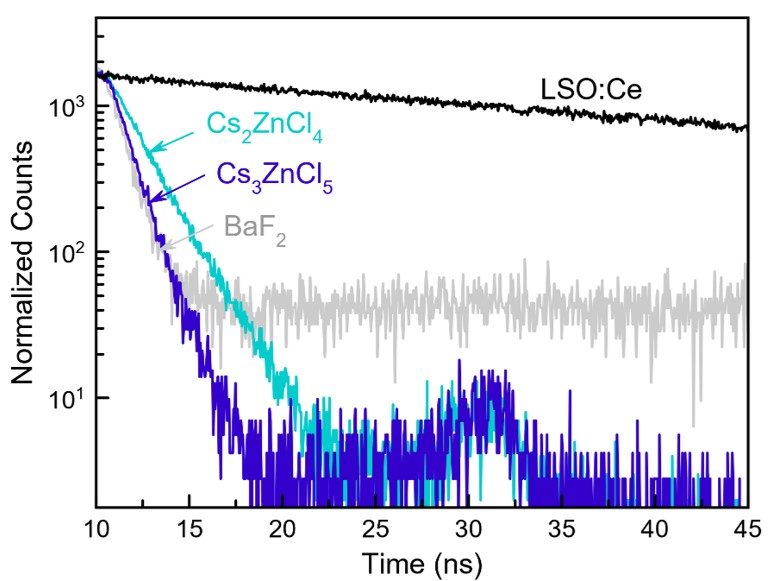}}
    \caption{Scintillation decay time profiles of Cs\textsubscript{2}ZnCl\textsubscript{4} and Cs\textsubscript{3}ZnCl\textsubscript{5} compared to BaF\textsubscript{2} and LSO:Ce. Crystals were measured using the Bollinger-Thomas single photon counting method. The feature around 30 ns is an instrumental artifact. Adapted from \cite{b28}.}
    \label{figUTK3}
\end{figure}

Several properties make Cs\textsubscript{2}ZnCl\textsubscript{4} an attractive candidate for further investigation. It is non-hygroscopic, has a single-component decay time around 1.7 ns, and has longer wavelength emission than BaF\textsubscript{2} \cite{b24,b28,b31,b32}. As a result of recent improvements to crystal quality, better performance has now been achieved with Cs\textsubscript{2}ZnCl\textsubscript{4}. Specifically, light yield as good as 1,980 ph/MeV (at 662 keV) has been measured for small crystals with approximate dimensions of 5 × 5 × 5 mm\textsuperscript{3} \cite{b28}, surpassing that of the CVL component of BaF\textsubscript{2} (1,400 ph/MeV). The CTR has so far been measured to be as good as 136 ps FWHM for two Ø7 mm × 3 mm thick slabs of Cs\textsubscript{2}ZnCl\textsubscript{4} measured in coincidence using SiPMs (Broadcom, AFBR-S4N44C013) \cite{b28}. This value is expected to improve with optimization of the measurement setup. Another promising new Zn-based CVL scintillator is Cs\textsubscript{3}ZnCl\textsubscript{5}, which has a 0.82 ns decay time \cite{b28}. A comparison of the decay profiles of Cs\textsubscript{2}ZnCl\textsubscript{4} and Cs\textsubscript{3}ZnCl\textsubscript{5} with BaF\textsubscript{2} is shown in Fig.~\ref{figUTK3} to illustrate their ultrafast timing characteristics and lack of slow decay components. 

Various strategies are currently being explored in an effort to improve the performance of radiation detectors for fast timing applications. For scintillator-based detectors, the performance is ultimately limited by the scintillator’s decay time, therefore, there is a strong push toward the discovery and development of ultrafast materials that may overcome the limitations of existing technology. Exploiting fast emission processes and the concept of metascintillators (or heterostructures) are two approaches in particular that have received much attention in the last few years. The compositional space in which CVL materials exist has not yet been exhausted, and continued efforts devoted to searching for and developing novel CVL materials are necessary in order to find suitable alternatives to existing ultrafast inorganic scintillators such as BaF\textsubscript{2}. Table 1 summarizes physical and scintillation properties of some promising new candidate materials. Deeper investigation into impurity-induced or impurity-enhanced CVL may also be a pathway for discovering new ultrafast scintillators and is a relatively unexplored area. If they are to be utilized in HEP experiments, radiation hardness of newly developed CVL scintillators needs to be assessed, and future work should also prioritize denser materials.

\begin{table*}[htbp]
\caption{Summary of physical and scintillation properties of promising new CVL scintillators compared with those of commercially available BaF\textsubscript{2}.}
        \begin{tabular}{|l|c|c|c|c|c|c|c|}
            \hline
            Composition & LY (ph/MeV) & Decay constant (ns) & CVL emission wavelengths (nm) & Hygroscopic & Density (g/cm\textsuperscript{3}) & Z\textit{\textsubscript{eff}} & Ref.\\[5pt]
            \hline
            Cs\textsubscript{2}ZnCl\textsubscript{4} & 1,980 & 1.66 & 285, 379 & no & 3.35 & 45.3 & \cite{b28,b32} \\[3pt]
            \hline
            Cs\textsubscript{3}ZnCl\textsubscript{5} \textsuperscript{\textit{a}} & 1,460 & 0.82 & 240, 289, 404 & no & 3.44 & 46.6 & \cite{b28} \\[3pt]
            \hline
            Cs\textsubscript{2}BaCl\textsubscript{4} \textsuperscript{\textit{a}} & 1,700 & 1.2 & 260, 290 & slightly & 3.76 & 49 & \cite{b23,b25} \\[3pt]
            \hline
            CsCaCl\textsubscript{3} & 1,371 & 2.47 & 260, 290 & slightly & 2.95 & 42 & \cite{b25} \\[3pt]
            \hline
            CsMgCl\textsubscript{3} & 1,113 & 2.36 & 260, 290 & slightly & 3.10 & 43 & \cite{b25} \\[3pt]
            \hline
            BaF\textsubscript{2} & 1,400 & 0.6 - 0.8 & 195, 220 & no & 4.88 & 53 & \cite{b34,b35} \\[3pt]
            \hline
            \multicolumn{8}{l}{\footnotesize{\textsuperscript{\textit{a}}Discovered in the last three years.}}
        \end{tabular}
        
\end{table*}

\subsection{Nanostructures \label{sec:nano}}
Structured scintillators are also called heterostructured scintillators, metascintillators, or metapixel, the idea is to combine materials with complementary properties to achieve functions better than any of the individual materials, see Sec.~\ref{sec:meta} above and references therein for additional information on fast timing applications. The application of structured scintillators is broader than fast timing apparently. Phoswich scintillators have been described for particle identification in a number of works, as summarized in Table.~\ref{scint:tab}. Scintillator innovation through structure engineering has also been recognized elsewhere~\cite{LLY:2021}.

In comparison with bulk scintillators, nanocomposite scintillators have features  potentially include enhanced light output, reduced cost, and greater size scalability~\cite{MDJ:2007}.
Optimisation of monolithic
nanocomposite and transparent
ceramic scintillation detectors for
positron emission tomography~\cite{WAA:2020}.

Patterning by plastic deformation or {\it nano-imprint} was
developed in the 1990s~\cite{CKR:1996}. It permits nanometer
patterns in clean ambient air and without complex optics. C. Cerna et al. [CNRS/U. Bordeaux] tested nanoscale structured plastic scintillators for better light extraction. By using
empirical methods for structuring commercial scintillating
polymers surfaces, up to 50\% more light was extracted by patterned surfaces. Additional results  related to the Purcell effect, photonic crystals were reported by various authors and institutions, see Table.~\ref{scint:tab} for additional examples.

\subsection{Micrometer thin films \label{sec:duj}}

 Two-dimensional (2D) X-ray imaging and three-dimensional (3D) microtomography with sub-micrometer resolutions can be achieved by using thin scintillators in the 3rd generation synchrotron facilities, when the X-ray flux can exceed 10$^6$ ph$\cdot$s$^{-1}$$\cdot $$\mu$m$^{-2}$~\cite{Koch:1998,TUT:2002}. In a microscope set-up that uses a thin scintillator screen to convert X-rays into visible light, the spatial resolution depends on the thickness of the screen, the depth of focus (or defect of focus), the optical aberrations, and the camera electronic noise. A spatial resolution of 0.8 $\mu$m fwhm (1000 line pairs/mm with 10\% contrast) was reported~\cite{Koch:1998}. Ce-doped crystalline YAG film of 5 $\mu$m thick was deposited on undoped YAG substrate (170 $\mu$m thick). LSO:Ce less than 10 $\mu$m was used in another case. X-ray absorption is weak in these rather thin scintillators. High material density is thus desired for high X-ray absorption efficiency, particularly at high x-ray energies. 
 
One promising growth technique for such range of thicknesses with high optical qualities is the Liquid Phase Epitaxy (LPE), which allows single crystal film deposited on single crystal substrate. Several materials such as doped Lu\textsubscript{2}SiO\textsubscript{5} (LSO), and many garnets doped with cerium, europium or terbium, Lu\textsubscript{2}O\textsubscript{3}:Eu\textsuperscript{3+} have been developed as thin scintillating films~\cite{Martin:2009, Zorenko:2012, Riva:2016}. Double layers (LSO:Tb\textsuperscript{3+} and LSO:Ce\textsuperscript{3+}) screens emitting at different wavelengths combined with double read-out systems spectrally filtered has been proposed to compensate the weak x-ray absorption~\cite{Doui:2010}. At low X-ray energies, the absorption edges of the absorption films play a crucial role, and the composition may be adapted for the specifics energies as exemplified by Riva et. al.~\cite{Riva:2016b}.  

\begin{figure}[htb!]
\centering\includegraphics[width=0.9\linewidth]{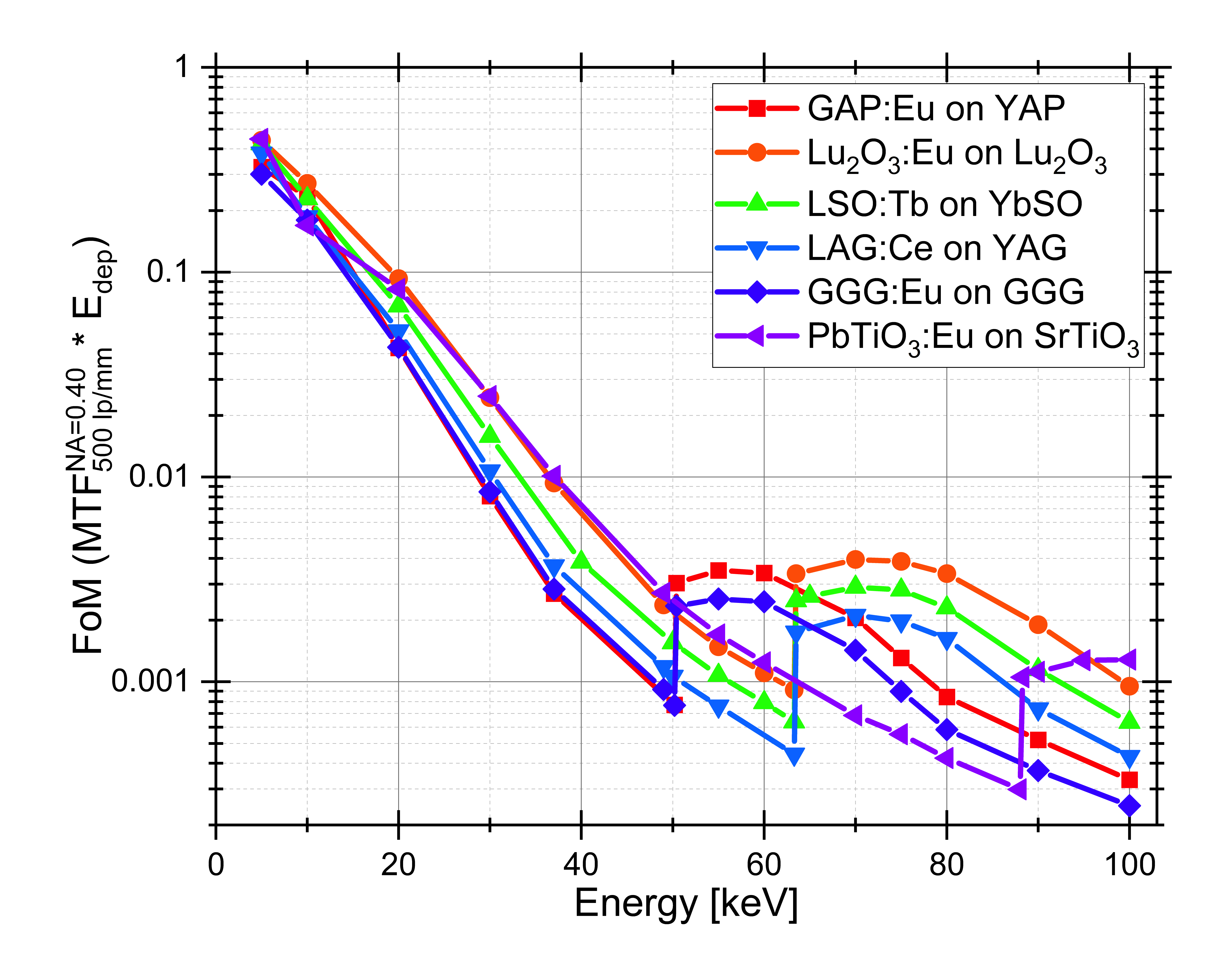}
\caption{Figure of Merit (FoM) Calculated from the contrast in the MTF blurred by optics (NA = 0.40) at 500 lp/mm and the energy deposited in the SCF. Values are extracted frm simulations at X-ray energies from 5-100 keV for 5 $\mu$m SCFs supported by 150 $\mu$m substrates. Used with permission from RSC.}
\label{fig:Du1}
\end{figure}

The substrate is also of crucial importance and has several severe requirements. First, it has to be compatible for the epitaxial growth, i.e. a showing the same crystalline structure and a weak lattice mismatch. Second, the X-ray absorption in the substrate being very large as compared to the scintillating film, the substrate has to be non scintillating. Indeed, even a weak scintillation leads to an image out of the focus plane of the objective contributing to blur the image. Finally, it has been recently shown using monte-carlo simulation and confirmed by experiments, that the secondary X-ray emission from the substrate may significantly affect the MTF of the overall device, and that effect strongly depends on the X-ray energy. This effect becomes very critical when ultimate spatial resolution are foreseen~\cite{Wollensen:2022}. It is demonstrated that based on a figure of merit combining the MTF at 500 lp/mm and the effective energy deposition in the active film that the best combination film/substrate is strongly changing in the range of the X-ray absorption edge of them, Fig.~\ref{fig:Du1}. Because the performances are pushed to the limit, it suggests that the scintillating screens tend to become very specific to each energy range, even to small change when approaching the absorption edge of their constituents. 

\subsection{Bulk composite scintillators \label{sec:bren}}

Scintillator detectors for pulsed fast neutron sources play a vital role in nuclear safeguards, material inspections and fundamental science~\cite{BW3,BW4}. While the combination of homogenous, monolithic scintillators (e.g. Cs$_2$LiYCl$_6$:Ce$^{3+}$, GS20)~\cite{BW5,BW6} with moderator material (e.g. poly-ethylene) can meet some of the current requirements, difficulties arise from operating a large volume neutron detector in the harsh radiation environments with (1) the ubiquitous gamma ($\gamma$) ray backgrounds and (2) neutron count rate limited by the size and geometry of the detector material or readout electronics. 

Detecting neutrons is unique due to its electric charge neutrality and the isotopically dependent absorption cross sections of neutron, see Fig.~\ref{fig:nPlastic} above for the example cross section of a plastic scintillators rich in $^1$H. Some other neutron converter isotopes are $^3$He, $^6$Li, $^{10}$B, $^{32}$S, $^{35}$Cl and $^{238}$U~\cite{BW1, BW2}. Time and energy resolved fast neutron detection requires efficient neutron detection in a relatively large volume  (and therefore low cost) of neutron-sensitive materials~\cite{WM:2013}. Additional requirements or highly desirable properties include  tolerance to radiation degradation and damage for a long period of use, particle or energy discrimination against background such as $\gamma$-rays. For event based or neutron counting detection, short response time is often needed to improve temporal resolution, event statistics, and to prevent event pileups. Generally, desirable detector attributes come with significant trade-offs due to  the lack of an `ideal' scintillator for neutrons in practice.

The organic-inorganic material subset can be viewed as an effort to merge desirable neutron response characteristics from both material spaces. The organic-inorganic composite subset represents a large group of diverse material combinations; however, successful examples of this material subset can be shown within the literature~\cite{BW7,BW8,BW9, BW10, BW11, BW12, BW13}. Organic-inorganic composite materials take advantage to neutron interactions primarily within the organic matrix and energy transfer within a scintillating material incorporated within the organic matrix. While neutron interaction rates with organic constituents can be relatively high compared to higher atomic number constituents, significant scintillation light scattering can occur due to the refractive index mis-match between the organic matrix and the scintillator material; ultimately, this optical property difference can lead to scintillation photons possessing a shorter mean free path, contributing to optical losses within the bulk~\cite{BW13,BW14}.  While the scintillation process is well studied for homogenous scintillators, the optical composite parameter space can be vast, notably, the main design criteria for this material subset focuses on scintillation light transport. The volume fraction of the individual constituents and the effective optical absorption coefficient of the resulting mixture can play a significant role in end user application constraints; fortunately, this unique multi-parameter space can also enable flexibility for targeted application specifications~\cite{BW7,BW8}. (i.e. neutron interaction rate, temporal response and light output to the photo-detector)  The key breakthroughs for composite-based detector technologies rely heavily on transparency and exploring every aspect of scintillation light transport.

\begin{table*}[htbp]
\caption{A Summary of scintillating particle composites.}
\centering
        \begin{tabular}{|l|c|c|c|c|c|}
            \hline
            Organic Matrix Material & Inorganic scintillator & Structured composite  & Peak scintillator emission ($\lambda$, in nm) & Decay time  (ns) & Ref. \\[3pt]
            \hline
            EJ-290 & Li$_6$Gd(BO)$_3$:Ce & No & 416 & 200 & ~\cite{BW9}  \\[3pt]
            \hline
            EJ-290NS & Li$_6$Gd(BO)$_3$:Ce & No & 416 & 200 & \cite{BW9}  \\[3pt]
            \hline
            Poly Vinyl Toluene & KG2 & No & 395 & 62 & \cite{BW11}  \\[3pt]
            \hline
            Drakeol 9 LT & GS20 & Yes & 395 & 57 & \cite{BW12,BW13}  \\[3pt]
            \hline
            Acrylic \textsuperscript{\textit{a}} & GS20 & Yes & 395 & 57 & \cite{BW15}  \\[3pt]
            \hline
            Acrylic +BPEA \textsuperscript{\textit{a}} & GS20 &Yes & 510 & 57 & \cite{BW15} \\[3pt]
            \hline
            Acrylic + NOA 71 \textsuperscript{\textit{a}} & GS20 & Yes & 395 & 57 & \cite{BW14}  \\[3pt]
            \hline
            \multicolumn{6}{l}{\footnotesize{\textsuperscript{\textit{a}} A recent composite scintillator development.}}
        \end{tabular}
\label{Wiggins:tb}        
\end{table*}

\begin{figure}[htbp]
    \centerline{\includegraphics[width=0.9 \columnwidth]{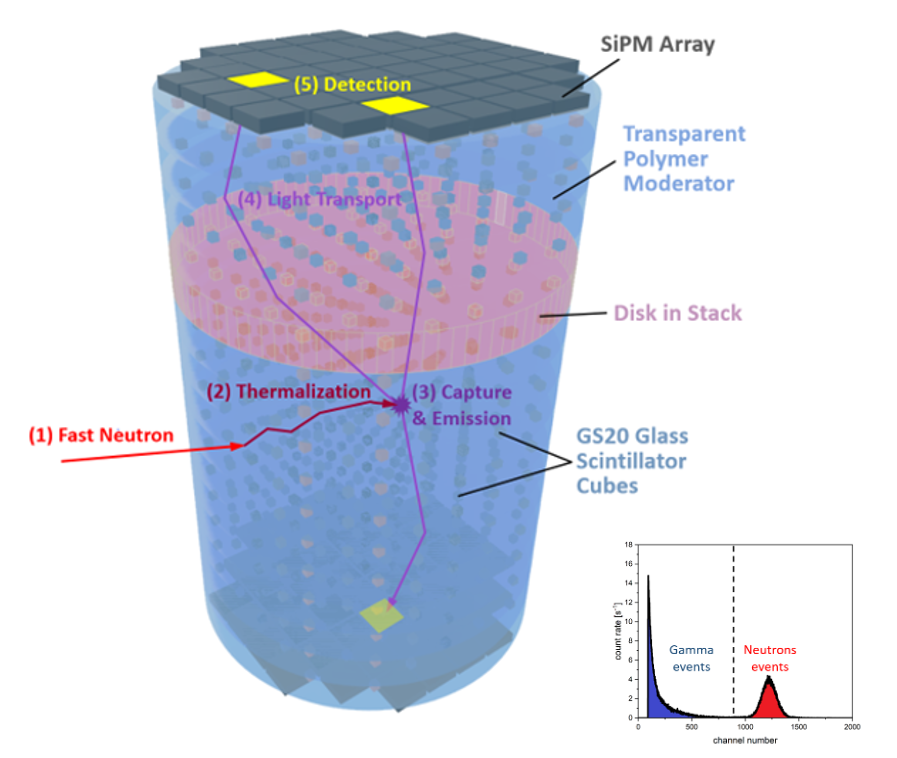}}
    \caption{ The ideal pulsed fast neutron detection scheme for an arranged, transparent composite detector; here, fast neutrons interact primarily with organic constituents, reducing the kinetic energy of the neutron, in order to increase the probability of neutron capture within the neutron-sensitive scintillating particle (e.g. GS20 scintillator cubes) followed by scintillating light transport. The refractive index difference between GS20 and the transparent organic moderator needs to be sufficiently small (less than 0.02) to enable good scintillation light transport to a photo-detector, represented as a silicon photo-multiplier array, with reduced scintillation light scattering.  (Insert at the low right corner) a representative pulse height spectrum for incident Cf-252 fission neutrons and gammas on an arranged, transparent composite detector. The dashed line illustrates that the maximum energy deposition for gamma related energy deposition events can be constrained with a threshold. }
    \label{fig:BW1}
\end{figure}

Recently, new developments in arranged, heterogeneous scintillating particle composite materials have demonstrated promising characteristics for applications that require compact, solid-state neutron detectors, see Table.~\ref{Wiggins:tb} for some examples; specifically, these detectors can potentially address the needs of good detection efficiency, large active volumes, fast timing and respectable radiation damage tolerances for reasonable cost investment.~\cite{BW14,BW15} Figure.~\ref{fig:BW1} demonstrates an illustration of the fast neutron detection scheme and a representative pulse height spectra under incident Cf-252 fission products. One of main advantages towards using an arranged, scintillating particle composite is that the gamma suppression performance relies solely on the arrangement of small neutron-sensitive scintillating particles within an organic matrix; here, the creativity landscape is vast and opens the door to many new possibilities within the hybrid scintillator particle landscape (e.g. scintillating particles with wavelength shifting coatings for improved light transport and performance stability or segmenting composite detector designs for unique application volumes)   

\section{Additional trends and opportunities \label{sec:TO}}
Quite a few trendy topics and opportunities have been covered in Sec.~\ref{sec:css} and Sec.~\ref{scin:new} above. In Sec.~\ref{sec:AT}, we discuss the novel application of neutron imaging for optimization of crystal growth. Sec.~\ref{sec:gan} is on the emerging opportunities associated with machine learning and data science. Space applications of scintillator, Sec.~\ref{sec:man}, are expect to grow rapidly with the expansion of human and artificial intelligent activities beyond the Earth. We conclude this section with multimodal RadIT, Sec.~\ref{sec:sky}, which will certainly require versatile scintillator technology to maximize the information output from dynamic and time-transient events in the laboratory, industry and medicine.

\subsection{Optimization of crystal growth with in-situ neutron imaging \label{sec:AT}}
Discovery of new promising scintillator or semiconductor materials is typically conducted with very small grains of synthesized samples. Once many important characteristics have been measured for these small samples a proper crystal growth recipe has to be developed for the production of large crystals needed to meet the requirements of specific applications, especially where large volumes are required by the relatively long attenuation length (as in case of gamma detection), or a large quantity production. It is that step in the development process which is often limits the transition of a particular promising material into an industrial scale manufacturing. These crystals need to be grown reproducibly, with high yield and at an affordable cost. Most of the time multiple trial-and-error crystal growth attempts are conducted and the grown materials are characterized ex-situ by various non-destructive and destructive techniques. The number of such optimization runs is limited by a relatively long time required for each attempt. If various crystal growth parameters could be monitored during the growth in real time the search of proper growth parameters would be much easier and a path from discovery to production would be much faster and cheaper. However, during crystal growth of many materials only a limited number of parameters can be measured in-situ without disturbing the growth process. Remote sensing of growth parameters is most of the time obscured by the equipment used for the crystal growth and by the opacity of grown materials themselves for many conventional probes (such as photons and electrons). It has been demonstrated recently that energy-resolved neutron imaging can provide unique possibilities to monitor in-situ various growth parameters such as elemental distribution within the solid material and the melt, the location and the shape of liquid-solid interface, mosaicity of solidified material, segregation and diffusion of dopant elements during growth, the presence of defects and others~\cite{AT1, AT2, AT3, AT4, AT5}. Although these experiments were conducted for a specific growth technique (Bridgeman crystal growth process), this relatively novel in-situ diagnostics approach can be extended to various other techniques due to the unique capability of neutrons to penetrate many materials including some metals. Although the number of facilities where such experiments can be conducted is very limited at the present time, the crystal growth optimization technique described here is only intended for the search of a better crystal growth recipe which then can be transferred to the industry. 

\begin{figure}[htb!]
\centering\includegraphics[width=0.9\linewidth]{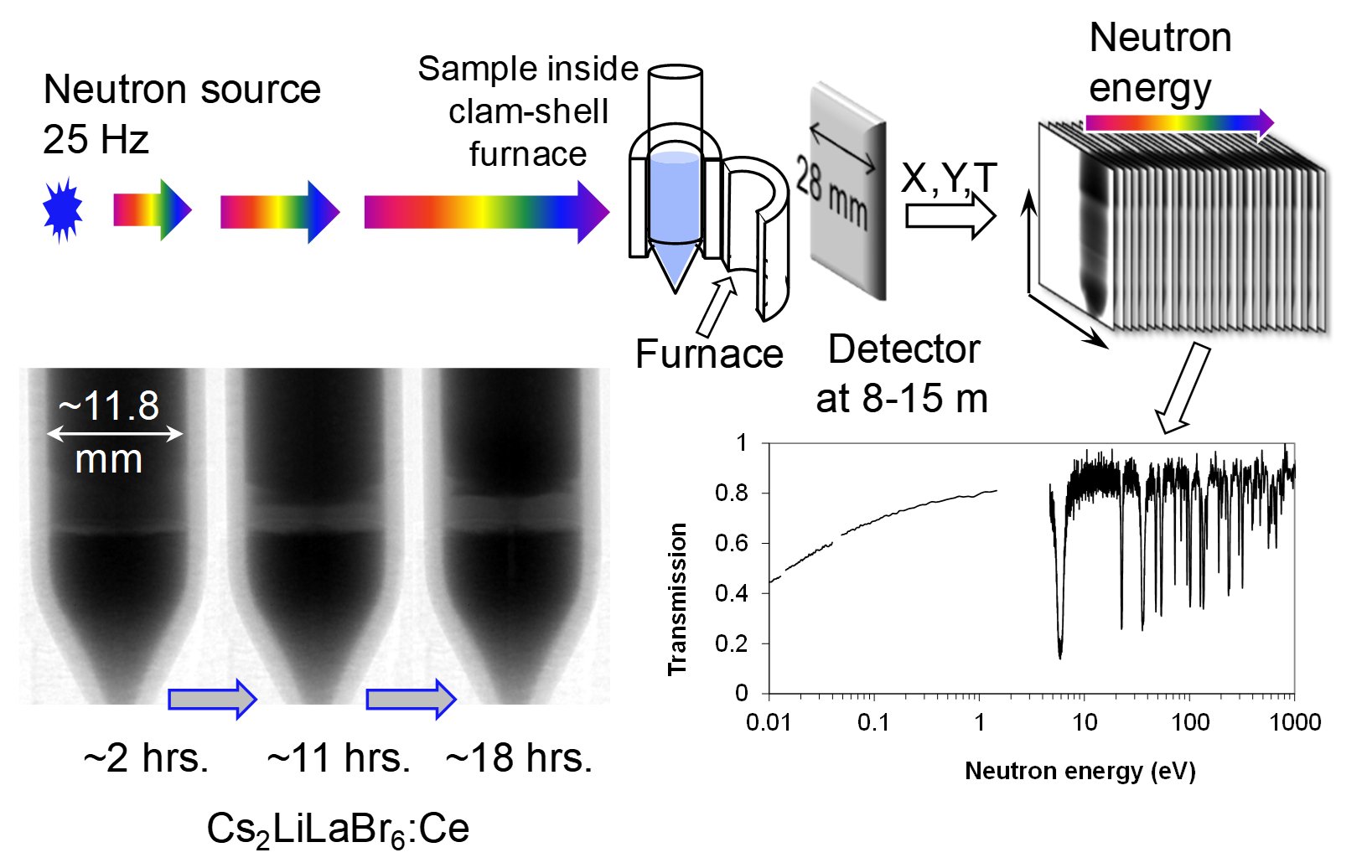}
\caption{Schematic diagram of the energy-resolved neutron imaging experiment. A crystal growth furnace is placed directly in the neutron beam in front of a neutron counting detector. Thousands of images are acquired simultaneously, each corresponding to a specific neutron energy. Neutron transmission spectrum is thus measured for each pixel of the transmission dataset. Reproduced with permission from~\cite{AT3}.}
\label{fig:Anton1}
\end{figure}

To measure some characteristics of in-situ crystal growth process with the help of neutron imaging the furnace needs to be placed in the direct neutron beam, in front of an imaging detector, as shown in Fig.~\ref{fig:Anton1}. Some growth parameters (such as the location and the shape of liquid-solid interface, qualitative uniformity of elemental composition, location of defects) for some materials can be measured with a regular neutron imaging where a white spectrum transmission is measured for each pixel. A wider range of parameters can be investigated at a pulsed neutron beam where transmission spectra can be measured for each pixel of the resulting dataset. 

The analysis of transmission spectra enables quantification of elemental composition, mapping of crystal orientation for multiple grains, mosaicity and others across the entire field of view of the neutron counting detector. The possibility to measure neutron transmission spectra in a wide range of energies at spallation neutron sources enables reconstruction of elemental maps during the crystal growth without any disturbance to the growth volume, as it was demonstrated in recent studies~\cite{AT2,AT3}. It should be noted that not all elements can be probed by this technique, only the ones which have sufficiently large neutron attenuation cross section. 

\begin{figure}[htb!]
    \centering
    \begin{subfigure}{0.4 \textwidth}
    \includegraphics[width= \textwidth]{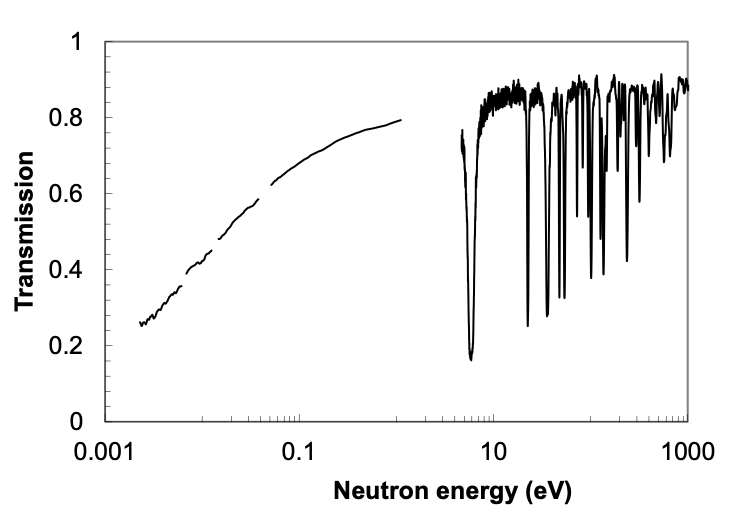}
    \caption{}
    \label{}
    \end{subfigure}
    \hfill
    \begin{subfigure}{0.4 \textwidth}
    \includegraphics[width= \textwidth]{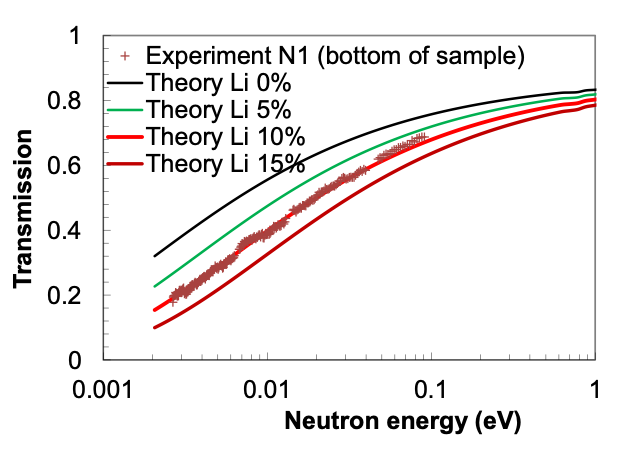}
    \caption{}
    \label{}
    \end{subfigure}
\caption{Neutron transmission spectra measured for Cs\textsubscript{2}LiLaBr\textsubscript{6}:Ce during crystal growth. (a) Entire energy range measured in the experiment, with many neutron resonance absorption dips seen above $\sim$5 eV. (b) Measured and calculated transmission spectra for different Li concentration. Reproduced with permission from~\cite{AT3}.}
\label{fig:Anton2}
\end{figure}

One of the strengths of energy resolved neutron imaging is its capability to use neutron resonance absorption to separate the contribution to the attenuation by different elements/isotopes and thus to map the elemental composition for several elements~\cite{AT7,AT8,AT9,AT10,AT11}. An example of measured transmission spectrum for Cs\textsubscript{2}LiLaBr\textsubscript{6}:Ce shown in Fig.~\ref{fig:Anton2} exhibits multiple resonance dips at the neutron energies above 5 eV, which correspond to resonances of Cs, La and Br, while attenuation by Li does not show any resonances in the measurable energy range, by its attenuation dominates at the lower neutron energies.

\begin{figure}[htb!]
\centering\includegraphics[width=0.9\linewidth]{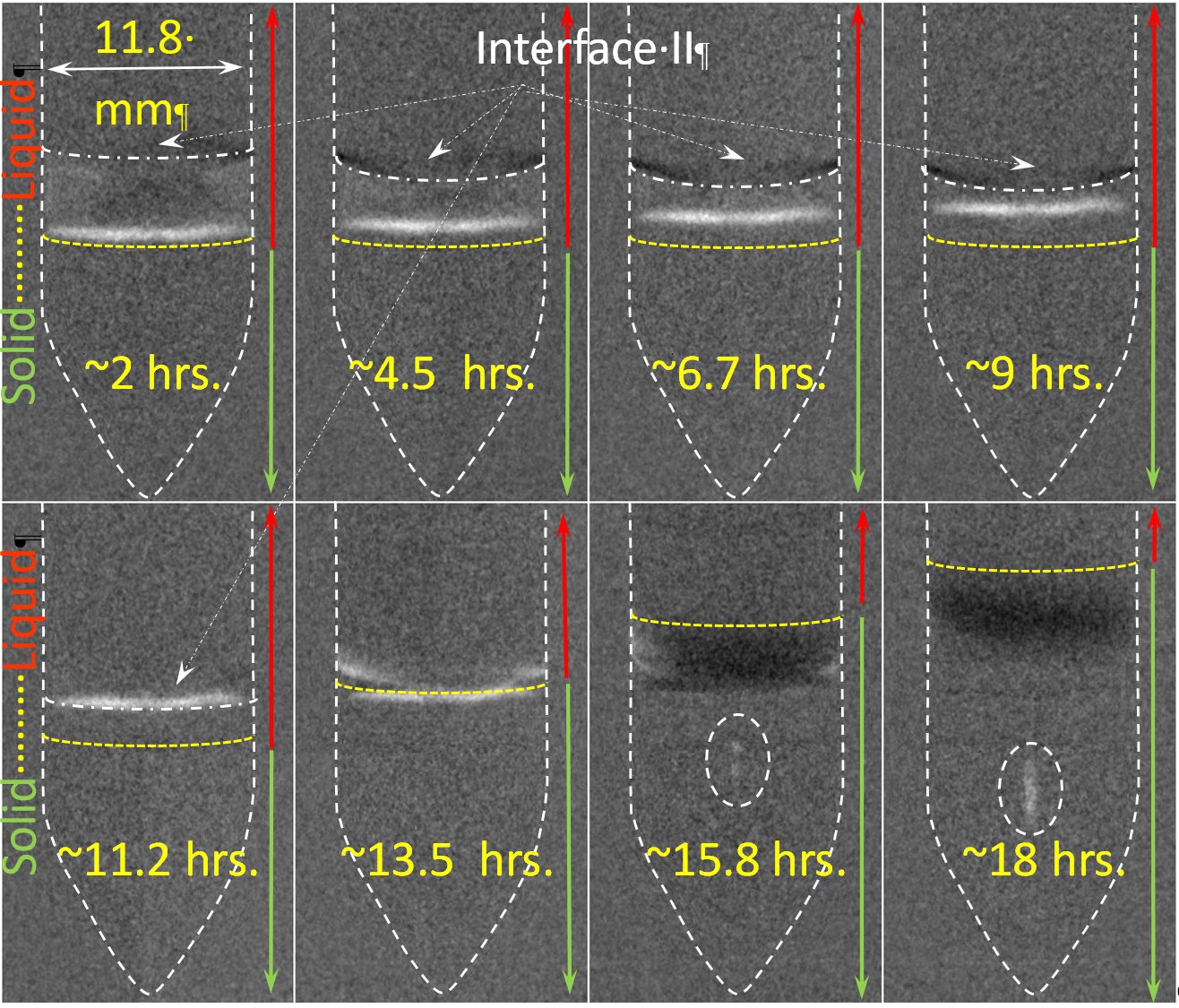}
\caption{White spectrum neutron transmission images of Cs\textsubscript{2}LiLaBr\textsubscript{6}:Ce charge sealed in quartz ampule during in-situ crystal growth experiment. Each image is normalized by the image acquired during preceding time period. A small change in the elemental composition between solid and melt forms the white/dark bands in these images at the location of the interface between solid and liquid phases. First 11 hours the temperature profile was held steady, followed by the crystal growth. The yellow dashed line represents the interface between the solid and liguid phases, and the white dashed line represents the interface between two phases formed within the melt. Dashed ovals indicate the area where some diffusion of Cs-rich blob within the solid phase was observed.. Reproduced with permission from~\cite{AT3}.}
\label{fig:Anton3}
\end{figure}

In addition to concentration within both solid and liquid phase, the pressure of certain gaseous phases can be measured remotely within sealed ampules for some elements, as it was demonstrated in the experiment non-related to crystal growth~\cite{AT10}.
During the growth of BaBrCl:Eu and Cs\textsubscript{2}LiLaBr\textsubscript{6}:Ce crystals it was also observed that some elements (Eu in the former and Cs in the later crystals) exhibited migration, in some cases against the concentration gradient within the solid phase~\cite{AT3, AT5}, Fig.~\ref{fig:Anton3}. As a result of this migration within the solid crystal the Eu dopant concentration profile along the grown crystal depends on the speed of ampule translation and the cooling profile during Bridgman growth process. The possibility to measure remotely the dynamics of this diffusion within the solid phases at high temperatures is another unique capability provided by energy-resolved neutron imaging, which can be used for the optimization of crystal growth processes.

\begin{figure}[htb!]
\centering\includegraphics[width=0.9\linewidth]{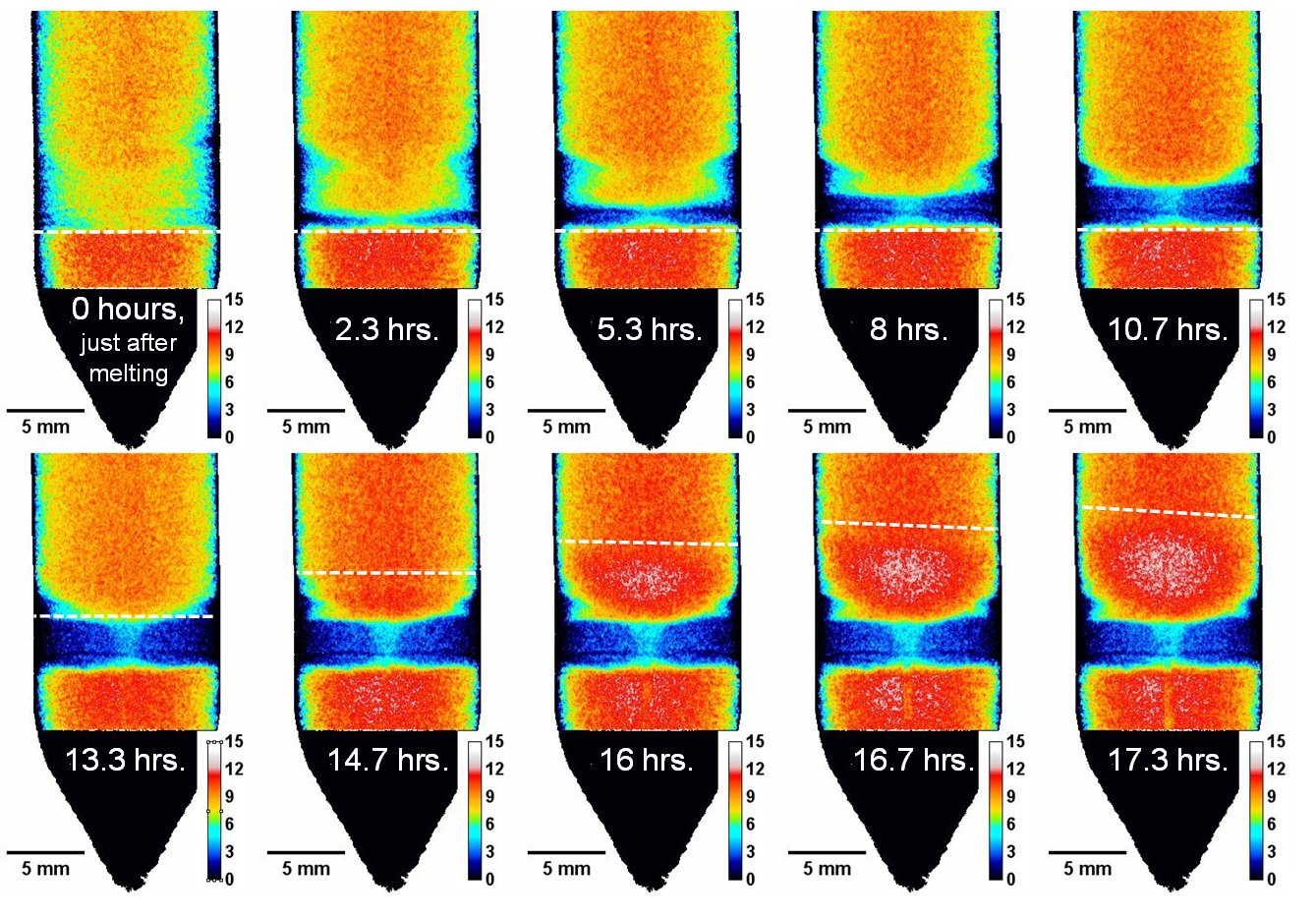}
\caption{Maps of Li concentration within the growth volume of Cs\textsubscript{2}LiLaBr\textsubscript{6}:Ce charge. Dashed lines represent the location of solid/liquid interface. The first 11 hours the temperature profile and the ampule location were kept steady. Cs-righ/Li-deficient phase formed within the melt. The color bar represents the Li concentration in atom \%. Reproduced with permission from~\cite{AT3}.}
\label{fig:Anton4}
\end{figure}

It is well known in crystal growth community that the shape, stability, and location of liquid/solid interface plays an important role in determining the quality of the grown material. In many cases a convex interface is desired. Observation of the interface in some cases is enabled by the presence of dopant segregation, which was used in several studies ~\cite{AT2,AT3,AT5,AT12,AT13}. With the help of neutron imaging, the shape and the location of liquid-solid interface and the speed of ampule translation could be optimized in real time by adjusting the temperature profile and the speed of translation. Moreover, formation of two phases within the melt was observed during in-situ growth of Cs\textsubscript{2}LiLaBr\textsubscript{6}:Ce crystal, as shown in Fig.~\ref{fig:Anton4}. In the very first in-situ growth of this crystal it was found that Cs-rich/Li-deficient layer forms in the melt just above the solid phase. It was found that good scintillator material was only grown after the process of formation of Cs-rich layer is settled over ~11-hour period. Quantitative maps of Li concentration, shown in Fig.~\ref{fig:Anton4} for different times of crystal growth process demonstrate the formation of this Cs-rich layer at a steady state temperature distribution. Once that process of phase separation within the melt stabilized, the temperature profile was gradually changed, and a good crystal could be grown. Discovering the need for that stabilization blindly, as it was done by the industry before that experiment was conducted, was obviously a much longer process with multiple hypothesis-driven trial-and-error runs. This experiment can be considered as one of the examples where neutron imaging can substantially reduce the cost and time of the transition from the discovery of new candidate materials to industrial growth of bulk crystals. 

Another non-destructive scintillator crystal characterization technique used recently was the dual mode imaging, where 3-dimensional transmission tomographic reconstruction of crystals was obtained simultaneously for both neutron and gamma radiations ~\cite{AT14}. The large differences in the attenuation coefficients between neutrons and gamma photons provide information on the location of defects in the materials and their relation to the uniformity of the elemental composition, all done non-destructively for the materials still sealed in growth ampules. 
One other very attractive application of neutron imaging in the optimization of crystal growth is the verification and refinement of computer simulation models, which can substantially improve our understanding and quantification of processes occurring during crystal growth of various materials~\cite{AT12, AT13} and their subsequent optimization.

\subsection{Data Science and Machine Learning for Scintillators \label{sec:gan}}
Over the past decade resounding success of machine learning (ML) and artificial intelligence (AI)
based methods in other fields, including finance, marketing, healthcare, networking and
transportation, have had a significant impact in physical sciences including materials science~\cite{Ram:2017,Mog:2020, Pil:2021, Mog:2022, Agr:2019}.The field of scintillator design and development is not an exception. In fact, in recent past data-
enabled methods have been actively applied to expedite the development and optimization of
luminescent materials. In particular, the ability of ML-based algorithms to efficiently encode
chemical similarity and interpolate across high dimensional feature spaces to identify hidden
trends and patterns in functionality across chemistries are harnessed to not only screen potentially
new scintillator compounds but also to develop predictive models for their performance estimation.

The overall performance of a scintillator is characterized by a number of metrics. These include,
but are not limited to, the speed of response, the light output per incident MeV of radiation, the
emission wavelength, long-term stability under radiation, density and effective atomic number as
well as absence of defect/trap states. While some of these metrics describe raw performance of a
scintillator material, others, such as emission wavelength, need to be tuned to best couple the
material to a given detector system. Although practically much desired, a first principles-based
approach for the entire scintillator property portfolio prediction remains beyond the scope of state-
of-the-art-computations and, for this reason, several recent studies have focused on the alternative
data-enabled route of using ML for scintillator property predictions. A majority of the research in
this direction has focused on predicting one or more scintillator performance metrics, such as light
output or response time, utilizing a prescribed set of features or descriptors that are largely
selected based on the domain knowledge. This surrogate model development process for efficient
property predictions mainly relies on implementation the following key steps: (i) selection of easily
accessible attributes or design variables (also referred to as features or descriptors) that are
expected to be causally related to the target property of interest, (ii) integration of the identified
variables in a ML model to establish a mapping between the materials and the target properties
and (iii) assessment and analysis of the predictive power and generalizability of the developed
models and identified design rules using unseen data. This framework has been applied to predict
a range of properties that are either directly or indirectly connected to the performance of
luminescent materials, including scintillators~\cite{Zhou:2018, Zhou:2020, Zhuo:2020B,Pil:2019, Park:2021}.

\begin{figure}[htb!]
\centering\includegraphics[width=0.9\linewidth]{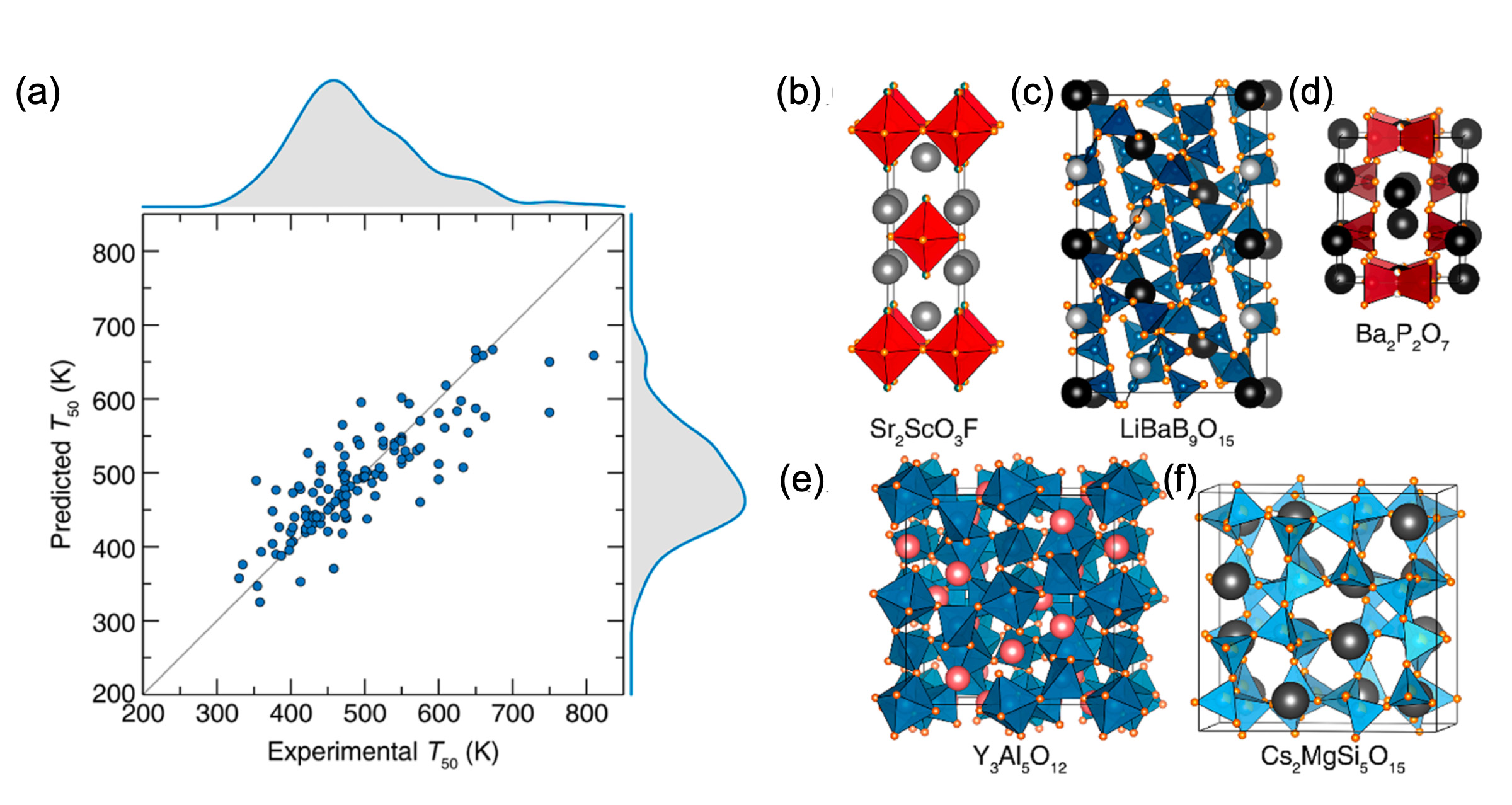}
\caption{(a) A parity plot comparing cross validated ML predictions of thermal quenching temperatures for 134 compounds with the corresponding experimental values. Crystal structures of new host compounds considered for experimental verifications following the ML predictions are shown: (b) Sc\textsubscript{2}ScO\textsubscript{3}F, (c) LiBaB\textsubscript{9}O\textsubscript{15}, (d) Ba\textsubscript{2}P\textsubscript{2}O\textsubscript{7}, (e) Y\textsubscript{3}Al\textsubscript{5}O\textsubscript{12}, and (f) Cs\textsubscript{2}MgSi\textsubscript{5}O\textsubscript{15}. Sc, B, P, Al, and Si (Mg) occupy the center of the polyhedra. Sr, Li, Ba, and Cs are colored in different shades of gray. F is in green, Y is in pink, and O is in orange. Adapted from Ref.~\cite{Zhuo:2020B}.}
\label{fig:GP1}
\end{figure}

For instance, Zhuo et al.~\cite{Zhou:2020} employed a tree-based ensemble learning algorithm along with elemental features (such as the average electronegativity, average polarizability), coupled with local configurational information and the relative dielectric permittivity of the host medium to train a ML model that could reliably predict 5d level centroid shift of Ce$^{3+}$ substituted inorganic phosphors. A quantity that is critical in predicting the light yield and thermal response of rare-earth substituted inorganic luminescent materials. In a different study, Zhuo et al.~\cite{Zhuo:2020B} developed a ML regression model using a set of 134 experimentally-measured temperature-dependent Eu\textsuperscript{3+} emission spectra to rapidly estimate the thermal quenching temperature — defined as the temperature when the emission intensity is half of the initial value — and subsequently used the model to screen more than 1000 potential oxide Eu\textsuperscript{3+} doped host compounds to select five candidates with predicted thermal quenching temperatures $>$ 423 K (see Fig.~\ref{fig:GP1}). These compounds were eventually synthesized for validation of this informatics approach. Closely following along the similar lines of research, Park et al.~\cite{Park:2021} reported set up of an integrated ML platform, consisting of 18 different learning algorithms, to evaluate and compare the performance of different models towards predicting the band gap as well as the excitation, and emission wavelengths of Eu\textsuperscript{2+}-activated luminescent materials. 

\begin{figure}[htb!]
\centering\includegraphics[width=0.9\linewidth]{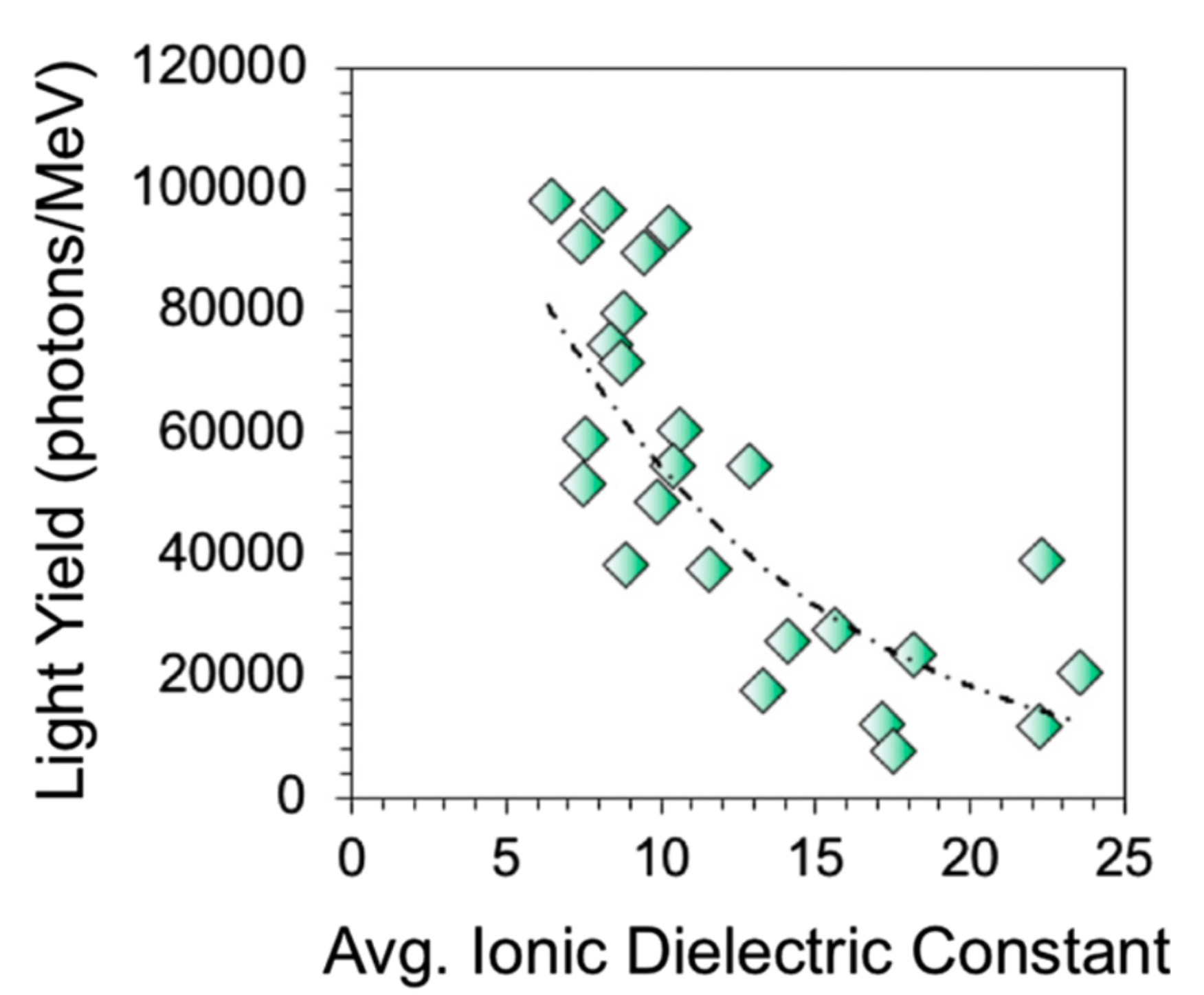}
\caption{Strong correlation between the experimentally reported light output of several Ce and Eu doped scintillators with averaged ionic part of the dielectric constant of the host over a wide range of chemistries. Adapted from Ref.~\cite{Pil:2019}.}
\label{fig:GP2}
\end{figure}

Going beyond the development of surrogate models for predictions of scintillation-related properties that are otherwise expensive to compute or time and resource-intensive to measure directly, the data-enabled approach has also been applied to extract new insights and practically useful design parameters from scintillator materials databases. As an example, in a recent study, Pilania et al.~\cite{Pil:2019} employed a curated dataset of scintillation light yield and response time measurements for twenty-five Ce- or Eu-doped scintillator compounds to discover a strong correlation between the lattice contribution to the dielectric constant and the light yield, regardless of the specific composition or crystal structure of the host material, as depicted in Fig.~\ref{fig:GP2}. This trend was then rationalized via identification a direct mechanistic connection between the light output and the efficiency of germinate recombination process through which hot charge carriers recombine to form excitons at an early stage of the energy absorption and thermalization process. At this stage, charge carriers multiply via impact ionization while settling down to the conduction and valence band edges by losing their energy to phonons and the dielectric permittivity plays an important role in modifying the carrier Coulombic interactions via dielectric screening~\cite{Pil:2019}.

A major limitation of these surrogate models for scintillator performance prediction is that they cannot readily be applied to discover potentially new scintillator while traversing through large chemical spaces. Since these models are predominantly trained on scintillators and other closely related luminescent compounds, with either no or a very few examples of non-scintillators, they generally fail in correctly classifying scintillators and non-scintillators when faced with large chemical spaces where a vast majority of compounds are, in fact, non-scintillators. A major roadblock that has hindered the progress in this direction is the absence of a universally accepted and sufficiently general metric that can be used to distinguish scintillators from non-scintillating compounds. 

\begin{figure}[htb!]
\centering\includegraphics[width=0.95\linewidth]{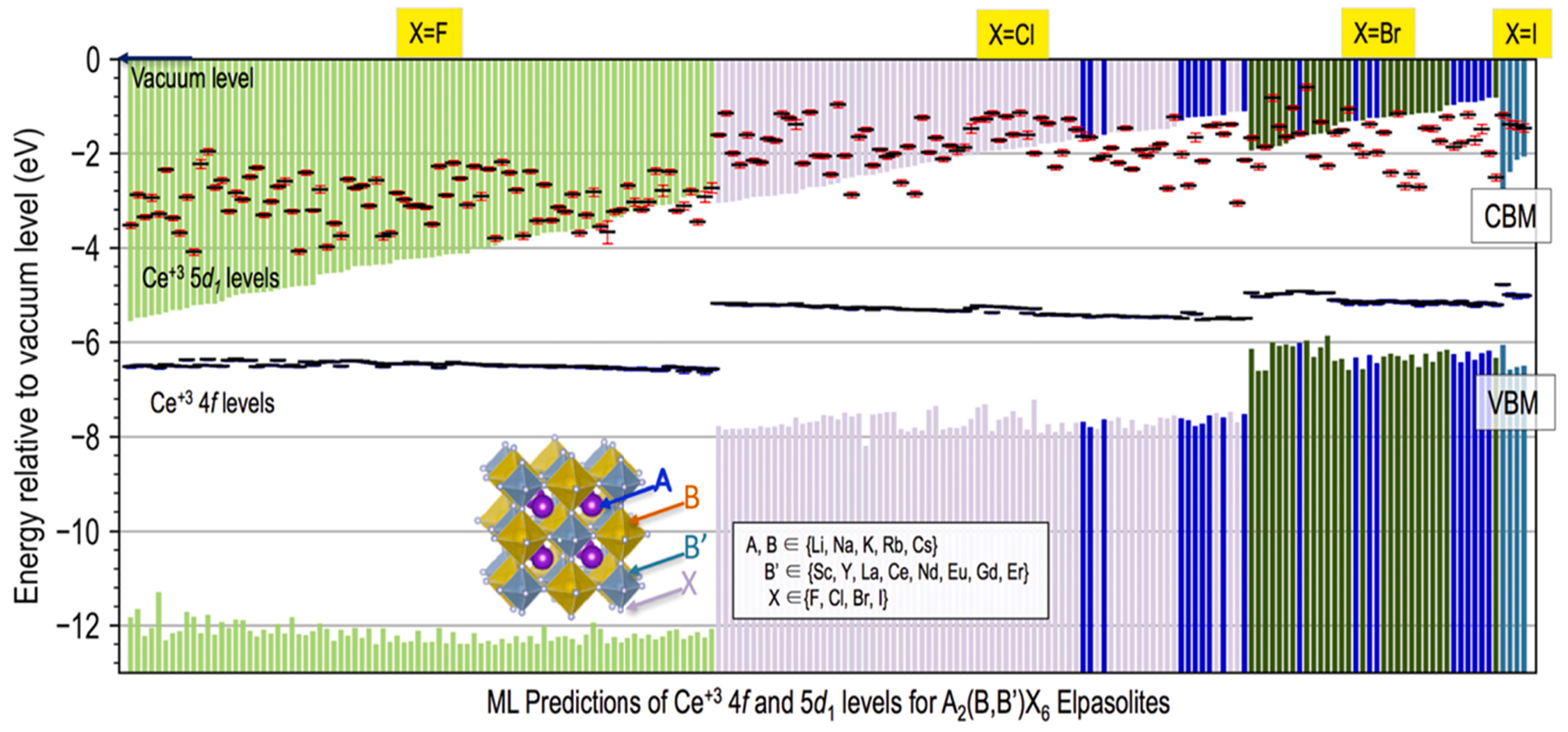}
\caption{DFT-computed relative valence and conduction band edge alignments and the ML-predicted vacuum-referred binding energies for Ce\textsuperscript{3+} activator’s 4$f$ and lowest 5$d$ levels for double Perovskite halides or Elpasolite compounds. The compounds are grouped according to the anion chemistries. Previously known scintillating compounds are highlighted with blue bars. Reproduced from Ref.~\cite{GP11} with permissions. }
\label{fig:GP3}
\end{figure}

In the quest of addressing the classification problem for Lanthanide-doped (in this particular case Ce) inorganic scintillators, a recent study by Pilania et al.~\cite{GP11} focused on the positions of 4$f$ and 5$d$\textsubscript{1} activator levels relative to the host valence and conduction band edges, respectively, as the main electronic structure signature dictating whether a given compound can be a viable scintillator or not. If the activator levels are buried in the band edges (i.e., 4$f$ in the valence band edge and/or 5$d$\textsubscript{1} in the conduction band edge, respectively), the charge carriers generated due to high energy radiation cannot localize at the activator sites to yield scintillation light via radiative recombination. On the other hand, if either of the 4$f$ or 5$d$\textsubscript{1} levels land too far off from the band edges, situated deep in the bandgap of the host, then again charge carriers will have to dissipate the excess energy required to bridge the gap between the activator states and the band edges via a nonradiative process, such as energy transfer to phonons, before localizing at the activator sites. This again is an undesirable situation because it not only increases the response time but also decreases the overall efficiency of the scintillation process. Ideally, a configuration where the 4$f$ or 5$d$\textsubscript{1} levels are situated close to the band edges, but not too close that thermal vibrations at the operating temperature can excite the localized charge carriers back to the delocalized host bands, is desired. With this domain-knowledge-informed criterion for scintillator versus non-scintillator classification, two different regression models were trained and validated using a database of accurate experimental measurements on two key spectroscopic quantities, namely the U and the D parameters~\cite{GP12,GP13}. The U parameter represents a quantitative measure of e-e repulsion in the localized 4$f$ shell of lanthanide ions, and therefore is directly related to the electronic binding energy in the 4$f$ shell. The D parameter, the other hand captures the relative shift of the lowest d level of a lanthanide ion in a given host environment with respect to that of the isolated ion in the vacuum and better known as the spectroscopic redshift. Knowledge of these two parameters, when combined with the accurate host bandgap computations using the Dorenbos chemical shift model 145 allows one to locate the relative position of the activator states with respect to the host band edges. This framework can be efficiently used to make predictions for potentially novel scintillators during a high throughput screening effort. The efficacy of this approach towards practically identifying new compounds was further demonstrated using a case study on Elpasolites or double Perovskite halides of A\textsubscript{2}BB$'$X\textsubscript{6} type. This class is known to harbor many known scintillator compounds and the physics-based classification approach was shown to correctly identify all the known scintillators within the target chemical space (Fig.~\ref{fig:GP3})~\cite{GP11}.

In addition to the materials-centric aspects discussed above, use of ML-based techniques for signal processing and analysis remains another promising and active area of research~\cite{GP14,GP15}.  As an example, Yoon et al.~\cite{GP14} employed a deep-learning model, relying on a convolution neural network framework, to discriminate signals induced by neutrons and gamma-rays in organic scintillation detectors. The pulse-shape discrimination performance for the conventional charge comparison method was compared against the convolution neural network discriminating algorithms for two different detectors to conform a superior performance of the deep-learning model. 

\begin{figure}[htb!]
\centering\includegraphics[width=0.95\linewidth]{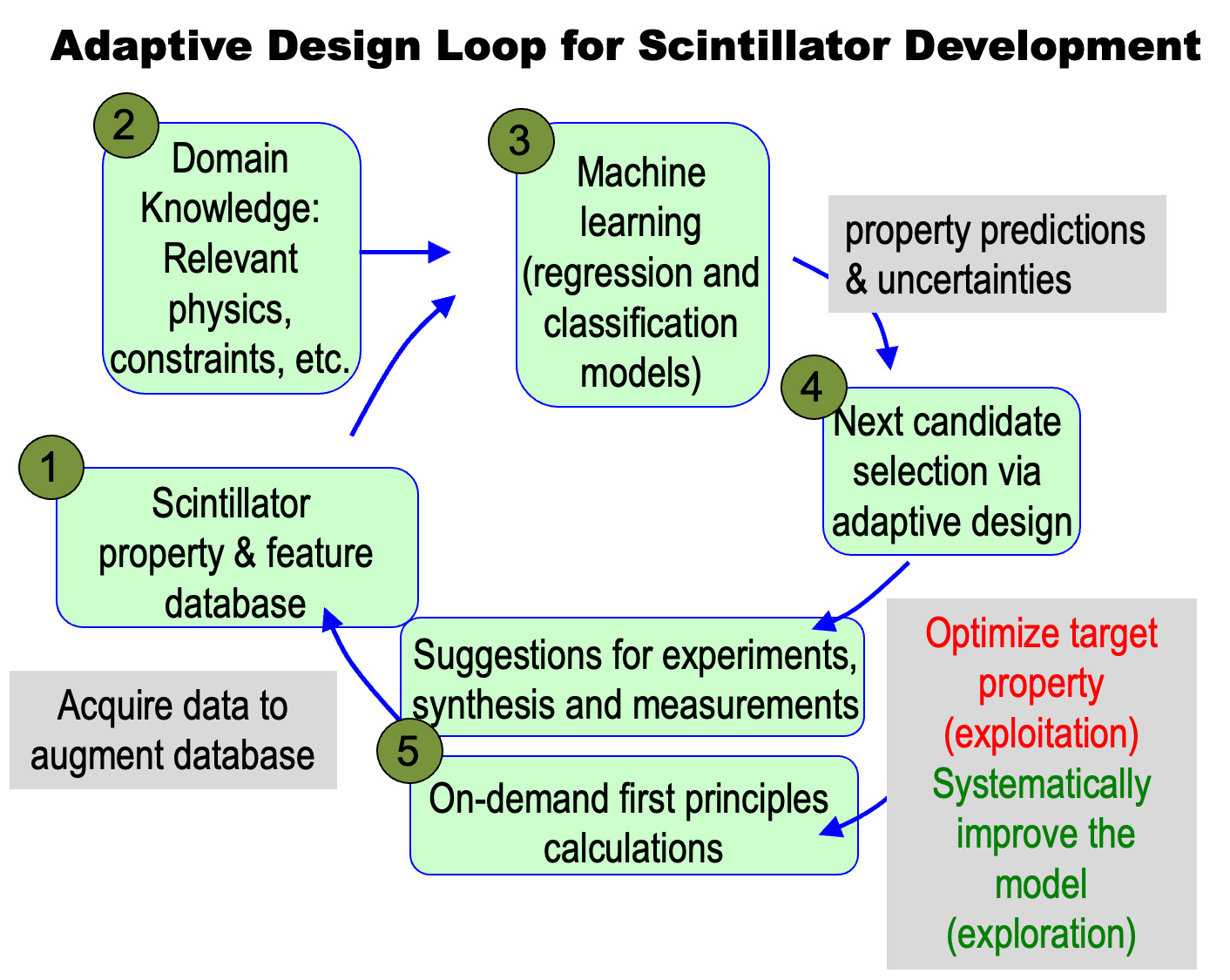}
\caption{Schematic illustration the close-loop adaptive design and statistical inference approach for targeted scintillator discovery and development (please refer to the text for details). }
\label{fig:GP4}
\end{figure}

Despite the considerable progress already made in the field of data-enabled scintillator (and related materials) design and development, many more exciting opportunities in terms of active learning and adaptive design for scintillator discovery and optimization remain largely unexplored.  Figure~\ref{fig:GP4} illustrates the essence of a closed-loop adaptive design approach~\cite{PR:2012,Look:2016} for expedited scintillator development in a target chemical space. This iterative feedback loop starts with the available assembled data on a set of key scintillator properties or performance metrics, which may be obtained either from accurate first principles computations or via direct measurements. Subsequently, existing materials knowledge in combination with advanced descriptor/feature selection tools can be employed to identify a set of physically meaningful and easily accessible descriptors for a targeted property. As a next step, an initial set of accumulated data is used to train a statistical inference model which estimates the property with associated uncertainties. A key aspect of the design loop is the uncertainty associated with the properties predicted from inference, which is often accessed through bootstrapping or other suitable model-specific routes such as Gaussian process regression~\cite{FHT:2001}. The uncertainties on the target properties play a key role in the adaptive experimental design which suggests the next material to be chosen for further calculation or experiments by balancing the tradeoffs between exploration and exploitation. That is, at any given stage several samples may be predicted to have given properties along with the associated uncertainties. The tradeoff is between exploiting the results by choosing to perform the next computation on the material predicted to have the optimal property or further improving the model by performing the calculation on a material where the predictions have the largest uncertainties. By choosing the latter, the uncertainty in the property is expected to (given the learning model) decrease, model will improve (and its domain of applicability will expand) and this will influence the results of the next iteration in the loop (i.e., exploration)~\cite{FHT:2001}. The new compounds proposed by the adaptive design strategy are synthesized, characterized and the new data is used to augment the training database. The loop repeats until one has identified a few materials, exploiting the trained models, that have the necessary performance and can serve as the starting point for further applied development or optimization. Note that a similar strategy can also be used during the optimization stages to further fine-tune a newly identified scintillator chemistry for a give application. 

In addition to the active learning and adaptive design, numerous other emerging opportunities in the quickly growing field of materials informatics and machine learning are expected to significantly change the ways in which functional materials’ discovery and development is going to be pursued. Going forward, increasingly efficient and improved ML methods integrated with advanced data infrastructure, automated and autonomous robotics for high throughput experimentations, generative design of materials with targeted properties and natural language processing for automated extraction of relevant information from text and over the web, are going to further push the boundaries of what is possible today with data-enabled routes for expedited development of novel luminescent materials.

\subsection{Space applications \label{sec:man}}
The  use of inorganic scintillators in space has a long history now spanning more than 60 years~\cite{NASA:1965}. Some of the familiar scintillators used in laboratory such as BGO, CeBr$_3$, CLLB:Ce, CLYC:Ce, CsI:Tl, GAGG:Ce, GSO:Ce, LaBr$_3$:Ce, NaI:Tl, PWO, SrI$_2$:Eu, YAP:Ce, and YSO:Ce have found their ways into space applications. Amongst the hundreds of missions, scintillator choice has been and continues to be influenced by many factors – e.g. particle of interest, efficiency, timing requirements, resolution, radiation background in space, radiation hardness requirements, and SWaP (Size, Weight, and Power). Mission- and measurement-specific requirements still greatly influence novel scintillator designs, R\&D, and implementation due to the lack of a `universal scintillator for all space missions'.

A few space applications of scintillators have been reported during the SCINT22 conference. NASA has recently sponsored ultrafast microcolumnar films development for high-speed x-ray imaging [Singh, in Table.~\ref{scint:tab}]. CeBr\textsubscript{3-x}I\textsubscript{x} is known to have high light outputs around 80 kph/MeV, an emission peak at 500 nm, a decay time of 36 ns, and virtually no afterglow. The spatial resolution up to 12 lp/mm was reported, depending on the CeBr\textsubscript{3-x}I\textsubscript{x}  film thickness. 
GAGG  scintillator is being considered  for the Flash Gamma-ray Spectrometer in future CNES-funded missions [Pailot, in Table.~\ref{scint:tab}]. With a target energy resolution of 10\% at 662 keV, the effects of wrapping and surface roughness on energy resolution are being investigated. A trade-off between the decay time and light yield may not be avoidable among different manufacturers’ options. 
Further developments for particle identification (ID) and discrimination are also noted. A phoswich detector combining a LaCl\textsubscript{3}+LiI:(Eu,Sr) and pure LaCl\textsubscript{3} crystals for discriminating thermal neutron, fast neutrons, from $\gamma$-rays, is under study [Sonu, in Table.~\ref{scint:tab}]. Another application is described for the quintuple discrimination of $\alpha$-particles, $\beta$s, $\gamma$s, thermal neutrons, and fast neutrons [Bertrand, in Table.~\ref{scint:tab}]. The composite organic scintillator detector is a three-layer phoswich. In still another application, novel oxide scintillators such Na$_2$W$_2$O$_7$ are investigated for dark matter search [Pandey, in Table.~\ref{scint:tab}]. 

\subsection{Multi-modal RadIT \label{sec:sky}}
Most RadIT methods summarized in Sec.~\ref{sec:css} can be characterized as a `single-modal' RadIT method in the sense that a monochromatic X-ray, $\gamma$ ray or a mono-energetic neutron, electron, proton is used or assumed as the source of illumination, and a single parameter such as the corresponding scintillator light intensity is measured and interpreted as signals. This simplified and idealized treatment of RadIT can be overly stringent for a number of reasons. From the illumination source perspective, a truly monochromatic X-rays or particle source is difficult to obtain experimentally. Very narrow energy bandwidth (0.1\% or so) X-ray free electron lasers are indeed available~\cite{PS:xx}. A synchrotron light source can also produce nearly monochromatic X-rays using Bragg crystals with a significant reduction in intensity. However, the number of such X-ray and particle facilities still limits the user access, and most of the RadIT measurements are carried out using broad band (`white' or `pink') X-ray or particle beams. The examples given above show that in practice, monochromatic photons or monoenergetic particles may not be necessary for many applications. From the X-ray (or particle) material interaction perspective, the X-ray (or particle) can change color (or energy) due to inelastic interactions or secondary particle production, {\it esp.} for MeV and higher energy photons (relativistic particle beams); therefore, the scintillator and photodetector collect both the transmitted primary beam signals as well as inelastically scattered and secondary particle signals that reach the scintillator and detector. From the measurement and signal processing perspective, if any scintillator and detector response other than the `ideal' signals are treated as the noise and have to be rejected, it is not only very challenging in signal discrimination, but also unncessarily information lossy and wasteful. In short, multi-modal IT that overcomes the above mentioned limitations of single-modal IT is an emerging branch of RadIT.

Multi-modal RadIT comes in different flavors. From the radiation source perspective, X-ray photons and particles with mass can be used together for multi-source RadIT. Two or more color (energy) X-rays (particles) have already been used as dual- or multi- color (energy) RadIT. From the signal collection perspective, light intensity (or particle counts) can be used with conjunction with energy, momentum, polarization and other light (particle) properties for multi-messenger RadIT. From the signal processing perspective, signals in real space can be augmented by signal collection or mathematical transformations to frequency for multi-dimensional RadIT. One example is the X-ray ptychography. 

\begin{figure}
\centering
\begin{tabular}{cc}
\begin{minipage}{0.22\textwidth}
\includegraphics[width=\textwidth]{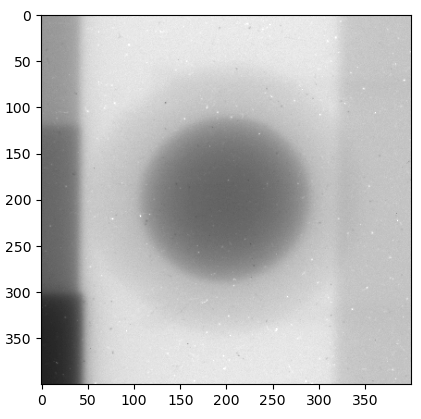}
\end{minipage}
&
\begin{minipage}{0.22\textwidth}
\includegraphics[width=\textwidth]{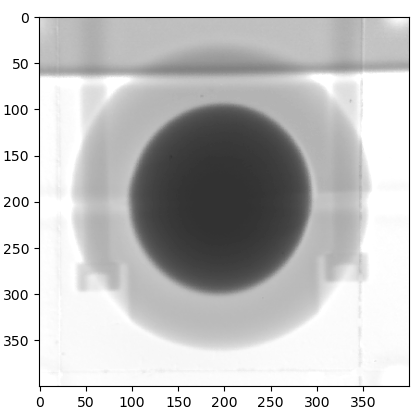} 
\end{minipage}
\end{tabular}
\caption{\label{fig:iqicomp} IQI images from the Microtron (left), LANSCE proton radiography (right).}
\end{figure}

A recent example by combining X-ray and proton radiographic methods is shown in Figure~\ref{fig:iqicomp}.  A static image quality indicator (IQI) was fielded at two LANL facilities.  Two images of the same IQI are shown in Figure~\ref{fig:iqicomp}.

\begin{figure}
\centering
\includegraphics[width=0.5\textwidth]{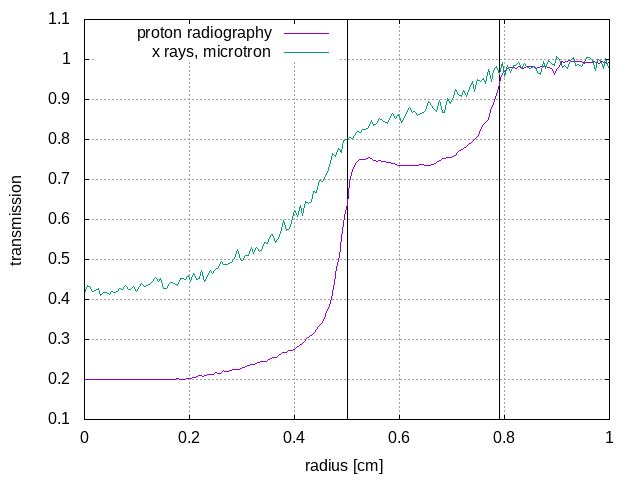}
\caption{\label{fig:iqicomp2} A comparison of the transmission intensities in Fig.~\ref{fig:iqicomp} based on their central columns.}
\end{figure}

The left image in the Figure~\ref{fig:iqicomp} was taken with a 20 MeV endpoint bremsstrahlung X-ray spectrum, converted into light with a 20 mm-thick LYSO scintillator.  It is at slightly lower magnification than the proton radiography image, taken using 800 MeV kinetic energy proton beam and a magnetic imaging system at the LANSCE~\cite{Mo:2}.  The scintillator was a 3$\times$2 tiled array of monolithic LSO crystals 1.9 mm thick~\cite{Mo:3}. Both images include the same static IQI, which includes a 1 cm-diameter sphere of gold inside a 1.58 cm outer diameter spherical shell if Ti-6Al-4V alloy \cite{iqidoc}.  The proton radiography image also includes an energy matching upstream object and 10 mrad collimation in the X3 magnifying lens.

Both images feature the same mounting apparatus.  In the proton image it is easy to see the screws holding together the hemispheres of Ti-6Al-4V alloy enclosing the gold sphere, with better signal to noise and more definitive discrimination between the two materials.  The transmission is higher in the X-ray image, with dynamic range remaining to measure larger or denser versions of this geometry, shown by comparing the transmission from central columns of both images corrected to the same magnification.


One of the primary drivers for different multi-modal RadIT approaches is to extract more information from an object. In addition to radiation source development such as light (particle) accelerators, broadband multi-species scintillator detectors are poised to grow together with the new scintillator discoveries and innovative uses of existing scintillators and structures towards better intensity linearity, noise reduction, energy resolution, and particle discrimination capabilities.

\section{Summary and Conclusion \label{sec:conc}}
Started by R\"ontgen and other pioneers at the dawn of 20th century, the interdisciplinary field of RadIT science and technology is now more than 100 years old (young). Scintillators played pivotal roles since the very beginning when the human eyes were the best photodetectors and continue to be enabling for RadIT. In addition to absorption-based X-ray radiography, there are many other RadIT forms or modalities such as phase contrast X-ray imaging, coherent X-ray diffractive imaging, high-energy X- and $\gamma-$ray radiography at above 1 MeV, X-ray computed tomography (CT), proton imaging and tomography (IT), neutron IT, positron emission tomography (PET), high-energy electron radiography, muon tomography, {\it etc}. The coexistence of many RadIT modalities opens door to multimodal RadIT.

More than 160 kinds of scintillators and applications are presented during the SCINT22 conference, as summarized in Table~\ref{scint:tab}. Recent work includes inorganic and organic scintillator composites or heterostructures, liquid phase synthesis of perovskites and single-crystal micrometer-thick films, use of multi-physics models and lately data science to guide scintillator development, structural innovations such as photonic crystals, nanostructured scintillators enhanced by the Purcell effect, novel use of existing scintillators through heterostructural innovations (fibers), and multilayer configurations.

Scintillator metrics such as light yield, decay time are discussed in light of RadIT metrics. RadIT, both photon and particle based, continue to aim for finer spatial, better temporal resolution, the highest possible efficiency in conjunction with advances in high luminosity X-ray and particle sources, photondetectors, and efficient algorithms for data processing. While X-ray and charged particle IT necessarily require faster, brighter scintillators, and the concerns with radiation damage are growing,  neutron IT on the other hand is currently limited by the neutron source intensity; and therefore, high efficiency scintillators with good spatial, energy resolution would be required for neutrons.  The scintillator requirements in RadIT overlap significantly with other applications such as in HEP. For example, the calorimeter applications at FCC at CERN, or CEPC in China will not only need excellent energy resolution, new scintillator functions include fast and precision timing (ps, driven by high signal rate above 10$^{34}$ cm$^{-2}$/s and corresponding high data rate and data sets above 300 fb$^{-1}$), outstanding radiation tolerance and finer
granularity or spatial resolution of the active elements. Dark matter search usually requires large volume and surface area of scintillators, which overlaps with requirements of RadIT in higher detection efficiencies, large field of view, and up to 4$\pi$ solid angle signal coverage.

Since there is no universal scintillator that can fit all needs or the bill, tradeoffs between, for example, cost and performance, spatial resolution and efficiency, light yield and decay time, are often necessary. Optimizing a scintillator for a specific application appears to be the next best option. Scintillator optimization can become a part of the `global' optimization strategy in RadIT applications, which include cradle-to-grave analysis of an ionizing photon or particle. In addition to a growing number of successful empirical approaches, a new approach is optimization through data science, and for the maximal information yield. For many years, the
discovery and design of new scintillator materials rely on a laborious, time-consuming, trial-and-error
approaches, yielding little physical insight sometimes, and leaving a vast space of potentially revolutionary
materials to be explored. A closed-loop machine-learning-driven
adaptive design framework based on data from literature, in-house experiments and first-principles (quantum
mechanical) calculations have recently been demonstrated for fast screening of perovsikites, garnets and elpasolites. There is no difficulty extending such a framework to, for example, high entropy scintillators, even though it is well recognized that computation can become a bottleneck.

Plenty of new opportunities exist that make RadIT and scintillator development mutually beneficial and dependent. Examples include  optimization of RadIT performance with reduced radiation dose, data-driven measurements, photon/particle counting and tracking methods supplementing time-integrated measurements, multimodal RadIT, and novel applications of RadIT for scintillator discovery.




\appendices

\section*{Appendix I: Scintillator list for SCINT22}
Table~\ref{scint:tab} summarizes different scintillators presented during the SCINT22 conference and their applications.

\clearpage
\onecolumn

\begingroup\footnotesize 

\begin{center}
\begin{longtable}{|l|c|c|c|c|c|c|}
\caption{\normalsize A summary of scintillators and intended applications based on the presentations during the SCINT22 conference. The scintillator metrics include light yield (LY), linearity/nonlinearity of the light yield as a function of energy deposition (NL), energy resolution ($\sigma_E$), single photon sensitivity (SP), spatial resolution ($\delta$), spectrum tuning capabilities (CL),  light emission anisotropy (EA), high density ($\rho$), high {\it effective} atomic number ($Z_{eff}$), efficiency of absorption, attenuation (Eff.), radiation hardness (RH), decay time ($\tau_d$), neutron/$\gamma$-ray discrimination (n/$\gamma$),
stability to environment (SE) such as temperature, humidity, environmental radiation,  adequacy for dose measurement (DS),  size (SZ) and cost. } \\

\hline
\multicolumn{1}{|c|}{\textbf{Composition}} & \multicolumn{1}{c|}{\textbf{Popular Name}} & \multicolumn{1}{c|}{\textbf{Material Phase}} &
\multicolumn{1}{|c|}{\textbf{Metric}} & \multicolumn{1}{|c|}{\textbf{Application}} & \multicolumn{1}{|c|}{\textbf{Institution}}  & \multicolumn{1}{|c|}{\textbf{Note}} \\ 

            &&&&& [First Author] & \\
         \hline
         A\textsubscript{2}XY\textsubscript{4}~\textsuperscript{\textit{av}} & & Perovskite & LY,  &  & LRS-PORT  &  \\
         {\it also} :Li doping  & & crystal~\textsuperscript{\textit{av}} & $\tau_d$& & [M. Sheikh] & \\
         \hline
            BaF\textsubscript{2} &  & crystal & $\sigma_E$, $\tau_d$ \textsuperscript{\textit{a}}  & HEP & CERN & Ref.~\textsuperscript{\textit{b}} \\
            &&&&&[R. Cala']& \\
	    &&&&	HEP/cal., &  Caltech [C. Hu] & Ref.~\textsuperscript{\textit{b}} \\
\end{longtable}

\end{center}

\endgroup

\clearpage
\twocolumn


%



\section*{Acknowledgment}

ZW wishes to thank the members of the local organizing committee, the international advisory committee, the session chairs, the Los Alamos National Laboratory (LANL) conference team, and many others (of course, all the presenters, sponsors, in person or online) for helping with the 16th International Conference on Scintillating Materials \& their Applications (SCINT22), September 19-23, 2022, Santa Fe, NM, USA; {\it esp.} Drs./Ms./Mr. Etiennette Auffray (CERN), Zane Bell (IEEE TNS), Gregory Bizarri (Cranfield University), Edith Bourret (Lawrence Berkeley National Lab.), Bruce Chai (Crystal Photonics, Inc.), Pieter Dorenbos (Delft University of Technology), Alex Gektin (National Academy of Sciences of Ukraine), Sarah Haag (LANL),  Luiz Jacobsohn (Clemson University), Kai Kamada (Tohoku University), Alexei Klimenko (LANL), Eva Mih\'okov\'a (Institute of Physics, Czech Academy of Sciences), Ross Muenchausen (LANL), Wanyi Nie (LANL), Bob Reinovsky (LANL), Rich Sheffield (LANL), Sergei Tretiak (LANL), Blas Uberuaga (LANL), Anna Vedda (University of Milano, Bicocca), Irena Villa (Institute of physics, the Czech Academy of Sciences, FZU),  Sven Vogel (LANL), 
Weronika Wolszczak (Lawrence Berkeley National Lab), Craig Woody (Brookhaven National Lab), Dmitry Yarotski (LANL), Liyuan Zhang (Caltech), Ren-yuan Zhu (Caltech), and Alex Zubelewicz (LANL). GP would like to acknowledge discussions with  Drs. K. J. McClellan, C. R. Stanek, and B. P. Uberuaga at LANL and funding support from LANL's Laboratory Directed Research and Development (LDRD) program  (award \#20190043DR). RP acknowledges support from Dynamic Materials Properties Campaign under Department of Energy National Nuclear Security Administration (DOE-NNSA). UTK work (DJR, MZ, CLM) is supported by the DOE-NNSA through the Nuclear Science and Security Consortium under Award Numbers DE-NA- 0003180 and DE-NA-0003996. DJR is supported under a DOE, Office of Nuclear Energy, Integrated University Program Graduate Fellowship. AST was partially supported by the U.S.
Department of Energy/NNSA/DNN R\&D and by Lawrence Berkeley National Laboratory under Contract No. AC02-05CH11231.

\end{document}